\documentclass[sigconf,natbib=false,balance=false]{acmart}
\usepackage{popets}

\usepackage{wasysym}
\usepackage{booktabs}
\usepackage{tabularx}
\usepackage{xspace}
\usepackage{multirow}

\usepackage[sorting=anyt,%
                giveninits=true,%
                maxbibnames=99,%
                citestyle=numeric-comp,%
                natbib=true,%
                backend=biber,
                mincrossrefs=1000]{biblatex}
\AtEveryBibitem{
    \clearfield{editor}
    \clearfield{pages}
    \ifentrytype{book}{
    }{
      \ifentrytype{inbook}{}{
        \clearlist{publisher}
        \clearname{editor}
        \clearfield{isbn}
        \clearfield{issn}
    }}
    \ifentrytype{inproceedings}{
      \clearfield{series}
      \clearfield{volume}
      \clearfield{edition}
      \clearfield{doi}
      \clearfield{isbn}
      \clearfield{url}
      \clearfield{pages}
      \clearfield{month}
    }{}
    \clearlist{location}
    \clearlist{address}
}

\usepackage{popets}
\setcopyright{popets}
\copyrightyear{2023}
\acmYear{2023}
\acmVolume{2023}
\acmNumber{2}
\acmDOI{XXXXXXX.XXXXXXX}
\acmISBN{}
\acmConference{Proceedings on Privacy Enhancing Technologies}
\usepackage[plain]{fancyref}
\usepackage[inline]{enumitem}
\usepackage[prologue,table]{xcolor}

\newcommand{\lastdate}[0]{October 2022\xspace}
\newcommand{\lastdatefull}[0]{Oct~31,~2022\xspace}

\newcommand{\CSBTTFMAXABSeduAMAZONGOOGLEMICROSOFT}[0]{227\xspace}
\newcommand{\CSBTTFMAXPERCeduAMAZONGOOGLEMICROSOFT}[0]{87.31\%\xspace}
\newcommand{\CSBTTFMAXABSukGOOGLEonly}[0]{62\xspace}
\newcommand{\CSBTTFMAXPERCukGOOGLEonly}[0]{53.91\%\xspace}
\newcommand{\CSBTTFMAXABSnlGOOGLEonly}[0]{10\xspace}
\newcommand{\CSBTTFMAXPERCnlGOOGLEonly}[0]{52.63\%\xspace}
\newcommand{\CSBTTFMAXABSdeSUM}[0]{40\xspace}
\newcommand{\CSBTTFMAXPERCdeSUM}[0]{49.38\%\xspace}
\newcommand{\CSBTTFMAXABSfrSUM}[0]{32\xspace}
\newcommand{\CSBTTFMAXPERCfrSUM}[0]{43.24\%\xspace}
\newcommand{\CSBTTFMAXABSatSUM}[0]{20\xspace}
\newcommand{\CSBTTFMAXPERCatSUM}[0]{58.82\%\xspace}
\newcommand{\CSBTTFMAXABSchSUM}[0]{12\xspace}
\newcommand{\CSBTTFMAXPERCchSUM}[0]{85.71\%\xspace}

\newcommand{\MAILMAXABSatsumonly}[0]{9\xspace}
\newcommand{\MAILMAXPERCatsumonly}[0]{26.47\%\xspace}

\newcommand{\MAILMAXABSfrsumonly}[0]{7\xspace}
\newcommand{\MAILMAXPERCfrsumonly}[0]{9.46\%\xspace}
\newcommand{\MAILMAXABSnlsumonly}[0]{13\xspace}
\newcommand{\MAILMAXPERCnlsumonly}[0]{68.42\%\xspace}
\newcommand{\MAILMAXABSthesumonly}[0]{59\xspace}
\newcommand{\MAILMAXPERCthesumonly}[0]{59.00\%\xspace}
\newcommand{\MAILMAXABSuksumonly}[0]{81\xspace}
\newcommand{\MAILMAXPERCuksumonly}[0]{70.43\%\xspace}
\newcommand{\MAILMAXABSchsumonly}[0]{3\xspace}
\newcommand{\MAILMAXPERCchsumonly}[0]{21.43\%\xspace}
\newcommand{\MAILMAXABSdesumonly}[0]{2\xspace}
\newcommand{\MAILMAXPERCdesumonly}[0]{2.47\%\xspace}

\newcommand{\MAILMAXABSedusumBIGonly}[0]{220\xspace}
\newcommand{\MAILMAXPERCedusumBIGonly}[0]{84.62\%\xspace}

\newcommand{\MAILMAXPERCukother}[0]{11.30\%\xspace}

\newcommand{\MAILMAXABStheProofpointAppliance}[0]{11\xspace}
\newcommand{\MAILMAXPERCtheProofpointAppliance}[0]{11.00\%\xspace}
\newcommand{\MAILMAXABStheProofpointHosted}[0]{12\xspace}
\newcommand{\MAILMAXPERCtheProofpointHosted}[0]{12.00\%\xspace}

\newcommand{\MAILMAXABSeduProofpointAppliance}[0]{22\xspace}
\newcommand{\MAILMAXPERCeduProofpointAppliance}[0]{8.46\%\xspace}
\newcommand{\MAILMAXABSeduProofpointHosted}[0]{30\xspace}
\newcommand{\MAILMAXPERCeduProofpointHosted}[0]{11.54\%\xspace}

\newcommand{\LMSMAXABSedusumBIGonly}[0]{196\xspace}
\newcommand{\LMSMAXPERCedusumBIGonly}[0]{75.38\%\xspace}
\newcommand{\LMSMAXABSuksumBIGonly}[0]{64\xspace}
\newcommand{\LMSMAXPERCuksumBIGonly}[0]{55.65\%\xspace}
\newcommand{\LMSMAXABSnlsumBIGonly}[0]{13\xspace}
\newcommand{\LMSMAXPERCnlsumBIGonly}[0]{68.42\%\xspace}
\newcommand{\LMSMAXABSthesumBIGonly}[0]{62\xspace}
\newcommand{\LMSMAXPERCthesumBIGonly}[0]{62.00\%\xspace}

\newcommand{\LMSMAXtheUS}[0]{37\xspace}
\newcommand{\LMSMAXtheUK}[0]{6\xspace}
\newcommand{\LMSMAXtheCA}[0]{3\xspace}
\newcommand{\LMSMAXtheAU}[0]{4\xspace}
\newcommand{\LMSMAXtheHK}[0]{2\xspace}
\newcommand{\LMSMAXtheSG}[0]{2\xspace}
\newcommand{\LMSMAXtheNL}[0]{6\xspace}
\newcommand{\LMSMAXtheSE}[0]{1\xspace}
\newcommand{\LMSMAXtheBE}[0]{1\xspace}
\newcommand{\LMSMAXtheSUM}[0]{25\xspace}

\newcommand{\ZOOMMAXABSnlSfBonly}[0]{12\xspace}
\newcommand{\ZOOMMAXPERCnlSfBonly}[0]{63.16\%\xspace}
\newcommand{\ZOOMMAXABSeduSfBonly}[0]{199\xspace}
\newcommand{\ZOOMMAXPERCeduSfBonly}[0]{76.54\%\xspace}
\newcommand{\ZOOMMAXABSukSfBonly}[0]{62\xspace}
\newcommand{\ZOOMMAXPERCukSfBonly}[0]{53.91\%\xspace}
\newcommand{\ZOOMMAXABStheSfBonly}[0]{71\xspace}
\newcommand{\ZOOMMAXPERCtheSfBonly}[0]{71.00\%\xspace}
\newcommand{\ZOOMMAXABSeduZoomonly}[0]{212\xspace}
\newcommand{\ZOOMMAXPERCeduZoomonly}[0]{81.54\%\xspace}
\newcommand{\ZOOMMAXABSeduWebexonly}[0]{71\xspace}
\newcommand{\ZOOMMAXPERCeduWebexonly}[0]{27.31\%\xspace}
\newcommand{\ZOOMMAXABSeduAdobeonly}[0]{130\xspace}
\newcommand{\ZOOMMAXPERCeduAdobeonly}[0]{50.00\%\xspace}
\newcommand{\ZOOMMAXABSedudiffPlatTWOonly}[0]{97\xspace}
\newcommand{\ZOOMMAXPERCedudiffPlatTWOonly}[0]{37.31\%\xspace}
\newcommand{\ZOOMMAXABSedudiffPlatTHREEonly}[0]{84\xspace}
\newcommand{\ZOOMMAXPERCedudiffPlatTHREEonly}[0]{32.31\%\xspace}
\newcommand{\ZOOMMAXABSedudiffPlatFOURonly}[0]{32\xspace}
\newcommand{\ZOOMMAXPERCedudiffPlatFOURonly}[0]{12.31\%\xspace}
\newcommand{\ZOOMMAXABStheZoomonly}[0]{79\xspace}
\newcommand{\ZOOMMAXPERCtheZoomonly}[0]{79.00\%\xspace}
\newcommand{\ZOOMMAXABSthediffPlatTWOonly}[0]{38\xspace}

\newcommand{\ZOOMMAXABSthediffPlatTHREEonly}[0]{29\xspace}

\newcommand{\ZOOMMAXABSthediffPlatFOURonly}[0]{12\xspace}

\newcommand{\ZOOMMAXABSatZoomonly}[0]{14\xspace}
\newcommand{\ZOOMMAXPERCatZoomonly}[0]{41.18\%\xspace}
\newcommand{\ZOOMMAXABSfrZoomonly}[0]{31\xspace}
\newcommand{\ZOOMMAXPERCfrZoomonly}[0]{41.89\%\xspace}
\newcommand{\ZOOMMAXABSnlZoomonly}[0]{13\xspace}
\newcommand{\ZOOMMAXPERCnlZoomonly}[0]{68.42\%\xspace}
\newcommand{\ZOOMMAXABSukZoomonly}[0]{51\xspace}
\newcommand{\ZOOMMAXPERCukZoomonly}[0]{44.35\%\xspace}
\newcommand{\ZOOMMAXABSchZoomonly}[0]{11\xspace}
\newcommand{\ZOOMMAXPERCchZoomonly}[0]{78.57\%\xspace}
\newcommand{\ZOOMMAXABSdeZoomonly}[0]{49\xspace}
\newcommand{\ZOOMMAXPERCdeZoomonly}[0]{60.49\%\xspace}

\newcommand{\ZOOMMAXABSatWebexonly}[0]{9\xspace}
\newcommand{\ZOOMMAXPERCatWebexonly}[0]{26.47\%\xspace}
\newcommand{\ZOOMMAXABSfrWebexonly}[0]{7\xspace}
\newcommand{\ZOOMMAXPERCfrWebexonly}[0]{9.46\%\xspace}
\newcommand{\ZOOMMAXABSnlWebexonly}[0]{2\xspace}
\newcommand{\ZOOMMAXPERCnlWebexonly}[0]{10.53\%\xspace}
\newcommand{\ZOOMMAXABSchWebexonly}[0]{5\xspace}
\newcommand{\ZOOMMAXPERCchWebexonly}[0]{35.71\%\xspace}
\newcommand{\ZOOMMAXABSdeWebexonly}[0]{32\xspace}
\newcommand{\ZOOMMAXPERCdeWebexonly}[0]{39.51\%\xspace}
\newcommand{\ZOOMMAXABSukWebexonly}[0]{19\xspace}
\newcommand{\ZOOMMAXPERCukWebexonly}[0]{16.52\%\xspace}

\newcommand{\ZOOMMAXABSatmso}[0]{2\xspace}
\newcommand{\ZOOMMAXPERCatmso}[0]{5.88\%\xspace}
\newcommand{\ZOOMMAXABSedumso}[0]{4\xspace}
\newcommand{\ZOOMMAXPERCedumso}[0]{1.54\%\xspace}
\newcommand{\ZOOMMAXABSfrmso}[0]{13\xspace}
\newcommand{\ZOOMMAXPERCfrmso}[0]{17.57\%\xspace}
\newcommand{\ZOOMMAXABSnlmso}[0]{2\xspace}
\newcommand{\ZOOMMAXPERCnlmso}[0]{10.53\%\xspace}
\newcommand{\ZOOMMAXABSthemso}[0]{8\xspace}
\newcommand{\ZOOMMAXPERCthemso}[0]{8.00\%\xspace}
\newcommand{\ZOOMMAXABSukmso}[0]{10\xspace}
\newcommand{\ZOOMMAXPERCukmso}[0]{8.70\%\xspace}
\newcommand{\ZOOMMAXABSchmso}[0]{0\xspace}

\newcommand{\ZOOMMAXABSdemso}[0]{3\xspace}
\newcommand{\ZOOMMAXPERCdemso}[0]{3.70\%\xspace}

\newcommand{\ZOOMMAXABSeduZoomDecInc}[0]{63\xspace}
\newcommand{\ZOOMMAXABSeduWebexDecInc}[0]{24\xspace}

\newcommand{\STUDIPMAXABSedumoodle}[0]{146\xspace}
\newcommand{\STUDIPMAXPERCedumoodle}[0]{56.15\%\xspace}

\newcommand{\STUDIPMAXABSdesumonly}[0]{65\xspace}
\newcommand{\STUDIPMAXPERCdesumonly}[0]{80.25\%\xspace}

\newcommand{\BBBMAXABSdebbb}[0]{58\xspace}
\newcommand{\BBBMAXPERCdebbb}[0]{71.60\%\xspace}

\PassOptionsToPackage{hyphens}{url}

\newcommand*{\fancyrefapplabelprefix}{app}
\Frefformat{plain}{\fancyrefapplabelprefix}{\MakeUppercase{\Freflstname}\fancyrefdefaultspacing#1#2}
\Frefformat{plain}{\fancyrefapplabelprefix}{Appendix~#1}

\newcommand{\mypar}[1]{\smallskip\noindent\textbf{#1}\xspace}

\newcommand{\us}[0]{U.S.\xspace}
\newcommand{\uk}[0]{U.K.\xspace}

\author{Tobias Fiebig}
\email{tfiebig@mpi-inf.mpg.de}
\affiliation{%
  \institution{Max-Planck-Institut für Informatik}
  \country{Germany}
  \city{Saarbrücken}
}
\author{Seda G{\"u}rses}
\email{F.S.Gurses@tudelft.nl}
\affiliation{%
  \institution{TU Delft}
  \country{The Netherlands}
  \city{Delft}
}
\author{Carlos H. Ga\~n\'an}
\email{C.HernandezGanan@tudelft.nl}
\affiliation{%
  \institution{TU Delft}
  \country{The Netherlands}
  \city{Delft}
}
\author{Erna Kotkamp}
\email{E.Kotkamp@tudelft.nl}
\affiliation{%
  \institution{TU Delft}
  \country{The Netherlands}
  \city{Delft}
}
\author{Fernando Kuipers}
\email{F.A.Kuipers@tudelft.nl}
\affiliation{%
  \institution{TU Delft}
  \country{The Netherlands}
  \city{Delft}
}
\author{Martina Lindorfer}
\email{martina.lindorfer@tuwien.ac.at}
\affiliation{%
  \institution{TU Wien}
  \country{Austria}
  \city{Wien}
}
\author{Menghua Prisse}
\email{M.M.G.C.Prisse@student.tudelft.nl}
\affiliation{%
  \institution{TU Delft}
  \city{Delft}
  \country{The Netherlands}
}
\author{Taritha Sari}
\email{taritha@cloudheads.net}
\affiliation{%
  \institution{No Affiliation}
  \country{The Netherlands}
}

\renewcommand{\shortauthors}{Fiebig et al.}

\keywords{Platformization, Platform Capitalism, Centralization, EdTech}

\begin{document}
\title[Heads in the Clouds? Measuring Universities' Migration to Public Clouds]{Heads in the Clouds? Measuring Universities' Migration to Public Clouds: Implications for Privacy \& Academic Freedom}

\begin{abstract}
With the emergence of remote education and work in universities due to
COVID-19, the `zoomification' of higher education, i.e., the migration of
universities to the clouds, reached the public discourse.  Ongoing discussions
reason about how this shift \emph{will} take control over students' data
\emph{away} from universities, and may ultimately harm the privacy of
researchers and students alike.  However, there has been no comprehensive
measurement of universities' use of public clouds and reliance on
Software-as-a-Service offerings to assess how far this migration has already
progressed.

We perform a longitudinal study of the migration to public clouds among
universities in the \us and Europe, as well as institutions listed in the Times
Higher Education (THE) Top100 between January 2015 and \lastdate.  We find that
cloud adoption differs between countries, with one cluster (Germany, France,
Austria, Switzerland) showing a limited move to clouds, while the other (\us,
\uk, the Netherlands, THE Top100) frequently outsources universities' core
functions and services---starting long before the COVID-19 pandemic.  We
attribute this clustering to several socio-economic factors in the respective
countries, including the general culture of higher education and the
administrative paradigm taken towards running universities. We then analyze and
interpret our results, finding that the implications reach beyond individuals'
privacy towards questions of academic independence and integrity.
\end{abstract}

\maketitle

\section{Introduction}
\label{sec:introduction}

Over the past decade, we have seen a shift in IT operations towards the use of
cloud infrastructures~\cite{mell2011nist,surbiryala2019cloud}.  Instead of
running IT services with on-site teams and on infrastructure owned by
organizations, services are now often deployed on public cloud infrastructure.
Especially for web services, the model of using Software-as-a-Service (SaaS)
has become prominent.  However, this operational paradigm shift also leads to a
change in control.  While, before, user data would remain on infrastructure
controlled by an organization, this data is now stored and processed by an
external operator.

For universities using cloud infrastructures, this leads to hard challenges,
stretching from limiting their ability to audit or implement privacy-by-design,
e.g., privacy guarantees ensured through technical means, or ensure
privacy-as-compliance, e.g., in terms of following privacy
regulations~\cite{dabrowski2019measuring,voigt2017eu}, to impacting a
university's ability to obtain meaningful informed consent when they employ
cloud operators.  Over the past year, for example, much debate surrounded the
use of Zoom as the now de facto standard for remote lectures.  Zoom only
started to systematically attend to privacy and security concerns raised by
educational institutions when pressure was handed down to the company from
investors~\cite{zoom-investors}.  Still, universities that adopted Zoom for
their lectures practically reduced students' consent choices to either using
Zoom, and having their personal data processed by Zoom, or not participating in
lectures.

The infrastructural and data control acquired by companies like Zoom have a
knock-on effect on academic freedom.  In 2020, Zoom ultimately prevented
faculty and students at New York University from conducting a guest lecture --
incidentally on censorship by Zoom and other tech companies -- using their Zoom
license~\cite{nyu-zoom}.  The question hence expands beyond \emph{`what private
data do universities share with cloud platforms,'} to include \emph{`in what
way can these cloud platforms use their infrastructural position and data
practices to influence academic processes in universities.'}

The adoption of educational technology (`\emph{EdTech}'), i.e., the use of
``market-facing digital technologies in education''~\cite{mirrlees2019edtech},
already prompted critical studies from the social sciences, warning about
blurring lines between public educational institutions and private corporations
as a threat to academic
self-governance~\cite{mirrlees2019edtech,Selwyn2020edtech,teras2020post,komljenovic2021rise}.
Despite these concerns, there are no comprehensive measurements of \emph{how}
reliant universities are on public cloud infrastructures.  We address this gap
by measuring cloud adoption in universities since 2015 in seven countries (the
\us, the \uk, Germany, Switzerland, Austria, the Netherlands, and France) and
in the Times Higher Education (THE) Top100.  We measure universities' hosting
on cloud platforms, their use of cloud-based email providers, cloud-based
learning management systems (LMS), and cloud-based video and lecturing tools.

We find that universities in the Netherlands, the \uk, the \us and in the THE
Top100 are significantly more prone to depend on cloud infrastructures, while
those in France and Germany rely far more on in-house services.  We attribute
these differences to a diverse set of socio-economic factors, including a
historically different understanding of \emph{what} higher education means, the
university functions (research, education, administration) the IT
infrastructure is aligned with, and the value placed on academic independence.
We further observe that universities' migration to centralized clouds
(Google/Amazon/Microsoft) does not show a clear pandemic effect as observed for
the Internet as a whole~\cite{feldmann2020lockdown}.  The notable exception are
video conferencing tools, where we see a clear uptick of adoption across the
board, \emph{except} for the \us, where especially Zoom adoption was on the
rise years before the pandemic.

\noindent
In summary, we make the following contributions:
\begin{itemize}[leftmargin=*]
\setlength\itemsep{0.25em}
	\item We are the first to map out the cloud dependence of universities in Europe, the \us, and the THE Top100, and find that it is an ongoing process predating the COVID-19 pandemic.
	\item We document and attribute \emph{differences} between countries to different paradigms for university IT and higher education.
	\item We find that data and infrastructure control have implications for privacy and beyond, also threatening academic freedom.
\end{itemize}

\section{University IT}
\label{sec:background}

Generally speaking, universities are \emph{organizations} with a \emph{purpose}
or \emph{function}, which can be supported by IT
pillars~\cite{versteeg2006business}.  Commonly, these major functions are:
\emph{Education}, \emph{Research}, and to enable these two,
\emph{Administration}~\cite{svensson2012establishing}.  While these functions
may seem intuitively discrete, they partially overlap, also in the tools and
applications used to address their needs.

Universities look towards cloud infrastructure as a way to reduce their own IT
investments, and potentially even a chance to free up and monetize assets,
e.g., IPv4 addresses~\cite{prehn2020wells,MITv4}.  While the use of specific
tools may lead the university to enter into agreements with a multitude of
companies, many of these tools themselves are hosted on one of the three
largest cloud platforms: Google Cloud, Amazon EC2, and Microsoft Azure.

\mypar{Education.}
IT infrastructure for education includes all tools that enable students to
learn.  Traditionally, this means all systems used for assessment and learning
management systems (LMS), e.g., Moodle~\cite{buchner2016moodle}.  While
educational software for remote teaching already received attention before the
COVID-19 pandemic in the context of blended learning and MOOCs (Massive Open
Online Courses), COVID-19 increased the importance of learning infrastructure,
like video chat and streaming solutions, as well as examination and proctoring
software.  In most universities, these tools are offered institution-wide as
centralized services, usually with the support of a central IT department.  In
addition, specific programs might need additional infrastructure, e.g., a
program on system and network engineering may also need dedicated server rooms
and networking labs~\cite{burgess2007master}, often offered in a decentralized
manner.

Several vendors offer cloud-based LMS, which allow universities to outsource
one of their largest systems (in terms of users).  Even though tools for
self-hosted remote lectures exist, the common perception associates remote
lecturing mainly with Zoom, and, to a lesser extent, other cloud-based
platforms like Microsoft Teams and Cisco WebEx.  Similarly, proctoring
solutions -- a concept of questionable
ethics~\cite{coghlan2020good,swauger:proctoring} -- are almost exclusively
provided as cloud-hosted services.

\mypar{Research.}
Research IT infrastructure is often more dependent on the individual needs of
researchers, and therefore tends to be decentralized.  Applications here range
from the, in our field, common experimental systems (IoT test labs, network
measurement infrastructure, and machines vulnerable to certain exploits) to IT
systems used to control a diverse set of research instruments, such as electron
microscopes or chemical processing lines.  In addition, super computing
capabilities~\cite{hager2010introduction}, data storage and open data
platforms~\cite{zuiderwijk2019sharing}, and research software that support
quantitative and qualitative methods, e.g., survey and statistics
tools~\cite{buchanan2009online} are often outsourced or centrally provided.

Cloud services can replace both types of research infrastructure.  Researchers
may use Platforms-as-a-Service for running measurement and experimental
systems, especially when using GPU-supported compute.  Furthermore,
universities may provide outsourced and cloud-hosted instances of survey and
interview platforms as a service for their researchers. Especially Amazon's
Mechanical Turk has become a common tool in human factors work, ranging from
social sciences to usable security and privacy~\cite{redmiles2019well}.

\mypar{Administration.}
The administrative function of a university entails all services and operations
needed to \emph{support} (not execute) its primary functions for education and
research.  This means budgeting and accounting tools, HR systems including
personnel management databases and applicant management systems, and also
student admissions.  Furthermore, this entails foundational services like
email, and the operation of a universities' network.  Similarly, telephony and
business communication tools -- before the pandemic tools like
Skype-for-Business (SfB), Microsoft Sharepoint, as well as Microsoft Teams and
other video chat solutions that now overlap with educational tooling --
\emph{traditionally} fall into this category.

Applications for specific use cases (hiring, student admission, finance and
accounting) are complex and highly business critical.  Hence, outsourcing
allows universities to reduce the needed local expertise to run these tools,
while shifting the responsibility in case they become inoperable.  Especially
for tasks like email or security management, cloud setups promise higher
reliability.

\section{Methodology Overview}
\label{sec:method}

We first describe our general methodology in terms of the dataset and selected
institutions.  We describe specific aspects for the individual services in the
corresponding sections (\S\ref{sec:overview} on cloud infrastructure,
\S\ref{sec:email} email, \S\ref{sec:lms} LMS, and \S\ref{sec:zoom} video
conferencing).

\subsection{Dataset}
We use Farsight's Security Information Exchange (SIE) dataset~\cite{farsight}
to measure \emph{(i)} to what extent universities depend on cloud
infrastructure, and, \emph{(ii)} how this dependency developed over time.
Farsight collects this dataset via recursive DNS resolvers of ISPs.
Collaborating ISPs can install a sensor, which sends all cache
misses~\cite{RFC1035,RFC8767}--see \Fref{tab:dns_terms} in
\Fref{app:method}--of their clients to Farsight.

Our dataset spans from January~1, 2015 to \lastdatefull in per-month slices.
Due to the nature of our dataset (see \Fref{app:method}) we focus on
determining \emph{if} an organization utilizes specific cloud resources, but
not \emph{how much} they utilize it.  We use a historic dataset of all cache
misses observed by participating DNS resolvers spanning from January~1, 2015 to
\lastdatefull in per-month slices.  A unique cache miss is defined by the tuple
of \texttt{<rrname, rrtype, bailiwick, rdata>} (see \Fref{app:method}).  As we
only receive cache misses, we cannot make statements about the
\emph{popularity} of domain names.  Therefore, we focus our analysis on
establishing a lower bound on the use of cloud resources, or, in more practical
terms, we determine \emph{if} an organization utilizes specific cloud
resources, but not \emph{how much} they utilize it.  We provide a comprehensive
description of DNS and the Farsight dataset in \Fref{app:method}.

Compared to actively collected large-scale DNS datasets, for example
OpenINTEL~\cite{hohlfeld2018operating,van2016high}, the Farsight dataset
enables us to look \emph{deeper} into the DNS tree of individual organizations.
As we see all names that were requested by clients behind DNS recursors
participating as sensors, we can see application-specific names (e.g.,
\verb+application.example.com.+) that are not part of the set of names gathered
by active measurement platforms (as they need a priori knowledge of these
names).  Specifically, these platforms request a known set of record types and
names for all domains listed in top-level-domain zone files to which they have
access~\cite{hohlfeld2018operating,van2016high}.

To illustrate this with a non-exhaustive example,  \texttt{example.com}, these
platforms will regularly request the \texttt{NS}, \texttt{MX}, \texttt{A}, and
\texttt{AAAA} record for \texttt{example.com}, as well as \texttt{A} and
\texttt{AAAA} records for \texttt{www.\-ex\-am\-ple.\-com}.  However,
\texttt{lms.students.example.com} will not be included, because the subdomain
\texttt{students.example.com} is not listed in the authoritative zone file.
Contrary, the Farsight SIE dataset contains data on
\texttt{lms.students.example.com}, if at least one client behind a sensor did
request that name during the measurement period, and data was successfully
returned.  For the limitations of this approach, please
see~\S\ref{sec:limitations}.

\mypar{Ethical Considerations.}
To not collect personally identifiable information (PII),  the Farsight passive
DNS dataset consists only of cache misses found at recursive DNS servers, and
does neither list the recursive resolver a record was seen on, nor the client
that requested it~\cite{farsight}.  Furthermore, we only process DNS entries
under universities' domains and under domains of major cloud services
(\verb+zoom.us+ etc.).  We followed established best practices for handling
passive datasets, as outlined by Allman et al.~\cite{allman2007issues}.  In our
analysis, we only look at \emph{specific} names under university domains (see
\S\ref{sec:overview}-\S\ref{sec:ger}), which are only related to services and
not individual users, and only investigate IP addresses of cloud platforms (see
\S\ref{sec:overview}).

A hypothetical scenario -- usually filtered for by Farsight -- that may still
leak PII is a university using dynamic DNS updates for user networks, see
RFC2136~\cite{RFC2136}.  For example, a user's machine with the hostname
`Firstnames-iPad' may obtain a DHCP~\cite{RFC2131} lease in a university's
WiFi, and subsequently use dynamic DNS to register an \texttt{A} record for the
IP address it just received under the name
\texttt{Firstnames-iPad.user-wifi.example.com}.  As we only search for
\emph{specific} names under universities' domains and not for this edge-case,
we cannot comment on its existence in the dataset.  Additionally Farsight
filters for these cases (see \Fref{app:method}).

Moreover, following the Menlo report~\cite{bailey2012menlo,dittrich2012menlo},
we conducted a harm-benefit analysis.  We found that, as we take additional
measures against accidentally handling PII -- as described above -- and we work
with a historic dataset that has been collected under the premise of not
containing PII, the benefits of using this dataset to investigate cloud
adoption, given its far reaching implications as discussed in
\S\ref{sec:discussion}, outweigh the limited and mitigated potential harm.

\subsection{Selection of Institutions}
\begin{table}[t!]
\centering
\caption{Count of selected Universities per country. See \Fref{app:domains} for a detailed list of institutions and domains.}
\label{tab:inst}
\footnotesize
\setlength{\tabcolsep}{2pt}
\begin{tabularx}{\columnwidth}{lrp{8em}lr}
\toprule
\textbf{Country} & \textbf{Count} & & \textbf{Country} & \textbf{Count}\\ \midrule
Austria & 34 & & The Netherlands & 19 \\
\rowcolor[HTML]{C0C0C0} 
France & 74 & & Times Higher Education Top100 & 100 \\
Germany & 81 & & United Kingdom & 115 \\
\rowcolor[HTML]{C0C0C0} 
Switzerland & 14 & & United States & 260 \\\bottomrule
\end{tabularx}
\end{table}
We focus on universities (PhD awarding institutions) in the global north,
specifically the \us, Germany, Switzerland, Austria, the \uk, the Netherlands,
and France, see \Fref{tab:inst}.  \Fref{app:domains} lists all institutions and
corresponding domains for each category. 

We are familiar with the laws and educational systems of these countries, which
we saw as a precondition to interpreting the data.  We also hoped to contrast
the effect of GDPR across countries, but found no conclusive evidence.  For
international comparison, we included institutions listed in the THE Top100 for
2020~\cite{the}. The universities we studied predominantly use services
dependent on dominant cloud providers from the \us.  An expansion of this study
to include universities and cloud providers from other parts of the world is
ongoing research with collaborators from those regions.

For the \us, we selected all R1~\cite{r1} and R2~\cite{r2} universities based
on the Carnegie Classification of Institutions of Higher Education (also listed
on Wikipedia~\cite{wp00}).  For the remaining countries, i.e.,
Germany~\cite{wp01}, the \uk~\cite{wp02}, Switzerland~\cite{wp03},
Austria~\cite{wp04}, France~\cite{wp05}, and the Netherlands~\cite{wp06}, we
rely on the Wikipedia pages listing universities.  We argue that this is a
sufficiently reliable source for this information, given its general nature. We
further manually investigated each listed university to identify their
associated domain name(s).  We do not claim completeness, but instead try to
estimate a lower bound with our measurements.  If a university uses multiple
domains, or used a different domain in the past (especially common in France
due to a history of reorganization of the university system), we check all
domains and aggregate the results under the name of the institution.

To ensure that our data is not influenced by institutions into which we only
have limited visibility, we excluded all institutions for which we did not see
at least ten distinct names\footnote{In \texttt{www.example.com}, \texttt{www}
is a name under \texttt{example.com}. If we talk about ten distinct names, we
mean seeing at least ten different names, e.g., \texttt{mail.example.com},
\texttt{www.example.com}, etc.} in at least one month within our seven-year
dataset. The institutions we filtered due to limited visibility are: 
\begin{enumerate*}[label=\emph{(\roman*)},itemjoin={{, }},itemjoin*={{, and }}]
\item in the Netherlands the Theological University Apeldoorn, a small topic-specific university with no considerable IT infrastructure
\item in France 16 domains, which are remnants from before the merging processes of universities in the late 1990s/early 2000s
\item in the \uk 28 domains, the Courtauld Institute of Art and 27 domains which belong to universities that are included in the dataset with other domains they predominantly use, e.g., \verb+ox.ac.uk.+ instead of \verb+oxford.ac.uk.+
\item in Austria four domains, one of which is a secondary domain for the University of Salzburg, which is included via its mainly used domain \verb+uni-salzburg.at.+, and three small private universities in Vienna.
\end{enumerate*}

\subsection{Visibility of Private Cloud Use}
\label{sec:visibility}
One question that may arise is whether we also observe private cloud use, e.g.,
an individual researcher using Gmail, or if such private cloud use from a
university campus adds noise to our data.  As the dataset we use is an
aggregate view of DNS requests made, and we only look at names \emph{under
universities' domains}, we cannot infer usage patterns on a campus or that of
individual users.  Specifically, a user visiting \texttt{mail.google.com} in
their browser -- no matter if on campus or not -- would have their browser make
a DNS request for \texttt{mail.google.com}.  The aggregation of \emph{all}
requests with the same data that month leads to this entry (simplified):
\begin{center}
	\begin{footnotesize}
	\vspace{-0.2em}
	\texttt{\{'count': 1234, 'rrname':'mail.google.com.',\\ 'rrtype':'A', 'rdata':['142.250.179.165']\}}
	\vspace{-0.2em}
	\end{footnotesize}
\end{center}
From this, it \emph{cannot} be inferred which user made this request, where
they made it, or when exactly they made it.  Hence, we are not even including
these requests in our work.  Instead, we select records \emph{under the domain
of a university}, to infer whether they are using Google services for their
\emph{inbound} email.  For example, for `Example University' using the domain
`example.com', we would find the following entry in our dataset:
\begin{center}
	\begin{footnotesize}
	\vspace{-0.2em}
	\texttt{\{'count': 42, 'rrname':'example.com.', 'rrtype':\\ 'MX', 'rdata':['5 alt1.aspmx.l.google.com.']\}}
	\vspace{-0.2em}
	\end{footnotesize}
\end{center}
This means that, during the month in which this DNS entry was observed, mail
for, e.g., \texttt{user@example.com} would have to be delivered to the mail
server \texttt{alt1.aspmx.l.google.com}.  For other analyses, we analogously
use different RRtypes \emph{under universities' domain names}.  Hence, private
use of cloud services \emph{does not show up in our dataset}.  To get a
perspective \emph{including} individual researchers' use of cloud products for
\emph{a single university}, a measurement study looking at a university's
network uplink would have to be conducted, similar to the concurrent study by
Karamollahi et al.~\cite{karamollahi2022zoomiversity}.

\subsection{Quantitative vs. Qualitative Approach}
In our work, we utilize a quantitative approach to measure cloud adoption
across a sample of over 600 universities.  Arguably, a qualitative approach
could have provided more in-depth insights into the organizational motives,
while being more direct, and allowing for a larger temporal sample, i.e.,
observations on situations before the start of our quantitative dataset.
Nevertheless, in order to attain a comparative perspective between regions,
collecting qualitative data from over 600 universities constitutes a
significant effort, and necessitates recruitment channels for these specific
universities as well as local language proficiency.  Qualitative investigations
of the observed differences can then become the subject of subsequent studies,
also reducing the number of interviews necessary, as these can be restricted to
more targeted research questions and observed differences in regions.
Furthermore, a qualitative approach requires linear effort for including
further regions and universities in future analyses.  In contrast, our
quantitative approach can be easily scaled beyond the list of universities we
studied.

\section{Cloud Use Overview}
\label{sec:overview}

In this section, we provide a first overview of universities' reliance on cloud
infrastructure of the `Big Three' (Amazon, Google, and Microsoft). We want to
understand to which extent names under universities' domains point toward these
infrastructures, regardless of their popularity.  This way, we do not only
capture the most frequented names -- for example the main website, or resources
commonly used by students -- but also capture, e.g., HR and administration
tools, along with systems used for research.  Hence, we look at whether
universities have \emph{at least one} name under one of their domains that
points to each of the three providers above.

\mypar{Service Deployment.}
We start by measuring whether at least one generic service under a university's
domain runs on cloud infrastructure.  Hosts and services on the Internet
generally need a name, and this name~\cite{RFC1123} usually points to an IPv4
address~\cite{RFC0791}, IPv6 address~\cite{RFC8200}, or both (dual
stack)~\cite{RFC4241}, commonly enabled by DNS~\cite{RFC1034}, via \texttt{A}
and \texttt{AAAA} resource records, respectively.  The servers, or hosts, a
service runs on can then be addressed by their corresponding IP address (IPv4
or IPv6).  IP addresses are commonly registered to an organization that uses
them~\cite{RFC7020} via their Regional Internet Registry (RIR).  When a service
is run on hosts in a cloud provider's infrastructure, the IP address via which
they are reachable will identify that cloud provider.

This means that if the infrastructure for
\texttt{studentadmin.ex\-am\-ple.\-com} runs on Amazon EC2 infrastructure, its
\texttt{A} and/or \texttt{AAAA} records will point to an IP address owned by
Amazon.  For this, it is not necessary that \texttt{studentadmin.example.com}
provides a web service (a service offered via HTTP(S)~\cite{RFC9110}), but it
could also be any other common network service like a file or authentication
server.  This also means that a university may use multiple cloud providers at
the same time, if \texttt{hrservice.example.com} runs on Microsoft
infrastructure and its \texttt{A} and/or \texttt{AAAA} correspondingly point
there.

\begin{figure}[t!]
	\begin{center}
		\includegraphics[width=\columnwidth]{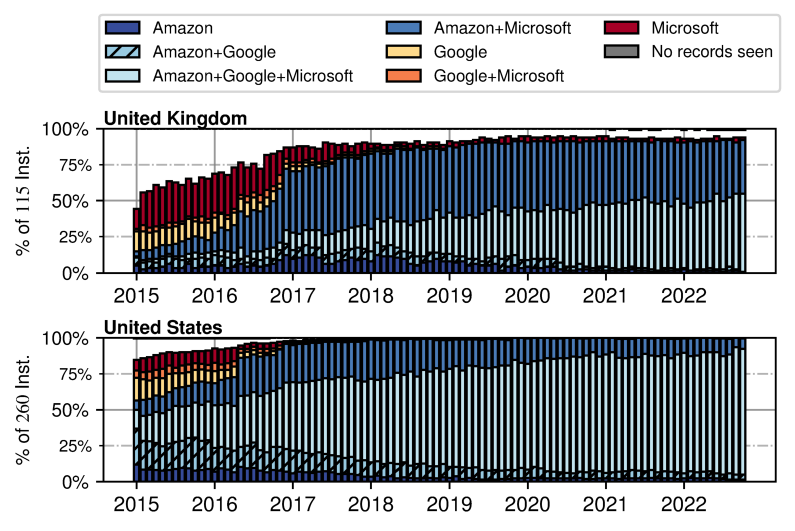}
		\caption{Universities' use of the `Big Three' cloud providers (Amazon, Google, Microsoft) in the \uk, the \us, and Germany from January 2015 to \lastdate. See \Fref{fig:full_public_clouds} in the Appendix for an overview of all measured countries.}
		\label{fig:public_clouds}
	\end{center}
\vskip-1.5em
\end{figure}

\mypar{Methodology.}
For each university, we collect all \verb+A+, \verb+AAAA+, and \verb+CNAME+
resource records (RRs) for its domains.  We then try to resolve all
\verb+CNAME+ RRs from the dataset of the corresponding month in which they were
observed.  If we are unable to resolve a \verb+CNAME+ to an IPv4 or IPv6
address, we match RRs for products regularly hosted in certain infrastructure
to IP addresses of the hoster.  For example, we consider \verb+CNAME+s like
\texttt{www.example.com. IN A
ec2-\-203-\-0-\-113-\-25.\-compute-1.\-amazonaws.\-com.} as hosted by Amazon.

We then use the AS59645 BTTF historic bulk whois service~\cite{streibelt2022we}
to identify the Autonomous System (AS) that has been announcing a specific IP
address during the month in the past for which we observed it in our dataset.
The AS59645 BTTF historic bulk whois service leverages several historic
datasets to provide information on the ASes that announced specific IP
addresses up to one-day resolutions, spanning the period from May 2005 (for
IPv4) and January 2007 (for IPv6) up until today.  Please see the BTTF whois
paper by Streibelt et al. for a detailed description of the
methodology~\cite{streibelt2022we}.

\mypar{Results.}
\Fref{fig:public_clouds} presents an overview of our findings.  On a
macroscopic level, we already see major differences between institutions from
different countries.  Having at least one system located at a major cloud
provider is common for the \us, the \uk, and the Netherlands.  The THE Top100
show a pattern similar to the \us.  Cloud usage in these three countries and
the THE Top100 shows a high share of using services hosted at Amazon.  We find
that the \us developed towards a situation where all of the three major
operators are used at universities at the same time, rising from 79
institutions (30.38\%) in January 2015 to \CSBTTFMAXABSeduAMAZONGOOGLEMICROSOFT
(\CSBTTFMAXPERCeduAMAZONGOOGLEMICROSOFT) in \lastdate.  For the Netherlands and
\uk, we see a lower share of Google over time, starting at 30 (26.09\%) of all
institutions for the \uk and 4 (21.05\%) for the Netherlands in January 2015,
reaching \CSBTTFMAXABSukGOOGLEonly (\CSBTTFMAXPERCukGOOGLEonly) and
\CSBTTFMAXABSnlGOOGLEonly (\CSBTTFMAXPERCnlGOOGLEonly), respectively, in
\lastdate.  Note that, in the \uk, the adoption of Amazon-hosted cloud services
took place between 2015 and 2017, with the largest adoption happening in 2016.
We conjecture that this is related to AWS being included in the \uk's
government cloud from late 2013 onwards~\cite{gcloud4}, as prior outsourcing
arrangements are unlikely to be quickly changed to new
offerings~\cite{goo2007investigation}.  Contrary to the \us, where commonly all
three major providers are used, a combination of Amazon and Microsoft is more
common in the Netherlands and \uk.

France, Germany, and Austria form a clear contrast to this picture.  All three
of these countries have a lower cloud usage, with less than 50\% of
universities relying on cloud providers for any services: 2 (2.47\%) to
\CSBTTFMAXABSdeSUM (\CSBTTFMAXPERCdeSUM) for Germany, 10 (13.51\%) to
\CSBTTFMAXABSfrSUM (\CSBTTFMAXPERCfrSUM) for France, and 0 to
\CSBTTFMAXABSatSUM (\CSBTTFMAXPERCatSUM) for Austria from January 2015 to
\lastdate. Switzerland, starting at 5 (35.71\%) in January 2015 and reaching
\CSBTTFMAXABSchSUM (\CSBTTFMAXPERCchSUM) in \lastdate developed from a middle
ground between these two clusters towards the first one.

Note that the uptick of Microsoft-related infrastructure in Germany in December
2020 relates to the occurrence of names like
\verb+(lync)autodiscover.example.com+ that point to Microsoft Azure addresses.
Without this increase, we observed at least 24 (29.63\%) of institutions in
Germany using public cloud infrastructure in November 2020.  We conjecture
(also see \S\ref{sec:zoom}), that this connects to the wider introduction of
Microsoft Office 365, or activation of new features in an existing MS Teams
installation (the specific name is a necessary condition for using
Skype-for-Business use, but \emph{may} also occur for an Office365 or Teams
deployment).  See, for example, an announcement of the Ruhr University
Bochum~\cite{rub2020o365}.

In general, we find that cloud infrastructure dependence across all sampled
countries is on the rise.  However, in the Netherlands, the \us, the THE
Top100, and the \uk, we find this increase from a high level, i.e., \us, \uk,
Dutch, and THE Top100 universities already frequently used cloud infrastructure
before January 2015.  Still, we find an increase in cloud usage for these
countries.

We note that we \emph{do not} find a `pandemic
effect'~\cite{feldmann2020lockdown} in the use of cloud infrastructure across
institutions.  Instead, the migration of higher education to the cloud seems to
be an ongoing process that started more than five years ago.  Furthermore, we
find that the use of cloud resources fundamentally differs between countries.
We revisit this pattern in \S\ref{sec:discussion}, as we can observe similar
effects for other cloud infrastructures as well.

\vspace{-0.2em}
\section{Cloud-based Email}
\label{sec:email}

Here, we investigate universities' use of cloud-based email infrastructure.
Email is arguably one of \emph{the} most essential services on the Internet for
professional communication.  It regularly carries significant PII, when
students have questions on courses, or seek advice in professional and personal
matters, and it transports grades and course assignments, but also job
applications, research data, academic discourse, and ideas.

Email is a common gateway for attackers to convince users to install malware or
redirect them to phishing
sites~\cite{google:pishing:imc,ho:spearphishing,chainsmith:eurosp}.  Hence,
spam and malware filtering are common services offered by outsourced email
platforms, and usually a significant selling point in moving to cloud-based
email providers~\cite{dada2019machine}.  However, as Patrick Breyer, a member
of the European Parliament, recently noted this also means that the operator is
in control of which emails are and which are not delivered to
users.\footnote{\textit{``Incredible: Microsoft decides which e-mail Members of
the European Parliament get to read in their inbox. It's called Outlook spam
filter and cannot be disabled.''}, Patrick Breyer, MEP,
\url{https://twitter.com/echo_pbreyer/status/1363854606132858882} (February,
22, 2021; archived: \url{https://archive.ph/5L00T}).} The strict inbound rules
of major providers, which can lead to false
positives~\cite{outlookmail,googlemail}, mean that universities using these
services outsource the decision which emails reach their faculty and students
along with the service.

\mypar{Service Deployment.}
Email is one of the most complex protocols currently used on the
Internet~\cite{flo2022simple}, and for a more in-depth explanation we refer to
related work, e.g., see Holzbauer et al.~\cite{flo2022simple} for details on
the configuration of modern email sending setups.

Here, we restrict ourselves to a description of inbound email handling.  To
receive email for a domain, one has to set \texttt{MX} records in that domain
that provide a (prioritized) list of names of servers that accept emails for
the domain.  When a cloud-based email service provider is being used, the
\texttt{MX} records of a domain point at email servers of said cloud provider.
So, when Exchange in the Cloud or Office 365 are being used at an institution,
the \texttt{MX} records point at servers with names under \texttt{outlook.com}.
In addition, there are various email security appliances that upload received
emails to cloud setups for, e.g., security and spam checks.  For these to work
effectively, additional DNS records for, e.g., \texttt{DMARC} have to be set to
direct information to other services of said operator. 

\begin{table}[t!]
\centering
\caption{Selected MX domains from cloud providers.}
\label{tab:mx-names}
\footnotesize
\setlength{\tabcolsep}{2pt}
\begin{tabularx}{\columnwidth}{lX}
\toprule
\textbf{Operator} & \textbf{MX Domains}                                                                                                                                                                                                                        \\ \midrule
Microsoft         & \tt outlook.com, \tt hotmail.com                                                                                                                                                                                                                   \\
\rowcolor[HTML]{C0C0C0} 
Google            & \tt google.com, \tt googlemail.com, \tt smtp.goog                                                                                                                                                                                                      \\
Proofpoint & \tt pphosted.com                                                                                                                                                                                                                               \\
\rowcolor[HTML]{C0C0C0} 
Cisco             & \tt iphmx.com                                                                                                                                                                                                                                  \\
Other             & \tt trendmicro.eu, \tt messagelabs.com, \tt schlund.de, \tt spamfighters.net, \tt mailcontrol.com, \tt spamhero.com, \tt emailsrvr.com, \tt mailspamprotection.com, \tt fireeyecloud.com, \tt mailanyone.net, \tt secureserver.net, \tt mailgun.org, \tt icritical.com,  \tt barracudanetworks.com\\ \bottomrule
\end{tabularx}
\end{table}

\mypar{Methodology.}
To identify whether universities use a cloud-based email service, we
investigate their \verb+MX+ records.  \texttt{MX} records are DNS entries that
determine the email servers responsible for receiving emails for a
domain~\cite{RFC5321}.  Hence, we only measure who handles \emph{inbound} email
for a university.  Their user email access and email storage may be handled
on-site or via another cloud-based solution.  Still, this means that all email
to this institution flows via the identified service operator.

To identify the used operators, we follow the methodology of Henze et
al.\@~\cite{henze2017tma}.  We first check if, for any of the second-level
domains (SLDs) of a university, any of the \verb+MX+ records points to a domain
associated with a cloud-based email provider (see \Fref{tab:mx-names}).  If we
do not find an \verb+MX+ record for any of the SLDs, we descend further down
the DNS tree.  This happens, for example, if an institution has dedicated
sub-domains for email, similar to using \verb+staff.example.com+ and
\verb+students.example.com.+ Hence, if a sub-domain points the \verb+MX+ record
at a cloud provider, we also consider the university to be using a cloud
provider for email.  Please note that the existence of an \texttt{MX} record
pointing to a cloud provider does not have interaction effects with the
measurements in \S\ref{sec:overview}, as we only utilize \texttt{A},
\texttt{AAAA}, and \texttt{CNAME} records, but not \texttt{MX} records there.
Nevertheless, use of a cloud email provider might lead to other services, e.g.,
webmail, having names allocated under a university's domain which then point
towards a cloud provider in the sense of what we measured in
\S\ref{sec:overview}.

We also check whether a university uses Proofpoint's email security solution,
which analyzes all incoming emails for an organization to filter out spam and
malicious emails.  It can either be used as a hosted solution where email is
redirected via Proofpoint's servers, similar to products from Cisco, or via an
appliance installed on-site that uploads emails and attachments for analysis to
the Proofpoint cloud.  We identify hosted setups via their \verb+MX+ records,
while measuring appliance usage indirectly by evaluating
\texttt{DMARC}~\cite{RFC7489} records.  If the \verb+rua+ or
\verb+ruf+\footnote{\texttt{rua} (Reporting URI(s) Aggregate) and \texttt{ruf}
(Reporting URI(s) Forensic) are entries in \texttt{DMARC} DNS records. If an
email server receives messages from a \texttt{DMARC}-enabled domain, it should
report information on the received messages (delivered, rejected, quarantined,
etc.) back to the addresses listed here.} of a university points to an email
address under \verb+emaildefense.proofpoint.com+, we assume that it uses
Proofpoint's appliance-based services.  If we do not find an \verb+MX+ record
that points to hosts under a cloud provider's domain, or a \texttt{DMARC}
record indicating the use of Proofpoint, we count the institution as
`Other/Private.' This approach may under-estimate the number of cloud providers
we find.  Furthermore, if we are unable to observe an \verb+MX+ record for an
institution included in our dataset for a given month, we mark this as `No MX.'

\mypar{Validation.}
We manually retrieved the \verb+MX+ records of universities found using
cloud-based email and verified that they indeed point to the identified cloud
provider.

\begin{figure}[t!]
	\begin{center}
		\includegraphics[width=\columnwidth]{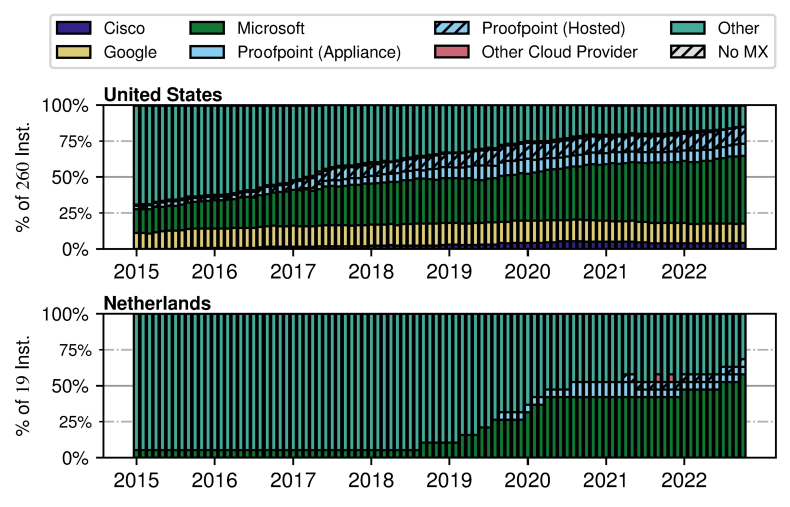}
		\caption{Email providers used by universities in the \us and the Netherlands from January 2015 to \lastdate. See \Fref{fig:full_mail} in the Appendix for all measured countries.}
		\label{fig:mail}
	\end{center}
	\vskip-1.5em
\end{figure}

\mypar{Results.}
When looking at the results of our measurements, see \Fref{fig:mail}, we find
that they align with our observations from \S\ref{sec:overview}.  The \us, the
\uk, the Netherlands, and the THE Top100 are again the countries with the most
frequent use of cloud-based email providers, reaching \MAILMAXABSuksumonly
(\MAILMAXPERCuksumonly) for the \uk, \MAILMAXABSthesumonly
(\MAILMAXPERCthesumonly) for the THE Top100, and \MAILMAXABSnlsumonly
(\MAILMAXPERCnlsumonly) for the Netherlands in \lastdate.

The Netherlands present an interesting case here, as we see an increase in
Microsoft-based email hosting between late 2018 and early 2020.  Manually going
through the websites of these universities revealed that they, either shortly
before this time posted news items announcing a plan to migrate to Microsoft
services, or directly announced a migration to Microsoft services at this time.
As for early adopters, Utrecht University had been using Gmail for its
students, while using a self-hosted solution for staff.\footnote{Note, that we
did not measure this specific use of Google's cloud email service, as it was
only deployed for the subdomain \texttt{students.uu.nl}, and we use the base
domain of universities to determine the predominantly used email provider, if
that domain is being used for email, i.e., in the case of Utrecht University,
we determined the email provider based on the one used for \texttt{uu.nl}.} In
2018, they then decided to migrate students' and staff's email to Microsoft to
create a common platform and -- as mentioned in their press release -- improve
`security'~\cite{uu-migration}.  We assume that this relates to concerns about
Google in the context of GDPR.  Similarly, Nyenrode University migrated to
Microsoft as part of a larger strategy to unify their IT
infrastructure~\cite{nyenrode-migration}.  The larger increase in 2019 then may
connect to a letter from the Dutch Ministry of Justice and Security sent to
parliament, that essentially notes that privacy concerns regarding Microsoft
cloud products have been resolved in negotiations~\cite{grapperhaus}.  This
letter is then, for example, explicitly cited by TU Eindhoven as a reason why
earlier concerns about privacy and security no longer apply, and they now
migrate their email infrastructure to Microsoft~\cite{tue-migration}.

In the \us, a total of \emph{five} companies control email services for
\MAILMAXABSedusumBIGonly (\MAILMAXPERCedusumBIGonly) of all R1 and R2
universities in 2020.  Again, Germany and France have a lower use of cloud
resources, with neither of those countries exceeding 20\% in \lastdate:
\MAILMAXABSdesumonly (\MAILMAXPERCdesumonly) for Germany and
\MAILMAXABSfrsumonly (\MAILMAXPERCfrsumonly) for France.  Both Austria and
Switzerland have a higher adoption of cloud-based email services than Germany
and France, with \MAILMAXABSchsumonly (\MAILMAXPERCchsumonly) for Switzerland
and \MAILMAXABSatsumonly (\MAILMAXPERCatsumonly) for Austria in \lastdate, both
of them staying well below 50\% adoption.  We see a slight upward trend in the
\us and the THE Top100, and a notable increase in the \uk from 67 (58.26\%) in
January 2015 to \MAILMAXABSuksumonly (\MAILMAXPERCuksumonly) in \lastdate.  For
the remaining countries, adoption of cloud email seems to stagnate over our
measurement period.

The two most prominent operators are Google, likely with their classroom
product, a work-suite containing email, documents and integration with Chrome
Books, as well as Microsoft with cloud-hosted Exchange/Office365/Teams.  Other
cloud providers only play a notable role in the \uk, where they occupy
\MAILMAXPERCukother of the market in \lastdate.  The most prominent smaller
cloud providers are FireEye and Trend Micro.  We find that Proofpoint is most
prominent in the \us and  the THE Top100, where we see the service being used
by \MAILMAXABSeduProofpointAppliance/\MAILMAXABSeduProofpointHosted
(appliance/hosted;
\MAILMAXPERCeduProofpointAppliance/\MAILMAXPERCeduProofpointHosted) and
\MAILMAXABStheProofpointAppliance/\MAILMAXABStheProofpointHosted
(\MAILMAXPERCtheProofpointAppliance/\MAILMAXPERCtheProofpointHosted)
institutions in \lastdate, respectively.  We also see Proofpoint moving in the
Dutch market, with first universities deploying their products in September
2019.

\section{Cloud-based LMS}
\label{sec:lms}

We now take a look at universities' use of cloud-based Learning Management
Systems (LMS), i.e., online tools that allow lecturers to manage and automate
courses, reaching from registration, via providing content, to assessment and
examination of enrolled students.  As such, these systems provide some of the
core functionality of what a university \emph{does}.  However, these systems
also hold the most sensitive data stored about students: Grades, deliverables,
and overall study performance.

Putting these systems into cloud infrastructure potentially provides access to
this confidential data to unauthorized entities, e.g., via the cloud
act~\cite{rojszczak2020cloud}.  At the same time, it also prevents students
from effectively consenting to their data being processed by cloud companies,
as an opt-out is only possible by \emph{not} studying at a university using one
of these products.  Furthermore, these systems are also especially susceptible
if a cloud provider decides to enforce their own policies and principles.  If,
for example, a \us-based LMS provider decides to enforce \us sanctions against
citizens of specific countries for an LMS, including customers outside the \us,
it can effectively dictate which students a university enrolls by controlling
the `means of study.' Given the precedent of GitHub restricting
accounts~\cite{gh-ban} for developers located in Crimea, Cuba, Iran, North
Korea, and Syria to comply with \us trade sanctions, this is by far no
hypothetical scenario.

\mypar{Service Deployment.} 
Cloud-hosted LMS are commonly run on servers operated by the company providing
the LMS as a SaaS.  However, to integrate these solutions with the organization
for which they are provided, they are commonly aliased to a name under a
university's domain name using a \texttt{CNAME} record, see, e.g., the
documentation of Brightspace~\cite{bscn} and Blackboard~\cite{blackboardcn}.
Hence, by having a \texttt{CNAME} of the form \texttt{lms.\-example.\-com IN
CNAME uni\-ver\-sity-\-name.\-bright\-space.\-com.}, users can use the LMS by
directing their browser at \texttt{lms.\-ex\-am\-ple.\-com}, providing a
consistent appearance for services used by an organization.

\mypar{Methodology.}
We focus on four large providers of cloud-based LMS: Brightspace (Desire2Learn,
\texttt{brightspace.com}), Courseleaf (\texttt{courseleaf.com}),  Blackboard
(\texttt{blackboard.com}), and Canvas (Instructure, \texttt{instructure.com}).
These tools provision their services by having a name in a university's zone
pointing a \verb+CNAME+ to their infrastructure, e.g., for Canvas
\texttt{canvas.example.com.} \texttt{IN CNAME example-com.instructure.com.} To
measure whether a university uses one of these LMS, we check whether we find a
\verb+CNAME+ with a target that is below one of the domains used by the above
cloud LMS.  Note that we also count a SaaS-hosted LMS with servers located with
Amazon, Google, or Microsoft, as a cloud-hosted service in
\S\ref{sec:overview}.  Naturally, we do not see whether a university uses an
on-site LMS, like Moodle, or a locally hosted version of Blackboard.

\mypar{Validation.} 
We manually went over matches for December 2021 and visited the identified LMS
sites to verify that these universities indeed run the cloud-based LMS.

\begin{figure}[t!]
	\begin{center}
		\includegraphics[width=\columnwidth]{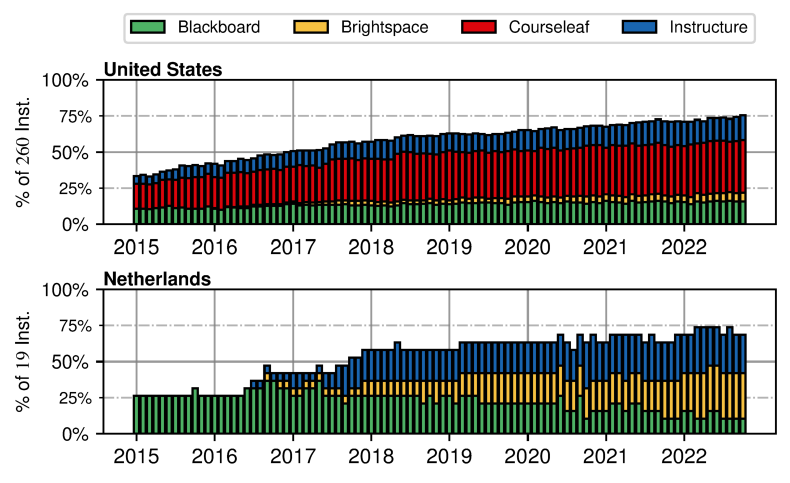}
		\caption{Cloud-hosted Learning Management Systems (LMS) used by universities in the \us and the Netherlands from January 2015 to \lastdate. See \Fref{fig:full_lms} in the Appendix for all measured countries.}
		\label{fig:lms}
	\end{center}
	\vskip-1.5em
\end{figure}

\mypar{Results.}
We find that cloud-hosted LMS are mostly relevant in the \uk, the \us, and the
Netherlands, for the latter two see \Fref{fig:lms}.  We find no instances of
cloud-hosted LMS in Germany and France, and only two in Austria.  In
Switzerland, we only find a single Canvas instance at the University of St.
Gallen, which has been in operation since January 2019.  We revisit the
question what universities in these countries are using \emph{instead} in
\S\ref{sec:discussion}.  For the THE Top100, the use of cloud-hosted LMS is
mostly due to \us universities.  In fact, \LMSMAXtheUS of the
\LMSMAXABSthesumBIGonly universities in the THE Top100 that use a cloud-based
LMS in \lastdate are \us universities, while \us universities only make up 40
universities in the THE Top100.  The remaining \LMSMAXtheSUM institutions using
cloud-based LMS in the THE Top100 are from the Netherlands (\LMSMAXtheNL), the
\uk (\LMSMAXtheUK), Canada (\LMSMAXtheCA), Australia (\LMSMAXtheAU), Hong Kong
(\LMSMAXtheHK), Singapore (\LMSMAXtheSG), Sweden (\LMSMAXtheSE), and Belgium
(\LMSMAXtheBE).  Courseleaf is exclusively catering to the \us market, as we
find no instances outside of the \us.  We also find a steady growth of the use
of cloud-based LMS over time between January 2015 to \lastdate: in the \us from
87 (33.46\%) to \LMSMAXABSedusumBIGonly (\LMSMAXPERCedusumBIGonly), in the \uk
from 15 (13.04\%) to \LMSMAXABSuksumBIGonly (\LMSMAXPERCuksumBIGonly), in the
Netherlands from 5 (26.32\%) to \LMSMAXABSnlsumBIGonly
(\LMSMAXPERCnlsumBIGonly), and in the THE Top100 from 24 (24.00\%) to
\LMSMAXABSthesumBIGonly (\LMSMAXPERCthesumBIGonly).  In line with
\S\ref{sec:overview} we find that in October 2022, Blackboard, Brightspace,
Courseleaf, and Instructure are hosted on Amazon EC2.

\section{Video \& Remote Lecturing Tools}
\label{sec:zoom}

Tools for video chatting and VoIP carry longstanding significance in
professional communications, for example, Skype-for-Business (SfB).  However,
with COVID-19, academic core activities -- teaching, meetings, and conferences
-- became dependent on them, a discussion often framed as the `zoomification'
of education.

Hence, here, we review universities' reliance on cloud-hosted video chat
solutions.  Following \S\ref{sec:background}, we look at common video call
tools like SfB, Cisco WebEx, Adobe Connect, and Zoom.  Furthermore, we estimate
the use of Microsoft Teams, but due to its implementation are limited to an
upper-bound estimate.

\mypar{Service Deployment.}
Cloud-based video chat tools are commonly hosted under the provider's domain
name, e.g., \texttt{zoom.us} for Zoom.  Large customers can, however,
authenticate their domain using a challenge response mechanic via \texttt{TXT}
records for their own domain, allowing the consolidation of users under that
domain~\cite{zoom1}.  In addition, organizational users can create custom
`vanity' sub-domains under the service's second-level domain.  This is
obligatory if the organization, as most universities, uses a Single-Sign-On
(SSO) system, which necessitates a custom landing page~\cite{zoom2}.  In
addition, there are also options where a part of the video communication
platform can be hosted on-premise by a customer, while account management
remains in the public cloud infrastructure~\cite{zoom3}.  Adobe Connect, WebEx,
and Zoom follow this approach.

MS Teams and SfB may also contain some on-premise components, requiring
specific DNS records to enable cloud integration of these
products~\cite{msdocs,msdocs2}.  Furthermore, organizations can authenticate
their domain to Microsoft using a dedicated \texttt{TXT} record.

\mypar{Methodology.}
To identify universities' use of centralized video chat solutions, we follow
three different approaches, based on the platform we are looking at.  For Zoom
(\texttt{zoom.us}), Cisco WebEx (\texttt{webex.com}), and Adobe Connect
(\texttt{adobeconnect.com}), we follow the naming scheme of these services for
clients under their domains, i.e., we check if a name exists whose first label
is \emph{(i)} the SLD of a university, \emph{(ii)} the SLD and TLD of a
university with hyphens in between, or \emph{(iii)} the SLD of a university
plus \verb+-live+ (see \Fref{tab:zoomperm}).  If we find a corresponding name
lookup in our dataset during a month, we consider a university as using this
service during that month.  Furthermore, we consider universities as using Zoom
that have a Zoom verification \verb+TXT+ record (see \Fref{app:method}).

To establish if a university uses SfB, we check for required DNS entries when
operating SfB~\cite{msdocs}, specifically
\texttt{lync\-dis\-cover.\-ex\-am\-ple.\-com}, with \texttt{ex\-am\-ple.\-com}
being replaced by a university's domain.  This overlaps with the prior product
name of SfB, Microsoft Lync. Finally, we check for universities which
\emph{may} be using Microsoft Teams.  Unlike SfB, Microsoft Teams does not
require special DNS entries that make its use uniquely
identifiable~\cite{msdocs2}, even though the voice component requires the same
DNS entries as SfB~\cite{msdocs}.  However, to be able to use Microsoft Teams,
an operator still has to set a Microsoft cloud verification token of the form
\verb+MS=ms12345678+.  Even though the presence of this record does not mean a
site \emph{does} use Microsoft Teams -- it may use other Microsoft products as
well -- we also count the number of sites using this token and report the
number of \emph{additional} universities that may be \emph{exclusively} using
Microsoft Teams, i.e., that \emph{do not} use any of the other tools (SfB,
Zoom, WebEx, or Adobe Connect).

\mypar{Validation.}
We manually verified all Zoom links for July 2021 by visiting the corresponding
Zoom page and ensuring that it forces users to log in via the related
universities' SSO systems.  If this was not the case, we used web searches to
identify whether the related university refers to this Zoom subdomain for
events on any of its websites.  From 363 Zoom links we verified, 12 (3.31\%)
were incorrectly attributed or could not be verified through other channels.
We excluded these false positives from our analysis.

\begin{table}[t!]
\centering
\caption{Example name permutations for assessing \texttt{zoom.us}, \texttt{webex.com}, and \texttt{adobeconnect.com} usage.}
\label{tab:zoomperm}
\footnotesize
\addtolength{\tabcolsep}{1em}    
\begin{tabularx}{\columnwidth}{p{1cm}p{2.10cm}l}
\toprule
\textbf{Input} & \textbf{Permutation} & \textbf{Checked Service Domain}                                                                                                                                                                                             \\ \midrule
\verb+example.com+ & Second Level & \verb+example.zoom.us+ \\
                   & Dot Substitution & \verb+example-com.zoom.us+\\
                   & SLD + \verb+-live+ & \verb+example-live.zoom.us+ \\\bottomrule
\end{tabularx}
\end{table}

\begin{figure}[t!]
	\begin{center}
		\includegraphics[width=\columnwidth]{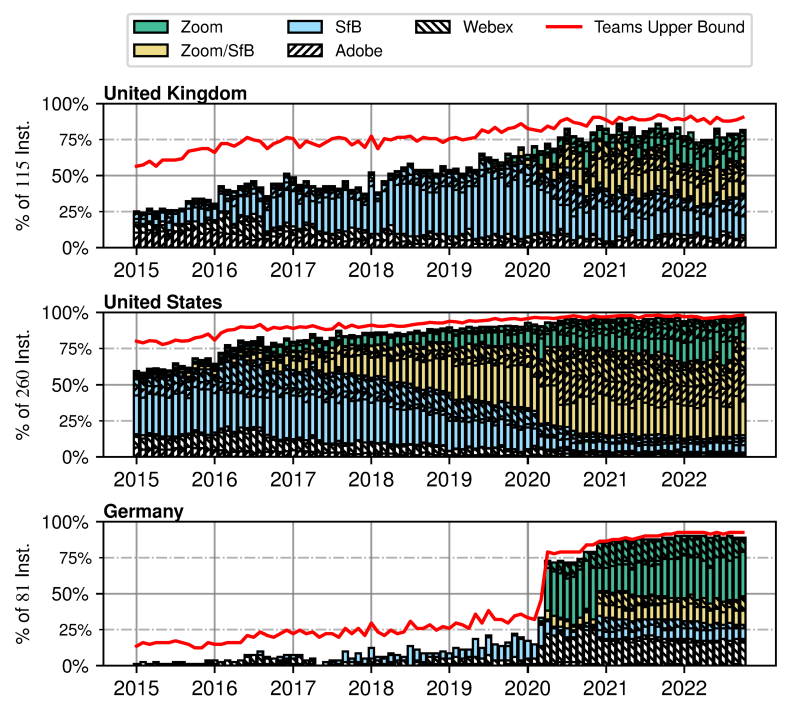}
		\caption{Video chat tools used by universities in the \uk, the \us, and Germany from January 2015 to \lastdate. See \Fref{fig:full_public_clouds} in the Appendix for all measured countries.}
		\label{fig:zoom}
	\end{center}
	\vskip-1.5em
\end{figure}

\mypar{Results.}
Taking a macroscopic look at our data, we again see a similar segmentation as
with the previous cases of general cloud usage, email, and cloud-based LMS (see
\Fref{fig:zoom}).  We see a heavy adoption of SfB (from 2015 to 2021) in the
Netherlands (one with a large increase mid-2015 to \ZOOMMAXABSnlSfBonly
(\ZOOMMAXPERCnlSfBonly)), the \us (110 (42.31\%) to \ZOOMMAXABSeduSfBonly
(\ZOOMMAXPERCeduSfBonly)), the \uk (9 (7.83\%) to \ZOOMMAXABSukSfBonly
(\ZOOMMAXPERCukSfBonly)) and the THE Top100 (30 (30.00\%) to
\ZOOMMAXABStheSfBonly (\ZOOMMAXPERCtheSfBonly)).  At the same time we see close
to no SfB instances in France, and limited adoption in the remaining countries:
5 (35.71\%) in Switzerland, 11 (32.35\%) in Austria, and 25 (30.86\%) in
Germany.  Note that in Germany we observed an increase of 20 institutions using
SfB between November and December 2020, most likely due to the introduction of
Microsoft Teams, which partially uses DNS entries overlapping with those for
SfB.  We conjecture that this overall picture connects to different operational
paradigms between universities in these two clusters, also in terms of
administrative centralization (see \S\ref{sec:background}).

When we look at the adoption of the other three platforms, we find an
interesting picture, also in relation to the COVID-19 pandemic.  In the \us, we
find that the adoption of Zoom and, to a slightly lower extent, WebEx has been
an ongoing process that already started back in 2016 leading to
\ZOOMMAXABSeduZoomonly (\ZOOMMAXPERCeduZoomonly) \us universities using Zoom
and \ZOOMMAXABSeduWebexonly (\ZOOMMAXPERCeduWebexonly) using WebEx in
\lastdate.  However, in comparison to December 2019, these numbers `only' rose
by \ZOOMMAXABSeduZoomDecInc from 144 for Zoom and by \ZOOMMAXABSeduWebexDecInc
from 68 for WebEx, meaning that the pandemic effect is not as large as in other
countries, mostly due to the already high adoption of Zoom in the \us.  Adobe
Connect, in general, has a market share similar to WebEx, with
\ZOOMMAXABSeduAdobeonly (\ZOOMMAXPERCeduAdobeonly) \us universities using it in
\lastdate.  \us universities seem to generally be using a multitude of video
chat solutions, with \ZOOMMAXABSedudiffPlatTWOonly
(\ZOOMMAXPERCedudiffPlatTWOonly) using two, \ZOOMMAXABSedudiffPlatTHREEonly
(\ZOOMMAXPERCedudiffPlatTHREEonly) three, and \ZOOMMAXABSedudiffPlatFOURonly
(\ZOOMMAXPERCedudiffPlatFOURonly) all four of the surveyed tools in \lastdate.

This effect can again be found to a similar extent in the THE Top100.  Please
note that only 40 universities in the THE Top100 are \us universities.  Here,
we also see a continuous adoption of Zoom starting in 2016, leading up to
\ZOOMMAXABStheZoomonly (\ZOOMMAXPERCtheZoomonly) institutions using Zoom in
\lastdate.  We also observe an apparent lack of a significant pandemic effect,
and a large diversity of employed tools across universities, with
\ZOOMMAXABSthediffPlatTWOonly using two, \ZOOMMAXABSthediffPlatTHREEonly three,
and \ZOOMMAXABSthediffPlatFOURonly all of the surveyed video chat solutions.

We see a pandemic effect among the remaining countries, especially in terms of
Zoom adoption.  While Zoom played essentially no role in European universities
before February 2020, its adoption quickly increased with the move to remote
teaching.  Interesting observations here are that most European universities
are more discrete in their choice of video teaching platform (either Zoom or
WebEx), and that the onset of these tools was sudden, i.e., within a month in
the beginning of 2020.  France shows a slower increase focused on Zoom,
contrary to other European countries where we also observe an increase in
WebEx.  In the end, we find that in \lastdate Zoom/WebEx use in German
universities is at \ZOOMMAXABSdeZoomonly
(\ZOOMMAXPERCdeZoomonly)/\ZOOMMAXABSdeWebexonly (\ZOOMMAXPERCdeWebexonly), in
the \uk \ZOOMMAXABSukZoomonly (\ZOOMMAXPERCukZoomonly)/\ZOOMMAXABSukWebexonly
(\ZOOMMAXPERCukWebexonly), in the Netherlands \ZOOMMAXABSnlZoomonly
(\ZOOMMAXPERCnlZoomonly)/\ZOOMMAXABSnlWebexonly (\ZOOMMAXPERCnlWebexonly), in
Austria \ZOOMMAXABSatZoomonly (\ZOOMMAXPERCatZoomonly)/\ZOOMMAXABSatWebexonly
(\ZOOMMAXPERCatWebexonly), in Switzerland \ZOOMMAXABSchZoomonly
(\ZOOMMAXPERCchZoomonly)/\ZOOMMAXABSchWebexonly (\ZOOMMAXPERCchWebexonly), and
in France \ZOOMMAXABSfrZoomonly (\ZOOMMAXPERCfrZoomonly)/\ZOOMMAXABSfrWebexonly
(\ZOOMMAXPERCfrWebexonly).

Looking at the possible upper bound for universities using Microsoft Teams
\emph{without} the SfB/voice and video chat component, we find that this number
is close to zero for the \us (\ZOOMMAXABSedumso/\ZOOMMAXPERCedumso), Germany
(\ZOOMMAXABSdemso/\ZOOMMAXPERCdemso), and Switzerland (\ZOOMMAXABSchmso) in
\lastdate.  In the \uk (\ZOOMMAXABSukmso/\ZOOMMAXPERCukmso), the THE Top100
(\ZOOMMAXABSthemso/\ZOOMMAXPERCthemso), Austria
(\ZOOMMAXABSatmso/\ZOOMMAXPERCatmso), and the Netherlands
(\ZOOMMAXABSnlmso/\ZOOMMAXPERCnlmso) we see a modest number of additional
institutions that \emph{might} be using Microsoft Teams.  France is the only
country where we find a comparatively large amount of potential Microsoft Teams
users who do not use \emph{any} of the other solutions of SfB, with
\ZOOMMAXABSfrmso (\ZOOMMAXPERCfrmso) institutions in \lastdate.  This
difference has been stable over the past years, and is likely not related to an
increase in Teams adoption by universities not \emph{already} using Microsoft
cloud services (or providing access to Microsoft software licenses to users
from their domain) in the beginning of 2020.

\section{Self Hosting in Germany}
\label{sec:ger}
The differences we observe from \S\ref{sec:overview} to \S\ref{sec:zoom} beg
the question what digital learning tools universities use instead of cloud
products, e.g., in Germany.  Hence, we look at the use of common self-hosted
alternatives for LMS (Moodle~\cite{buchner2016moodle} and
Stud.IP~\cite{appelrath2006einsatz,hamborg2014befunde}) and video chats
(BigBlueButton~\cite{bbb}) in Germany, which are reportedly deployed in 90\% of
higher education institutions~\cite{Taz-OSanUnis}.  Self-hosted tools may, by
default, not be necessarily more privacy preserving than offerings of large
cloud providers.  However, \emph{control over data} nevertheless remains with
the university hosting them, and they are able to audit and -- if necessary --
reconfigure and patch these tools to conform to privacy regulations and
requirements.  This could, for example, be seen with BigBlueButton, where the
user group around German universities made significant contributions towards
the privacy-preserving operation once privacy limitations in its design became
apparent~\cite{bbbpriv,bbbtickets}.

\mypar{Service Deployment.}
Self-hosted services are typically deployed on servers located in a
university's data-center.  As with all services, see \S\ref{sec:overview},
these systems and associated services need an IP address and a name to be
easily accessible by users.  Best practice for naming systems in a professional
setting is the use of a hybrid naming scheme, i.e., a scheme in which systems
are named partially in a functional way, e.g., \texttt{mail}, \texttt{survey},
or \texttt{lms}, in combination with a formularic component~\cite{liabook}.
With this hybrid scheme, a mail system might have the name \texttt{mail023},
for being the 23\(^{rd}\) mailserver, leading to the FQDN
\texttt{mail023.example.com}.  To make such systems more accessible to users,
frontend systems commonly also receive an additional functional alias via a
\texttt{CNAME}~\cite{liabook}.  In the above example, a frontend alias might be
\texttt{mail.example.com}, which may also be a load balancer to distribute load
and scale the mail setup horizontally~\cite{liabook}.  While technically not
advised as best practice, functional names are also commonly modeled after the
utilized software instead of the function of the software~\cite{liabook}.

Hence, for the three systems we study in this section, operators are not
restricted to specific naming.  Yet, common operational practice makes it
likely that systems are either provisioned under partially hybrid
functional-formularic names based on the utilized software stack, or at least
have an alias with a semi-functional naming scheme, i.e., a scheme that
utilizes the software name instead of the system's function for naming.  This
is also a practice we observed in \S\ref{sec:lms}, where, e.g., SaaS instances
of Brightspace commonly were aliased from \texttt{brightspace.example.com}
instead of the purely functional \texttt{lms.example.com}.

\mypar{Methodology.}
To estimate self-hosted LMS and BigBlueButton use in Germany and the \us, we
count the number of universities that either have Moodle/Stud.IP or
BigBlueButton related names under their domain.  For Moodle and Stud.IP, we
count a university as having Moodle or Stud.IP if there is at least one name
containing either \verb+moodle+ or \verb+studip+.  For BigBlueButton we count
all universities that have at least one name containing \verb+bbb+,
\verb+bigbluebutton+, \verb+scalelite+ (the load balancer component of
BigBlueButton), or \verb+greenlight+ (a common BigBlueButton frontend).

Note that our matching is fuzzy: we may overmatch on hostnames that contain
product names without running the associated service, while we may also
undermatch when universities host these tools under different names.  For
example, in Germany, we often found BigBlueButton systems being called
\texttt{kon\-fer\-enz}, the German word for conference, explaining the
difference between our measured \BBBMAXPERCdebbb and the 90\% reported in the
media~\cite{Taz-OSanUnis}.

\begin{figure}[t!]
	\begin{center}
		\includegraphics[width=\columnwidth]{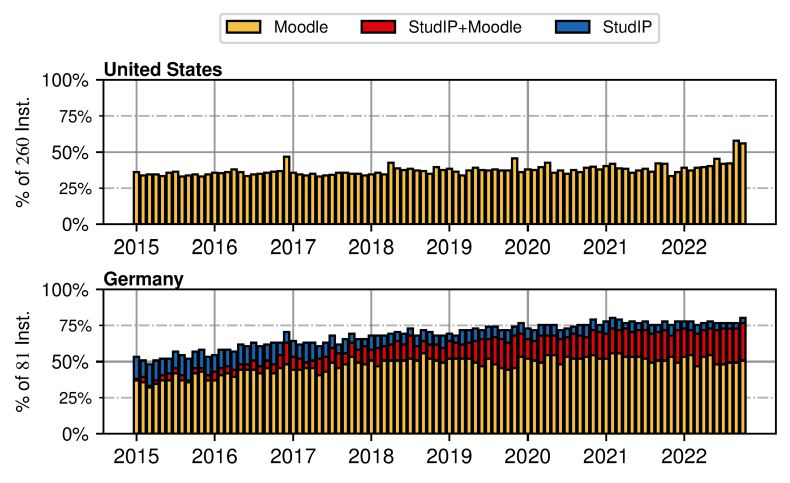}
		\caption{Universities with at least one name containing `moodle' or `studip' for Germany and the \us (January 2015--\lastdate). See \Fref{fig:full_studip} in the Appendix for an overview of all measured countries.}
		\label{fig:studip}
	\end{center}
	\vskip-1.25em
\end{figure}
\begin{figure}[t!]
	\begin{center}
		\includegraphics[width=\columnwidth]{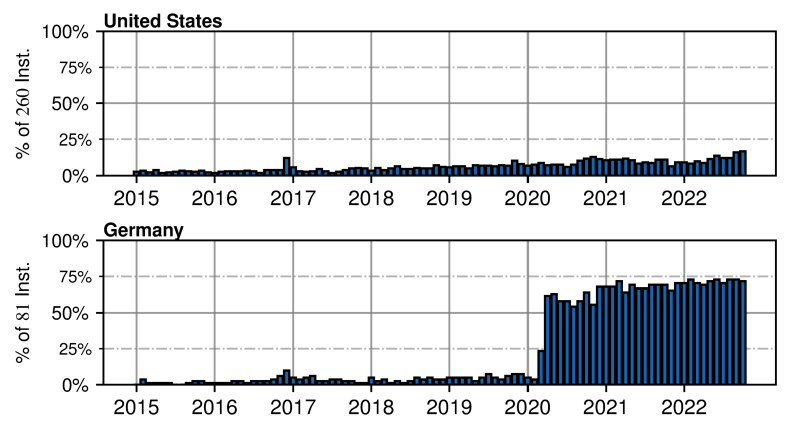}
		\caption{Universities with at least one BigBlueButton-related DNS entry for Germany and the \us (January 2015--\lastdate). See \Fref{fig:full_bbb} in the Appendix for an overview of all measured countries.}
		\label{fig:bbb}
	\end{center}
	\vskip-1.25em

\end{figure}

\mypar{Validation.}
We manually verified the use of BBB by visiting the BBB pages for the 59 German
universities observed using BBB in December 2021, checking if they run BBB
related software. We found that all of them did, indeed, ran BBB related
software.

\mypar{Results.}
In \lastdate \STUDIPMAXABSdesumonly (\STUDIPMAXPERCdesumonly)
universities in Germany have names related to Moodle or Stud.IP vs.
\STUDIPMAXABSedumoodle (\STUDIPMAXPERCedumoodle) in the \us Moodle.  
Similarly, we find \BBBMAXABSdebbb
(\BBBMAXPERCdebbb) universities in Germany having BigBlueButton related names
under their domain, while this is the case for around 10\% in the \us, see
\Fref{fig:bbb}.  We see a pandemic effect for BigBlueButton in Germany,
starting in February~2022.

\section{Discussion}
\label{sec:discussion}

\subsection{Cloud Infrastructures and Power}

The last decade has seen big tech companies honing in on cloud infrastructures
as an alternative source of growth~\cite{eurich2011revenue,van2020seeing}.
This growth relies on two effects: Realizing the value proposition of reducing
costs in terms of Capital Expenses (CapEx) and local Operational Expenses
(OpEx) for lower OpEx paid for cloud service charges, and -- for individual cloud
providers -- by attaining a market position making them `the default' platform
to be used~\cite{Waters2016Cloud}. 

The increasing dependency of big tech on cloud computing for their financial
success means that they use political, economic and technical resources to
ensure that the clouds are `the default' infrastructure in as many domains as
possible. Their political force is brought to bare using international
initiatives, e.g., New Pedagogies for Deep Learning is a global partnership
between the OECD, Gates Foundation, Pearson and
Microsoft~\cite{williamson2020commercialisation}; government partnerships,
e.g., the \uk government has incentivised schools to opt for platforms that are
both free to use and bundled up with govern\-ment-funded technical
assistance~\cite{williamson2020commercialisation}; and lobbying
efforts~\cite{nemitz2018constitutional}.  Cloud providers use economic
incentives by mounting the benefits of economies of scale, financing and
physically migrating data to the cloud, and by providing free services that can
bypass regular procurement rules. They can capture educational IT either by
providing storage, compute, communication platforms, or by becoming the default
infrastructure for smaller EdTech companies~\cite{Judge2020Zoom}. As our
results in \S\ref{sec:overview} show, universities increasingly use cloud
services provided by Amazon, Google, and Microsoft. The fact that we also see a
combination of different cloud services, hints at cloud platforms being
introduced through other EdTech services.

On the flip side, this migration leads to `cloud lock-in,' i.e., the dependency
on cloud services even when terms and conditions change.  For example,
Google~\cite{drive:policy} discontinued free unlimited cl\-o\-ud storage, limiting,
e.g., the University of Washington\footnote{Twitter thread by Julie Kientz:
\url{https://twitter.com/juliekientz/status/1400545039688736768} (June, 3,
2021; archived: \url{https://archive.ph/MYllj})} and University of
Utah\footnote{Twitter thread by Bryan W. Jones:
\url{https://twitter.com/BWJones/status/1490802506628145153} (February, 7,
2022; archived: \url{https://archive.ph/qCiCg})} to 100 TB of total shared data
for each university.

The trend of big tech monopolies shifting from \emph{``being mere owners of
information, ... to becoming owners of the infrastructures of
society''}~\cite{srnicek2017platform} has prompted an ongoing public discussion
about the implications of this `platform capitalism' on different aspects of
society~\cite{brodnig2019,christl2016networks,srnicek2017platform}, yet without
zooming in on its implications on higher education. At first sight, the
political economic advantages put forth by cloud companies make good fellows
with the economized management of universities.  However, this also comes with
power shifts.  Mirrlees and Alvi~\cite{mirrlees2019edtech} argue that
universities focus on cutting costs, while allowing the big five (Apple,
Alphabet/Google, Amazon, Microsoft, Facebook) and a growing ecosystem of
start-ups, e.g., in the area of MOOCs, to compete with, and ultimately replace,
public education.  Most universities do not have the economic or political
power to insert their own values and interests in such a market, unless they
coordinate on these issues.  The international initiatives these companies
support make up informal policy networks that increasingly dominate educational
policy~\cite{dijck2018platform}.  Aside from potential impact on democratic
societies and educational values, these networks are likely to promote certain
forms of education, e.g., the individualised pursuit of ‘mastery’ enacted
primarily through adaptive software, in favor of education that, e.g.,
promotes interpersonal dialogue and relations with
others~\cite{williamson2020commercialisation}.  In the bigger scheme of things,
there are also concerns about `platform imperialism:' \us-based companies could
use their global digital default infrastructure to exert `soft-power' and
economic control influencing global norms and values of digital
cultures~\cite{jin2015}, steering curricula and research activities.

\subsection{Cloud Use vs. Academic Freedom}

The dependence of universities on cloud platforms for teaching, communication,
and research that we observed has serious implications for academic freedom.
If education and research \emph{depend} on an external cloud service,
researchers may become bound to comply with requirements set by these
organizations.  We recently saw Google's handling of Timnit Gebru for a paper
not `deemed worthy for publication' by the company~\cite{timnit}, as well with
other instances of Google telling its researchers to put a positive spin on
`sensitive topics,\footnote{Sensitive topics include the oil industry, China,
Iran, Israel, COVID-19, home security, insurance, location data, religion,
self-driving vehicles, telecoms and systems that recommend or personalize web
content~\cite{google2020sensitive}.} or remove references to Google
products~\cite{google2020sensitive}.  One might argue that this concerns
\emph{employees} of Google, but it also begs the question whether cloud
operators could leverage their power to influence critical university research
in a similar way.  In fact, Google has already been in the spotlight for
sponsoring favorable research that is in line with its business and policy
interests~\cite{wsj:google,google2021techpolicy}, both in the
\us~\cite{accountability:us}, and
Europe~\cite{accountability:eu}.\footnote{Note that the reports published as
part of the \emph{Google Transparency
Project}~\cite{accountability:eu,accountability:us} have also drawn criticism
for being funded by Google's competitor
Oracle~\cite{accountability:background}.} More generally, Abdalla et
al.~\cite{abdalla:greyhoodie} discussed `Big Tech' funding for research on the
societal impact and ethics of AI, as a way to influence research questions (and
answers). In the area of EdTech research, Mirrlees and
Alvi~\cite{mirrlees2019edtech} observed a lack of critical research, likely
because of \emph{``...little incentive to `bite the hand that feeds'
''}~\cite{Selwyn2015}. 

Conceivably, a major cloud provider could simply indicate that a continued
business relation with a university may not be desirable in case the
institution and its researchers continue to voice positions critical of that
cloud provider.  That institution would then face the dilemma of either
`aligning' their researchers, or facing a costly migration of essential
services.  Such a migration could easily cost millions while severely
interrupting research and teaching.

Similar cases can be made for cloud operators enforcing their business rules in
terms of, e.g., global sanctions as in the case of GitHub~\cite{gh-ban}. This
may effectively put universities in a position where they either bar students
from sanctioned countries attending the university, or at least from using
their digital learning environment. Similarly, the reliance on platforms and
their policies might impede global research collaborations~\cite{acm:collab}.
Thus, this centralization of power may indeed inadvertently threaten core
functions of universities.  Hence, the question we have to ask as academics is
not whether cloud operators \emph{would} use these powers.  Instead, we have to
ask ourselves if we are willing to risk that they \emph{could}.

\subsection{Privacy and Academia}

The move to the cloud raises a number of concerns with respect to the
application of privacy by design or compliance.  Past studies show that
educational institutions do not fare well in making \textit{transparent} the
data collection and processing practices of cloud providers to their faculty,
staff and students~\cite{lindh2016information,marek2017privacy,jones2020we}.
This can, e.g., happen when a university implements a blanket privacy policy
for all digital tooling, including all cloud services.  Depending on the
diversity of data collection and processing these services entail, privacy
policies may become generic, potentially falling short of legal transparency
requirements~\cite{reidenberg2016ambiguity}.  It may also not be clear to the
university what data is going to the cloud.  Universities may evaluate and make
data agreements with cloud providers, but ensuring these are effective can be
challenging.  Besides vague privacy policies~\cite{kotal2020vicloud}, cloud
services come with the promise of being plug-and-play, and recursively, they
leverage the benefit of service architectures, and often bundle dozens of third
parties~\cite{gursesvanhoboken_2018}.  As a result, cloud service
providers may fail to make their data flows transparent.  The promise of
plug-and-play also means that university IT departments are often not given the
time or resources to evaluate services.  Even when performing privacy
evaluations, these stand against digital branding efforts of the university and
the partnerships between public institutions and cloud
providers~\cite{williamson2020commercialisation}.

When students, faculty and staff access these services, they are not
asked for explicit consent. Universities can, e.g., in the case of
GDPR~\cite{voigt2017eu}, use \textit{legitimate interest}, \textit{public tasks}
or \textit{performance of a contract} to justify data flows to cloud services.
Hence, students and faculty may not have a (meaningful) option to opt-out of
these services. When there is an opt-out process, people may be incentivized
not to use them, e.g., reserving them for `severe cases' creating time and
capacity burdens for faculty and staff. When incentive structures are set up
by-design and by-policy to push people onto cloud infrastructures, it is hard
to speak of choice. Hence, public education institutions may end up leveraging
their structures to on-board students, faculty, and staff as cloud service
consumers~\cite{lindh2016information}. 

If universities continue to outsource core functions to cloud platforms,
students will no longer have \emph{a choice} on whether they want to expose
some of their most private information to these major cloud providers.
Considering that these cloud services are economically under pressure to
monetize either the data they collect (e.g., by creating a recruiting
business~\cite{piazza2017}), or the infrastructural dependency they create, the
practices that are being established here are concerning. As Hasan and
Fritz~\cite{edtech-pets-22} argued, platforms, such as LMS (see
\S\ref{sec:lms}), can collect a wealth of sensitive behavioral and demographic
student data, which can be abused for advertisement or surveillance. They also
observed, even when this data is not collected directly, student demographics
can still be inferred, potentially through the combination with data from other
sources. Thus, universities may have to consider whether it is ethical or legal
to create an environment where informed consent to data collection is,
essentially, no longer possible.

\subsection{Universities as Enterprise Networks}

In \S\ref{sec:zoom} we observe that regions with major cloud adoption also saw
a major adoption of SfB early on.  Revisiting \S\ref{sec:background}, we noted
that tools like SfB would be expected in \emph{centralized} enterprise IT.
Hence, we argue that SfB adoption can serve as a proxy to assess the general
operational paradigm of a university, i.e., if it is run more like an
\emph{enterprise network} or a \emph{university network}.

This mechanic of administrative alignment of IT infrastructures with
administration leading to centralization and organizations \emph{behaving}
similarly is a well documented effect in the field of Information Systems (IS).
DiMaggio and Powell~\cite{dimaggio1983iron} discuss how
bu\-reau\-cra\-ti\-za\-tion via coercive, memetic, and normative processes
leads to a structural alignment of organizations in a market, see also Scott
for a more recent reflection on these theories~\cite{scott2008approaching}.
Avgerou~\cite{avgerou2000and} transferred this institutional perspective to the
introduction of IT systems and their connection to organizational change.  To
synthesize, the findings from IS indicate an effect in organizations where
administrative alignment leads to IT transformation as \emph{a goal in itself},
lacking \emph{``adequate legitimacy''~\cite{avgerou2000and}}, without any
\emph{``contribution to the process of organizational
change''~\cite{avgerou2000and}}.

Following SfB as a proxy, we conjecture that we observe an increased adoption
of cloud technology for countries in which the university system has seen a
stronger commoditization -- the \us, the \uk, the Netherlands, and THE Top100
-- as also discussed by Bosetti and Walker~\cite{bosetti2010perspectives}.  In
these countries, organizational alignment led to a situation where academic
leaders governing a body of scholars were replaced by administrators and
business managers \emph{overseeing} university operations.  These new managers
imported and integrated enterprise tools and culture into the heart of public
education, leading towards more cloud adoption.

\vspace{\baselineskip}
\subsection{Self-Hosting Challenges \& Future Work}
\label{sec:sh}
Concerning self-hosting vs. cloud infrastructures, there are opposing
perspectives which have to be discussed. On the one hand, there are positions
claiming that, for universities and other companies, the use of cloud
infrastructure provides a wealth of benefits.  The core idea of cloud
infrastructure is the tenant's ability to quickly increase and decrease
utilization based on actual demand, while only paying the resources they
actually use.  Following that point, there is an abundance of reports on
cloud-operators' websites describing that, in contrast to self-hosting, cloud
infrastructure enables more features and better adaptability~\cite{ntpp-cs-1},
transforming organizations \emph{``from "IT can't do that" to a 'can do'
situation''}~\cite{ms-cs-1}, as claimed in a Microsoft customer success-story.
Similarly, customer-stories from Amazon claim that cloud hosting decreases an
organization's carbon footprint and fiscal spend by reducing on-site personnel
and facilities, along with this increase in agility~\cite{amz-cs-1}.  Amazon's
customers even note that integrating cloud services in education also aids
students' future employ-ability~\cite{amz-cs-2}.

Self-hosting needs local expertise to build usable and secure infrastructure,
creating a conundrum between privacy issues of cloud infrastructure and their
comparatively higher staffing enabling better security and
reliability~\cite{fiebig202213}.  Similarly, self-hosted solutions may lack in
observed efficacy, stability, and usability. For example, email grew so
complex~\cite{flo2022simple} that even experts with decades of experience are
unable to get their self-hosted mail setup to deliver to major email providers
like Outlook.com and Gmail~\cite{fenollosa2022mail}.  Yet, Fenollosa attributes
this to these operators' strict filtering a\-gain\-st smaller operators in an
attempt to reduce spam~\cite{fenollosa2022mail}, see \S\ref{sec:email}.

Similarly, in terms of efficacy of specific tools, while some work notes
benefits for classroom implementations provided by Google (Google Classroom),
more recent work finds non-application specific border conditions more relevant
for usability, i.e., availability of Internet access and lecturer's engagement
with a platform~\cite{haiduwa2022integrating}.  To the best of the authors
knowledge, there are no recent publications \emph{comparing} the usability of
open-source self-hosted digital teaching tools against their cloud-hosted
counterparts in established venues in security and privacy, human computer
interaction, and educational technology research.  Still, research from the
early to late 2000s on Linux desktop software notes structural root-causes for
limited usability, e.g., due to lacking in user research~\cite{paul2009survey},
and community-based initiatives lacking the organizational structure needed for
common approaches to
usability~\cite{bach2007usability,bach2010future,benson2004professional}.
Furthermore, looking at decision makers' perspectives via a qualitative study
focused on the perceived benefits of cloud migrations, Lal et al. find that
clouds are seen as providing better scalability, higher flexibility, and more
usable interfaces~\cite{lal2016understanding}.  Still, concerning cost savings,
executives at Hey.com claim staffing reduced in engineering has to be added in
other areas~\cite{hey-cs-1}.  

Hence, operating self-hosted infrastructure is certainly neither easy, nor
guaranteed to be successful. Looking at cases where self-hosting was
successful, we find greater adaptability to be frequently mentioned.  For
example, the head of IT at the University of Osnabr{\"u}ck -- a university
committed to self-hosting and open-source for decades-- notes that their
self-hosting approach was ultimately cheaper and allowed them to react to the
COVID-19 pandemic much more seamlessly than other universities who procured
cloud products~\cite{npol}.  Similarly, the authors of Blacklight, an
open-source literature search and indexing software for university libraries
initially started at the University of Virginia, explicitly note adaptability
as a major reason to start the project~\cite{gilbert2013breaking}.

However, both cases highlight that preserving universities' digital sovereignty
-- especially given a reduction of local competencies in favor of cloud
infrastructure -- is not an easy task, but a long-term policy and resource
commitment.  In both cases, the concerned universities made a long-term
commitment, essentially driving their infrastructure and relevant open-source
software like in-house applications at large organizations, thereby creating
the organizational structure found lacking for solely community-based open
source projects in the
past~\cite{bach2007usability,bach2010future,benson2004professional}.  Hence,
\emph{current} cost savings and efficacy, e.g., in Osnabr{\"u}ck or for
organizations using Blacklight~\cite{nowviskie2007adapting}, stem from a
decade-long investment in the accumulation of competences by implementing a
long-term strategy in which IT is seen as a support facility for teaching and
research~\cite{npol}.  With no tangible return-on-investment in the short-term,
changes -- as all changes -- potentially showing a (perceived) reduced system
efficacy at first~\cite{liabook}, and benefits potentially taking decades to
materialize, this approach can be challenging to justify and sustain.

\vspace{\baselineskip}
\noindent\textbf{Future Work.}
Given our observations on a split perspective on self-hosting vs. cloud
infrastructure above, several directions of future research emerge.  The
challenges and benefits of self-hosting as well as cloud infrastructures have
to be critically and analytically evaluated.  Corporate promotion material is
as little of a comparative source for benefits of cloud infrastructures, as is
a single case of a counter example proof for the feasibility of self-hosting.
Especially in the latter case, scrutiny will have to be applied to the question
which combination of factors, including local and state policy, led to current
success \emph{in that isolated case}, and if and how these factors can be
facilitated more generally.  Similarly, the scientific community should execute
\emph{structured} evaluations of, e.g., public digital infrastructures'
usability, to provide an independent empirical basis for discussions on the
efficacy of proprietary and cloud-based vs. open-source and self-hosted
systems.  Furthermore, business, organizational, and societal factors in the
progressing adoption of cloud infrastructure have to move into the focus of
future research, see, e.g., the work of Srnicek~\cite{srnicek2017platform}.

\section{Limitations}
\label{sec:limitations}

The Farsight SIE dataset may not contain all cloud-related names, if these are
not queried from a client behind a sensor.  While instances of cloud hosting we
identify are certainly there, more universities may be using major cloud
providers without it showing up in the dataset.  Similarly, we cannot make
statements on the popularity of names, as Farsight SIE only collects DNS cache
misses~\cite{farsight}.  Furthermore, the number of universities among the
surveyed countries differs (14 in Switzerland, 260 in the \us), which may
amplify the effect of individual institutions' choices in smaller countries.
Our work partly relies on heuristics and the automated analysis and
classification of historic data, e.g., in the identification of Zoom, WebEx,
and Adobe Connect domains, the use of Proofpoint, and the estimation of Moodle,
Stud.IP, and BigBlueButton instances.  Hence, we manually revisited our results
and verified our findings against live data, as documented in a validation
paragraph for each methodology section.

Given the large effect sizes, the alignment of ratio chan\-ges between
different countries, our additional spot-checks, and our coverage of domain
names, we are confident that our results paint an accurate picture of
universities' cloud use since January 2015.

\section{Related Work}
\label{sec:related-work}

\mypar{Cloud Infrastructure Measurements.}
Similar to us, Borgolte et al.\@~\cite{borgolte2018cloud-strife} use the
Farsight SIE dataset to identify domains pointing at cloud infrastructure.
Jacquemart et al.~\cite{Jacquemart2019tma} performed active DNS measurements on
the most popular domains according to Alexa to measure the adoption of cloud
services from 2013--2018.  Portier et al.\@~\cite{portier2019security} and van
der Toorn et al.\@~\cite{van2020txting} identify cloud service usage via
\verb+TXT+ records.  Streibelt et al.\@~\cite{streibelt2013exploring} and
Calder et al.\@~\cite{calder2013mapping} use the \verb+EDNS0+ extension to map
cloud infrastructure.  Doan at al.~\cite{doan22webconsolidation} combined
active DNS measurements with crawling and rendering webpages to measure the
consolidation of web resource hosting by Content Delivery Networks (CDNs) and
cloud hosts. Henze et al.~\cite{henze2017tma} focused on the adoption of
cloud-based email services and identified them based on email headers on a
dataset collected from mailing lists, spam traps, and volunteer users.  Vermeer
et al.\@'s provide a general taxonomy of asset discovery techniques similar to
our targeted asset discovery using a passive dataset~\cite{vermere2021asset}.

\mypar{COVID-19 and the Internet.}
With COVID-19, it became apparent that the continued lock-down situation would
have an extended effect on the Internet.  As such, several researchers studied
this effect, including the increased utilization of cloud-based services.
Feldmann et al.\@~\cite{feldmann2020lockdown} studied the impact of the
COVID-19 pandemic through the lens of a major Internet Exchange Point from a
European perspective, while Liu et al.~\cite{Liu2021pam} performed a similar
study on changes in network traffic patterns in the \us.  Boettger et
al.\@~\cite{boettger2020internet} provided a similar perspective from the
vantage point of the Facebook social network.  Along the same lines, Lutu et
al.\@~\cite{lutu2020characterization} investigated the impact of COVID-19 on
mobile network traffic.

\mypar{Educational Technology in the Cloud.}
Cohney et al.~\cite{cohney2020virtual} perform a study into the privacy
implications of virtual classroom technology.  Contrary to us, they root their
evaluation of technology use in a self-reported study among 49 instructors and
administrators in \us universities, obtaining results similar to our Internet
measurement data.  In addition, they also analyze privacy policies of common
virtual classroom tools, as well as 50 public Data Privacy Addenda (DPAs) in
which universities negotiate their own terms platform operators. However, they
also note that these terms only apply for institution-wide contracts, and that
individual instructors might use other platforms without being aware of the
privacy implications. From a student perspective, Balash et
al.~\cite{balash2021proctoring} focused on online proctoring services. They
observed an institutional power dynamic and students' implicit trust in the
tools selected by their university: the assumption that a university vets a
tool or platform before using it lends it credibility. Emami-Naeini et
al.~\cite{enaminaeini2021attitudes} studied user attitudes towards video
conferencing tools, including in educational settings.  They also noted the
participants' lack of agency when it comes to platform selection.

Similar to us, Komljenovic~\cite{komljenovic2021rise} theoretically analyzes
the implications of the progressing centralization and platformization of
educational technology, particularly noting the de-institutionalization of
public education accelerated by centralized platforms.  Zeide and
Nissenbaum~\cite{zeide2018learner} analyze (before the COVID-19 pandemic)
learner privacy in MOOCs and virtual education, finding it to often violate
established norms in terms of privacy and education, supporting our assessment
that the `zoomification' of education is a long-standing process predating the
COVID-19 pandemic.  Besides these major related publications, several
small-scale evaluations often limited to specific tools (usually Zoom) were
undertaken during the last years, and have been summarized by Cohney et
al.~\cite{cohney2020virtual}.

Finally, similar to us, Angiolini et al.~\cite{Angiolini:OpinioJuri} identified
data protection challenges of remote teaching from a legal perspective, noting
the conflict between platforms' business models and the public interest goals
of universities, as well as threats to academic freedom and the right to
education.

\section{Conclusion}
\label{sec:conclusion}

We investigated the reliance of universities on cloud infrastructure in seven
countries and in the Times Higher Education Top100.  We found that the move to
the cloud has been ongoing for the past several years, and, apart from video
lecturing tools, was not heavily influenced by the COVID-19 pandemic.  Our
results also highlight that university systems highly differ in their
susceptibility to migrate to the cloud.  We conjecture that this ties in with a
multitude of factors, including the academic and administrative culture, and
the history of university IT in the corresponding countries.  Furthermore, we
discuss the potential impact of this progressing development on the very
essence of academic freedom.

In the end, as academics, we have to ask ourselves: Now that we know, are we
content with this development, and can we live with the broader implications.
If not, we have to find ways to counteract these developments, by investing in
decentralized capabilities for independent research and teaching
infrastructure, learning from -- certainly not perfect -- cases like in
Germany.

\vspace{0.5em}
\noindent\textbf{Data \& Artifacts:}
We share our university domain list dataset in
\Fref{app:domains} and at:
\url{https://github.com/headsinthecloud/universities} (self-hosted:
\url{https://git.aperture-labs.org/Cloudheads/universities}). Measure cloud
hosting at your own university with our artifact:
\url{https://github.com/headsinthecloud/cloudheadschecker} (self-hosted:
\url{https://git.aperture-labs.org/Cloudheads/cloudheadschecker}).

\begin{acks} 
This work has been supported by the European Commission via the H2020 program
in project CyberSecurity4Europe (Grant No. \#830929).  Our work was enabled by
the use of Slack and Signal (both hosted on Amazon EC2), Overleaf (hosted on
Google cloud), GitHub (owned and hosted by Microsoft), as well as self-hosted
BigBlueButton and Gitea instances. We thank Florian Streibelt for collaborating
with us on implementing BTTF whois~\cite{streibelt2022we}, which we leverage in
\S\ref{sec:overview},and Farsight Security, Inc. (now DomainTools) for
providing access to the Farsight Security Information Exchange's passive DNS
data feed.  Without this data, the project would not have been possible.  The
authors express their gratitude to the anonymous reviewers and the shepherd of
this paper for their continuous input during the review and shepherding
process.  Their input was instrumental in shaping the flow of the discussion in
\S\ref{sec:sh} and motivating the creation of the BTTF whois service.  Any
opinions, findings, and conclusions or recommendations expressed in this
material are those of the authors and do not necessarily reflect the views of
the authors' host institutions, Farsight Security, Inc., DomainTools, or those
of the European Commission.
\end{acks}

{
\printbibliography
}

\appendix

\section{Farsight Methodology Overview}
\label{app:method}

In this section, we provide a primer on the Farisght dataset and aspects of the
Domain Name System (DNS) to make our methodology accessible to a wider group of
readers.  Our primer assumes that the reader is familiar with the concept of
IPv4/IPv6 addresses and the common analogy of DNS functioning as a form of
phone book to look up IP addresses.  We will first discuss DNS, common DNS
terminology, and DNS resolution, i.e., how a client uses the DNS to resolve a
name to a value.  There we will see that DNS is not only a tool for looking up
IPs, but instead a globally distributed error tolerant database used for
various forms of lookups.  Finally, we discuss the Farsight dataset and how it
is collected.

\subsection{DNS}

Here, we first introduce the basics of DNS.  Please see \Fref{tab:dns_terms}
for an overview of DNS related terms and abbreviations we use.

\begin{figure}[t!]
	\begin{center}
		\includegraphics[width=.95\columnwidth, trim=0cm 1.8cm 0cm 1.8cm, clip]{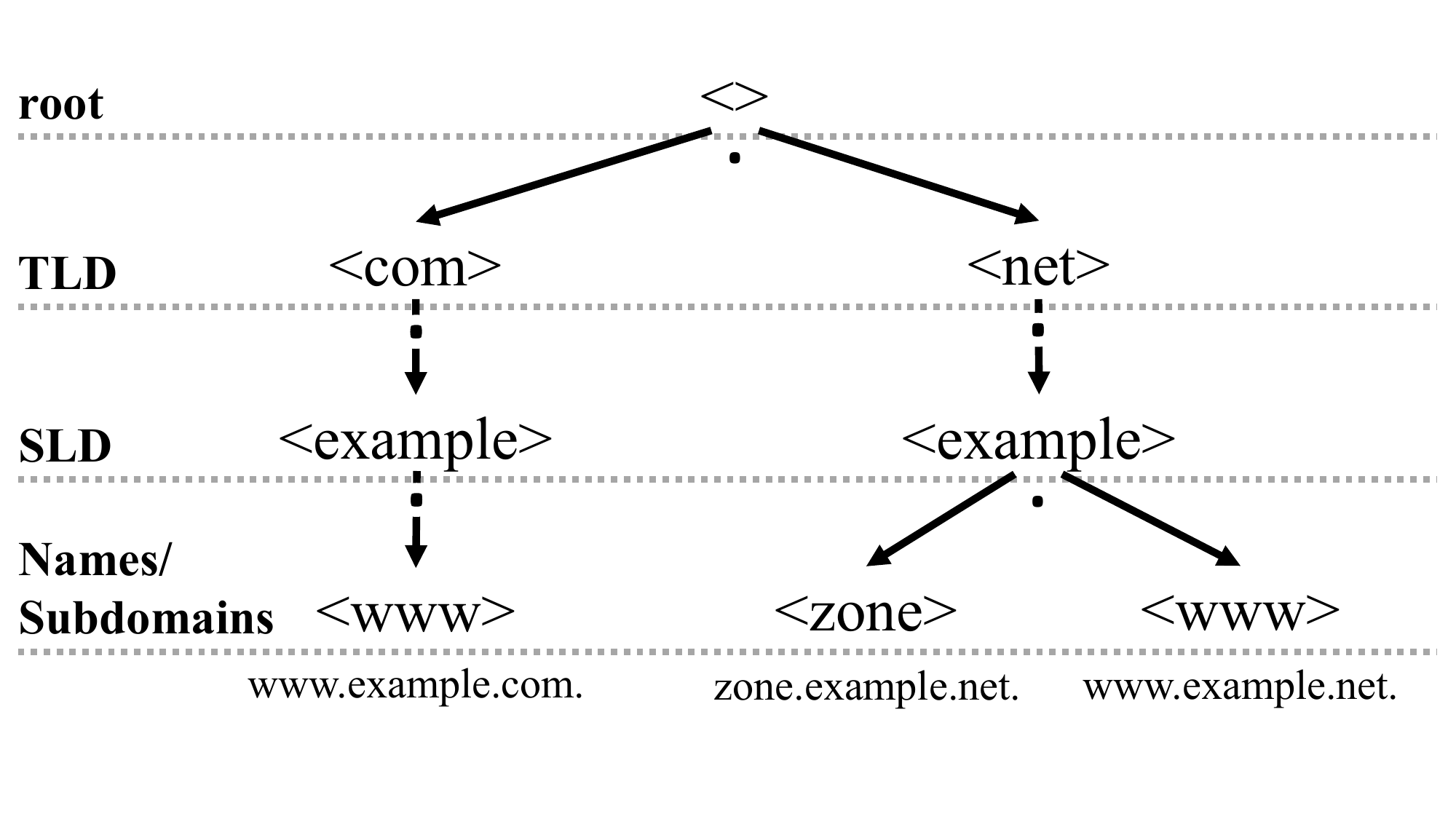}
		\caption{Simplified view of the DNS tree, starting from the root (empty label), for \texttt{www.\-ex\-am\-ple.\-com.}, \texttt{zone.example.net.} and \texttt{www.example.net.}, with all labels enclosed with `\texttt{<>}'.}
		\label{fig:dns_tree}
	\end{center}
\end{figure}

The DNS is a tree-shaped hierarchy for \emph{names}~\cite{RFC1034,RFC1035}
consisting of multiple labels delimited by dots~\cite{RFC1034}, with the root
of the tree at the end of the name, see also \Fref{fig:dns_tree}.  Names that
reach up to the root, i.e., have a right-most label that is empty, are also
called FQDNs (Fully Qualified Domain Names).  The final `\texttt{.}' separating
the empty root label is usually omitted when spelling out FQDNs~\cite{RFC2396}.
One most regularly encounters names when included in a URI on the
web~\cite{RFC2396}, i.e., in the form of \url{https://www.example.com/}.  A
zone can contain names (as leaf nodes) that form RRsets consisting of the name
and all resource records of one specific RRtype for that name, and a name can
have multiple RRsets for different RRtypes~\cite{RFC1035}.  Similarly, a zone
(parent) can contain a delegation to one or multiple other `authoritative DNS
server' for a zone below itself (child), creating a zone-cut~\cite{RFC1035}.  A
zone is `authoritative' for the names within itself or below if no zone-cut
takes place~\cite{RFC1035}.  We list and describe the most commonly used
RRtypes in \Fref{tab:rrtypes}.

\subsection{DNS Resolution}

The process of retrieving the RRset for a name in the DNS is called `resolving'
that name~\cite{RFC1034}.  DNS servers that recursively iterate through the DNS
tree to retry a reply~\cite{RFC1034} are called `recursors' or `recursive
resolvers.' Operating systems usually contain a so-called `stub'
resolver~\cite{RFC4074}, which simply forwards DNS resolution requests to a
configured recursive resolver, for example one provided by the end-users'
Internet service provider, or one of the well known public resolvers, e.g.,
\texttt{1.1.1.1} (Cloudflare), \texttt{8.8.8.8} (Google), or \texttt{9.9.9.9}
(Packet Clearing House/IBM).  These then resolve a requested name for a given
RRtype (together: query) for the client and return the answer, i.e., the
retrieved RRset~\cite{RFC1035}.

\begin{table*}[t!]
\centering
\caption{List of common DNS terms and abbreviations. See RFC8499 for a comprehensive list and as standard reference~\cite{RFC8499}.}
\label{tab:dns_terms}
\footnotesize
\setlength{\tabcolsep}{2pt}
\begin{tabularx}{\textwidth}{rrlX}
\toprule
&\textbf{Abbrev.} & \textbf{Term}                                                               & \textbf{Description}      \\ \midrule
\textbf{A}&&ADDITIONAL & Additional information in a DNS response, may consists of one or multiple RRsets~\cite{RFC1035}.\\
\rowcolor[HTML]{C0C0C0} 
\cellcolor[HTML]{FFFFFF}&&ANSWER & The part of a DNS response that contains one or multiple RRsets that hold the answer to the query. Commonly only if the queried server is authoritative for the QNAME, or a recursive resolver~\cite{RFC1035}.\\
&&Apex & All RRsets whose RRname is equal to the zone are `at the apex' of a zone. \\
\rowcolor[HTML]{C0C0C0} 
\cellcolor[HTML]{FFFFFF}&&Authoritative DNS Server & DNS server to which a zone is delegated, who can answer queries based on its local zone file.\\
\textbf{B}&&Bailiwick & Names either below or matching a zone are `in-bailiwick' for that zone.\\
\rowcolor[HTML]{C0C0C0} 
\cellcolor[HTML]{FFFFFF}&&BIND notation& The common notation of RRs for a zone in the form of \texttt{<FQDN> <CLASS> <RRtype> <RDATA>}. The syntax is more complex, but we will use this most simple form throughout the paper.\\
\textbf{C}&&Cache & A local temporary storage on recursive resolvers populated with earlier retrieved RRsets whose TTL has not yet expired~\cite{RFC1035,RFC2308}.\\
\rowcolor[HTML]{C0C0C0} 
\cellcolor[HTML]{FFFFFF}&&Caching & The process of committing retrieved RRsets to a cache, but also serving answers from this cache. \\
&&Cache-Hit & A query for which a recursive resolver is able to provide an answer from its cache. \\
\rowcolor[HTML]{C0C0C0} 
\cellcolor[HTML]{FFFFFF}&&Cache-Miss & A query for which a recursive resolver is unable to provide an answer from its cache, and has to perform recursion instead.\\
&&Child & A zone that has been delegated by a parent, i.e., a zone that is deeper in the tree than its parent. \\
\rowcolor[HTML]{C0C0C0} 
\cellcolor[HTML]{FFFFFF}&&CLASS & The DNS class. This is essentially always \texttt{IN} for Internet~\cite{RFC1035}, even though other classes (\texttt{CH} for Chaos~\cite{RFC2929}, \texttt{HS} for Hesiod~\cite{RFC2929}, \texttt{NONE}~\cite{RFC2136}, and \texttt{ANY}~\cite{RFC1035}) do exist.\\
\textbf{D}&&Delegation & The process of pointing to different authoritative servers for a child of the current zone. \\
\rowcolor[HTML]{C0C0C0} 
\cellcolor[HTML]{FFFFFF}&DNS & Domain Name System & A system to resolve names to a variety of data points, which replaced \texttt{/etc/hosts}~\cite{RFC0953}. \\
\textbf{E}&&Expire& A value in \texttt{SOA} records, instructing secondary authoritative servers how long (in seconds) they should wait after a failed zone transfer until they stop being authoritative for a zone.\\
\rowcolor[HTML]{C0C0C0} 
\cellcolor[HTML]{FFFFFF}\textbf{F}&FQDN & Fully Qualified Domain Name & An FQDN is a name, i.e., see below, containing all labels from the terminal label to the root (the empty label above the TLD)~\cite{RFC1035}. Hence, all FQDNs are names, but not all names are FQDNs.\\
\textbf{G}&&Glue & \texttt{A} and \texttt{AAAA} RRsets send along with \texttt{NS} that are in-bailiwick for a delegated zone by the authoritative NS for the parent in response to a recursive resolver trying to resolve a name in or below the child zone, to enable said recursive resolver to reach the in-bailiwick NS, as their authoritative \texttt{A} and \texttt{AAAA} records would have to be provided by themselves~\cite{RFC1912}.\\
\rowcolor[HTML]{C0C0C0} 
\cellcolor[HTML]{FFFFFF}\textbf{N}&&Name & A `\texttt{.}'-delimited set of labels, ordered by the distance to the root of the DNS tree from left (greatest) to right (smallest).\\
&&Negative response caching TTL& Value provided in \texttt{SOA} records that instructs a recursive resolver on how long it should cache the non-existence of records~\cite{RFC2308}.\\
\rowcolor[HTML]{C0C0C0} 
\cellcolor[HTML]{FFFFFF}&NS & Nameserver & A server implementing the DNS protocol, commonly used for authoritative DNS servers.\\
\textbf{P}&&Parent & A zone which is above a child in the DNS hierarchy, that delegates the child by reporting the NS authoritative for the child zone. \\
\rowcolor[HTML]{C0C0C0} 
\cellcolor[HTML]{FFFFFF}&&Primary& The authoritative server of a zone that holds the primary copy of the zone file and distributes it to secondaries via zone-transfers.\\
\textbf{Q}&QNAME & Query Name & The FQDN in a DNS query. \\
\rowcolor[HTML]{C0C0C0} 
\cellcolor[HTML]{FFFFFF}&&Query & A DNS request either from a stub to a recursive resolver or from a recursive resolver to an authoritative server. \\
&&QUESTION & The part of a DNS query or response that contains the combination of RRtype and RRname which was the initial query. \\
\rowcolor[HTML]{C0C0C0} 
\cellcolor[HTML]{FFFFFF}\textbf{R}&RDATA & Response Data & The typed value associated with an RRtype for an RRname. \\
&&Record & See Resource Record. \\
\rowcolor[HTML]{C0C0C0} 
\cellcolor[HTML]{FFFFFF}&&Recursion & Process of traversing the DNS hierarchy to retrieve the answer to a query from a NS that is authoritative for that zone.\\
&&Recursive Resolver & A DNS server performing recursion for clients. \\
\rowcolor[HTML]{C0C0C0} 
\cellcolor[HTML]{FFFFFF}&&Recursive DNS Server & See Recursive Resolver.\\
&&Refresh Time& A field in SOA records that communicates the frequency of zone transfers to secondary authoritative servers of a zone.\\
\rowcolor[HTML]{C0C0C0} 
\cellcolor[HTML]{FFFFFF}&&Reply & The DNS packet sent in response to a query.\\
&&Request & See Query.\\
\rowcolor[HTML]{C0C0C0} 
\cellcolor[HTML]{FFFFFF}&&Resolution & See Recursion.\\
&&Resolve & See Recursion. \\
\rowcolor[HTML]{C0C0C0} 
\cellcolor[HTML]{FFFFFF}&&Resolver & See Recursive Resolver.\\
&&Response & See Reply. \\
\rowcolor[HTML]{C0C0C0} 
\cellcolor[HTML]{FFFFFF}&&Retry& A value in \texttt{SOA} records, instructing secondary authoritative servers how long (in seconds) they should wait after a failed zone transfer to initiate a new transfer attempt.\\
&&Root & The top label of the DNS tree, i.e., the \emph{root} of the tree.\\
\rowcolor[HTML]{C0C0C0} 
\cellcolor[HTML]{FFFFFF}&&Root-Server & DNS servers that are authoritative for names at the root, i.e., TLDs~\cite{RFC2870}.\\
&RR& Resource Record & An entry at a node (label) within the DNS, consisting of the RRname, CLASS, RRtype, TTL, and its RRdata.\\
\rowcolor[HTML]{C0C0C0} 
\cellcolor[HTML]{FFFFFF}&RRname & Resource Record Name & The FQDN associated with a specific RR.\\
&RRset & Resource Record Set & The set of all RR that have the same RRtype and RRname.\\
\rowcolor[HTML]{C0C0C0} 
\cellcolor[HTML]{FFFFFF}&RRtype & Resource Record Type & The type of a RR, see \Fref{tab:rrtypes}.\\
\textbf{S}&&Serial& An identifier for the version of a zone in the \texttt{SOA} record. This is an integer, and \emph{must} be monotonously increasing. Commonly, the syntax for this value is YYYYMMDD00 for the first serial created on a day, continuously incremented over the day. This seeing the same serial on two authoritative servers for one zone means that the zone files \emph{should} be in sync, and no zone transfer is needed.\\
\rowcolor[HTML]{C0C0C0} 
\cellcolor[HTML]{FFFFFF}&& Secondary & A server that is authoritative for a zone, but receives the zone via a zone-transfer from a primary.\\
&SLD & Second Level Domain & A domain that is a child of a TLD.\\
\rowcolor[HTML]{C0C0C0} 
\cellcolor[HTML]{FFFFFF}&&Stub & See Stub Resolver. \\
&&Stub Resolver & A DNS server that does not perform recursion but instead just forwards queries it receives to a recursive resolver. \\
\rowcolor[HTML]{C0C0C0} 
\cellcolor[HTML]{FFFFFF}&&Subdomain & Generally, a domain below a parent, similar to a child, but only used for zones that are at least below SLDs. \\
\textbf{T}&TLD & Top Level Domain & Domains that are children of the root.\\
\rowcolor[HTML]{C0C0C0} 
\cellcolor[HTML]{FFFFFF}&TTL & Time-to-Live & The time a received RRset may be used to answer queries for its RRname and RRtype from the cache. \\
\textbf{Z}&&Zone & A zone represents a part of the DNS tree and contains a collection of RRsets for which it is authoritative. This means that the names of these RRsets are in bailiwick of the zone, and authority for these names has not been delegated elsewhere, i.e., the zone contains the final--hence authoritative--answer for queries for these names. For references to corner cases, see RFC8499, Section~7~\cite{RFC8499}. \\
\rowcolor[HTML]{C0C0C0} 
\cellcolor[HTML]{FFFFFF}&&Zone file & While nowadays database backed DNS servers are more common, Zones used to be stored in a single text file in BIND notation. Hence, this name is still being used for the data store of zone data~\cite{RFC1033}.\\
&&Zone-Cut & A zone-cut is the point in an FQDN where the authority is delegated from one zone to the other.\\
\rowcolor[HTML]{C0C0C0} 
\cellcolor[HTML]{FFFFFF}&& Zone Transfer & Traditionally, there were primary and secondary authoritative servers. Changes would be made on the primary and then distributed to secondaries via zone transfers. Using only DNS, this can be done using an \texttt{AXFR} RRtype query, to which the primary replies (to authorized secondaries) with a copy of its zone file, i.e., all RRsets of a zone are within the ANSWER section. Alternatively, an \texttt{IXFR} can be used, where the secondary provides the primary with its current \texttt{SOA} serial, and the primary then only sends the difference between the zone file with the primary's serial and the one with the secondary's serial~\cite{RFC5936}. \\
\bottomrule

\end{tabularx}
\end{table*}

\begin{figure}[t!]
	\begin{center}
		\includegraphics[width=.95\columnwidth, trim=13cm 3cm 4cm 2cm,clip]{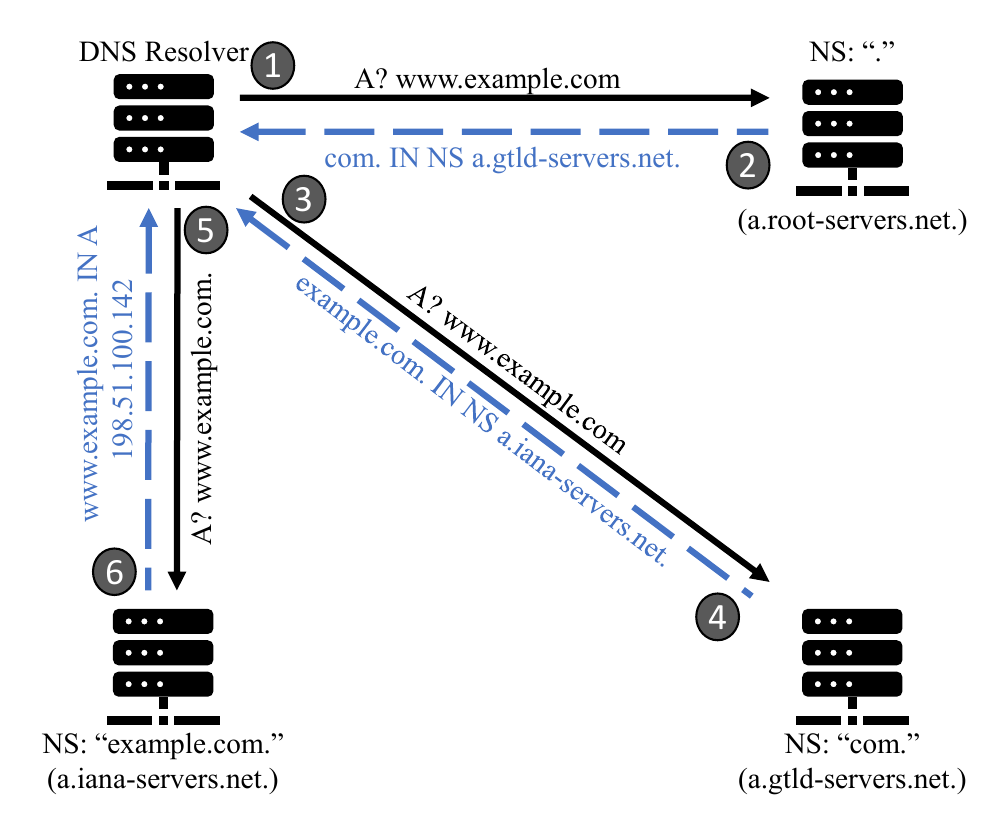}
		\caption{Simplified overview of DNS resolution for \texttt{IN A www.\-ex\-am\-ple.\-com.}: (1) The resolver asks the root-servers for \texttt{www.\-ex\-am\-ple.\-com.}, (2) The resolver is redirected to the authoritative NS for \texttt{com.}, (3) The resolver asks the \texttt{com.} NS for \texttt{www.\-ex\-am\-ple.\-com.}, (4) The resolver is redirected to the authoritative NS for \texttt{example.com.}, (5) The resolver asks the \texttt{example.com.} NS for \texttt{www.\-ex\-am\-ple.\-com.}, (6) The resolver receives the \texttt{A} RRset for \texttt{www.\-ex\-am\-ple.\-com.} in response.}
		\label{fig:dns_res}
	\end{center}
\vskip-1.5em
\end{figure}

\vspace{0.5em}
\noindent\textbf{Resolution.}
Recursion takes place by the recursive resolver asking at least one
authoritative DNS servers for all zones on the path to the root for the name,
starting at the root, see RFC1034 and RFC1035~\cite{RFC1034,RFC1035}.  A
recursive resolver usually contains a static `root hint' that lists the IP
addresses of the DNS servers authoritative for the root (`\texttt{.}').  When a
recursive resolver asks an authoritative server for an RRset in a zone for
which the authoritative server is not authoritative, while being authoritative
for a zone which contains a delegation for a child that is closer to the
requested name, it returns the name of that zone and the responsible
authoritative DNS servers.\footnote{We are skipping the concepts of QNAME
minimization~\cite{RFC7816}, \texttt{NS} hardening, and GLUE
records~\cite{RFC1033} for simplification here.} For example, if a recursive
resolver has to resolve the IPv4 address for \texttt{www.\-ex\-am\-ple.\-com.},
it will first ask the root-servers for \texttt{A www.\-ex\-am\-ple.\-com.}.  As
these are not authoritative for \texttt{example.com.}, they will reply with the
RRset containing the \texttt{NS} for \texttt{com.}.  The recursive resolver
will then ask these for \texttt{www.\-ex\-am\-ple.\-com.}, which will reply
with the RRset containing the \texttt{NS} authoritative for
\texttt{example.com.}.  Finally, the recursive resolve can then ask the
authoritative servers for \texttt{example.com.} for
\texttt{www.\-ex\-am\-ple.\-com.}.  As these are authoritative for the zone,
they will return the requested rdata, e.g., the IPv4 address of
\texttt{www.\-ex\-am\-ple.\-com.}, if \texttt{www.\-ex\-am\-ple.\-com.} is not
further delegated and an RRset for the requested RRtype at
\texttt{www.\-ex\-am\-ple.\-com.} exists.  Names either below or matching a
zone are `in-bailiwick' for that zone~\cite{RFC8499}.  Please see
\Fref{fig:dns_res} for an overview of this process.

\begin{table*}[t!]
\centering
\caption{List of common RRtypes. Most RRtypes can be listed multiple times for the same name, i.e., there can be an RRset with one RRname, one RRtype, and multiple rdata. Exceptions are noted below.}
\label{tab:rrtypes}
\footnotesize
\setlength{\tabcolsep}{2pt}
\begin{tabularx}{\textwidth}{rp{0.44\textwidth}X}
\toprule
\textbf{RRtype} & \textbf{Description} & \textbf{Example}      \\ \midrule
A & Provides an IPv4 address for a name~\cite{RFC1035}.& \texttt{www.\-ex\-am\-ple.\-com. IN A 198.51.100.11}\\
\rowcolor[HTML]{C0C0C0} 
AAAA & Provides an IPv6 address for a name~\cite{RFC3596}.& \texttt{www.\-ex\-am\-ple.\-com. IN AAAA 2001:db8:44c:d1a::2 } \\
CNAME & References another name, whose corresponding values for the requested RRtype should be used. May not be combined with other RRtypes for this name and only occurs once. If \texttt{www.\-ex\-am\-ple.\-com. IN CNAME test.example.net.} exists, as well as \texttt{test.example.net. IN A 198.51.100.11}, a recursive resolver resolving the \texttt{A} record for \texttt{www.\-ex\-am\-ple.\-com.} will receive \texttt{IN CNAME test.example.net.} as the response when querying the authoritative server for \texttt{example.com.}; It will then subsequently query \texttt{test.example.net.} for its \texttt{A} record. \texttt{CNAME}s can point to arbitrary FQDNs. If the NS is authoritative for the zone of the \texttt{CNAME} as well as the target, it should include the \texttt{A} record of the target in the ADDTIONAL section of the reply.~\cite{RFC1035} & \texttt{www.\-ex\-am\-ple.\-com. IN CNAME test.example.net.} \\
\rowcolor[HTML]{C0C0C0} 
DNAME & Similar to a \texttt{CNAME}, but mirrors a whole domain. Must be the only RRtype within a zone~\cite{RFC6672}. & \texttt{example.com. IN DNAME example.net.}\\
MX & Communicates the names of the mail servers responsible for handling inbound mails for a zone. May occur multiple times and has a preference for each record with the lowest value being most preferred. \texttt{MX} can point to arbitrary FQDNs~\cite{RFC1035}.& \texttt{example.com. IN MX 10 mail.example.net.} \\
\rowcolor[HTML]{C0C0C0} 
NS & Each zone must have at least one, by policy usually at least two for TLDs and SLDs, \texttt{NS} records identifying the authoritative DNS servers for this zone. These have to be set in the zone's apex as well as in the parent (creating the delegation). If the names used in \texttt{NS} records of a zone are within that zone, the parent must also provide \texttt{A} and/or \texttt{AAAA} records for these names. Even though the parent is not authoritative, it will send these RRsets as `ADDITIONAL' information along with QNAMES in or below \texttt{example.com.} so that recursive resolvers have a hint on where to find the nameservers for the domain with in-bailiwick \texttt{NS}. See `Glue' in \Fref{tab:dns_terms}. \texttt{NS} records \emph{must} point to names that have an \texttt{A} or \texttt{AAAA} record~\cite{RFC1035}. \texttt{CNAME}s are not allowed in \texttt{NS} records~\cite{RFC1035,RFC1912}. & \texttt{example.com. IN NS ns0.example.com.}\\
PTR & \texttt{PTR} or `pointer' records point to another part of the DNS tree. They are commonly used as `reverse pointers' for IPv4 and IPv6 addresses mapping these to FQDNs \emph{independent} of forward lookups~\cite{RFC5855}. For each IPv4 address, there is a representation below \texttt{in-addr.arpa.}~\cite{RFC1035,RFC2317}, and for each IPv6 address one under \texttt{ip6.arpa.}~\cite{RFC3152}, but the existence of a \texttt{PTR} pointing to, e.g., \texttt{web01.example.com.} does not imply or require the existence of a corresponding \texttt{A} or \texttt{AAAA} record. & \texttt{11.100.51.198.in-addr.arpa. IN PTR web01.example.com.}\\
\rowcolor[HTML]{C0C0C0} 
SOA & The `Start of Authority' record contains metadata for a domain. It contains, in order, the first authoritative NS of a zone, the email address of the responsible operator(s), with the first dot having to be replaced with an `@', the serial of the zone, an thereafter the refresh, retry, and expire timeouts, followed by the negative response TTL~\cite{RFC1035}.& \texttt{example.com. IN SOA ns0.example.com. hostmaster.example.com. 2022111701 28800 7200 604800 86400}\\
TXT & This record type is a free text record, which can contain an arbitrary set of up to \(2^{16}\) characters~\cite{RFC1464}. In practice, this record type is used for a variety of applications~\cite{RFC6763}. & \texttt{example.com. IN TXT "Hello world!"} \\
    & & \\
\multicolumn{3}{l}{\textbf{Uses of TXT records:}}\\
\midrule
\rowcolor[HTML]{C0C0C0} 
SPF & SPF, the Sender Policy Framework, allows domain owners to specify the hosts that are allowed to originate emails for a domain. This can be done via a set of mechanics, including specifying names (where the addresses referenced in \texttt{A} and \texttt{AAAA} records are then allowed to send emails), IPv4/IPv6 addresses and network prefixes, allowing the MXes of a domain, and including the SPF settings of another domain. Furthermore, a fall-back policy can be specified. While it used to have its own RRtype~\cite{RFC4408}, it now uses \texttt{TXT} records~\cite{RFC7208}.& \texttt{example.com. IN TXT "v=spf1 a:mail.example.com mx ip4:198.51.100.11 ip6:2001:db8:44c:d1a::2 include:test.example.com. -all"}\\
DMARC & DMARC, Domain Message Authentication Reporting, uses a \texttt{TXT} record to signal mail servers receiving emails from a domain how to treat emails that do neither validate according to the SPF of DKIM settings of a domain. It also clarifies how matching between the FROM header of a message and the DKIM/SPF policy has to take place. Furthermore, it allows operators to specify email addresses at which they want to receive reports about the number of accepted and rejected emails for their domain from remote servers, including the rejection reason. This can be done as aggregate reports (\texttt{rua=mailto:...}) or full forensic reports (\texttt{ruf=mailto:...}). Please note that this is a simplified description of DMARC, for further information, please see the corresponding RFCs~\cite{RFC7489}. & \texttt{\_dmarc.example.com. IN TXT "v=DMARC1; p=reject; rua=mailto:admin@example.com"}\\
\rowcolor[HTML]{C0C0C0} 
DKIM & DKIM, Domain Key Identified Mail, is an email security mechanism where mail server sign outbound mail for a domain using a private key, also providing a selector identifying the used key. The public key is then placed in the DNS under \texttt{<selector>.\_domainkey.example.com.}.~\cite{RFC6376} & \texttt{key01.\_domainkey.example.com. IN TXT "v=DKIM1; t=s; h=sha256; p=MIIBI... "}\\
Zoom Auth. & & \texttt{example.com. IN TXT "ZOOM\_verify\_5SD..."}\\
\rowcolor[HTML]{C0C0C0} 
MS Auth. & \cellcolor[HTML]{FFFFFF} & \texttt{example.com. IN TXT "MS=ms30654321"}\\
Google Auth. & & \texttt{example.com. IN TXT "google-site-verification=gzFR..."}\\
\rowcolor[HTML]{C0C0C0} 
Amazon SES Auth. & \cellcolor[HTML]{FFFFFF} & \texttt{example.com. IN TXT "amazonses:CyLQP..."}\\
Adobe Auth. & \multirow{-5}{0.44\textwidth}{\cellcolor[HTML]{FFFFFF} By placing a provided token at an RRname within a zone, one can proof ownership of that zone. This mechanic is used by several online and cloud services, including authorizing TLS certificates~\cite{RFC8555}. This list is not exhaustive.}& \texttt{example.com. IN TXT "adobe-idp-site-verification=6c3..."}\\
\bottomrule
\end{tabularx}
\end{table*}

\vspace{0.5em}
\noindent\textbf{Caching.}
Recursive resolution is a comparably long and latency dependent process.  As
such, recursive resolvers employ caching to improve their response time.  If a
resolver successfully retrieved an RRset, it will put this RRset into its local
cache.  If a subsequent request from a stub resolver for that RRset reaches the
recursive resolver, the resolver will not perform recursion, but instead reply
from cache.  The amount of time an RRset remains in a recursive resolver's
cache depends on the configured Time-To-Live (TTL) of that RRset.\footnote{We
are skipping the concept of using stale caches for resiliency
here~\cite{RFC8767}.} If a request can be answered from a recursive resolver's
cache, it is called a cache-hit, while cases where the RRset is not part of the
local cache are called cache-misses.

\subsection{The Farsight Dataset}
The Farsight Security Information Exchange (Farsight SIE) dataset,  is a
dataset of DNS requests shared by Farsight Inc. (now DomainTools) to allow
researchers and security professionals to handle digital threats and study the
Internet~\cite{farsight}.  In the past, it has been used to characterize
general use of the Internet,  characterize malware,  or study specific security
vulnerabilities.  Here, we describe how this dataset is being collected, and
what data it contains.

\begin{figure}[t!]
	\begin{center}
		\includegraphics[width=\columnwidth, trim=2.4cm 1.9cm 14.2cm 0cm,clip]{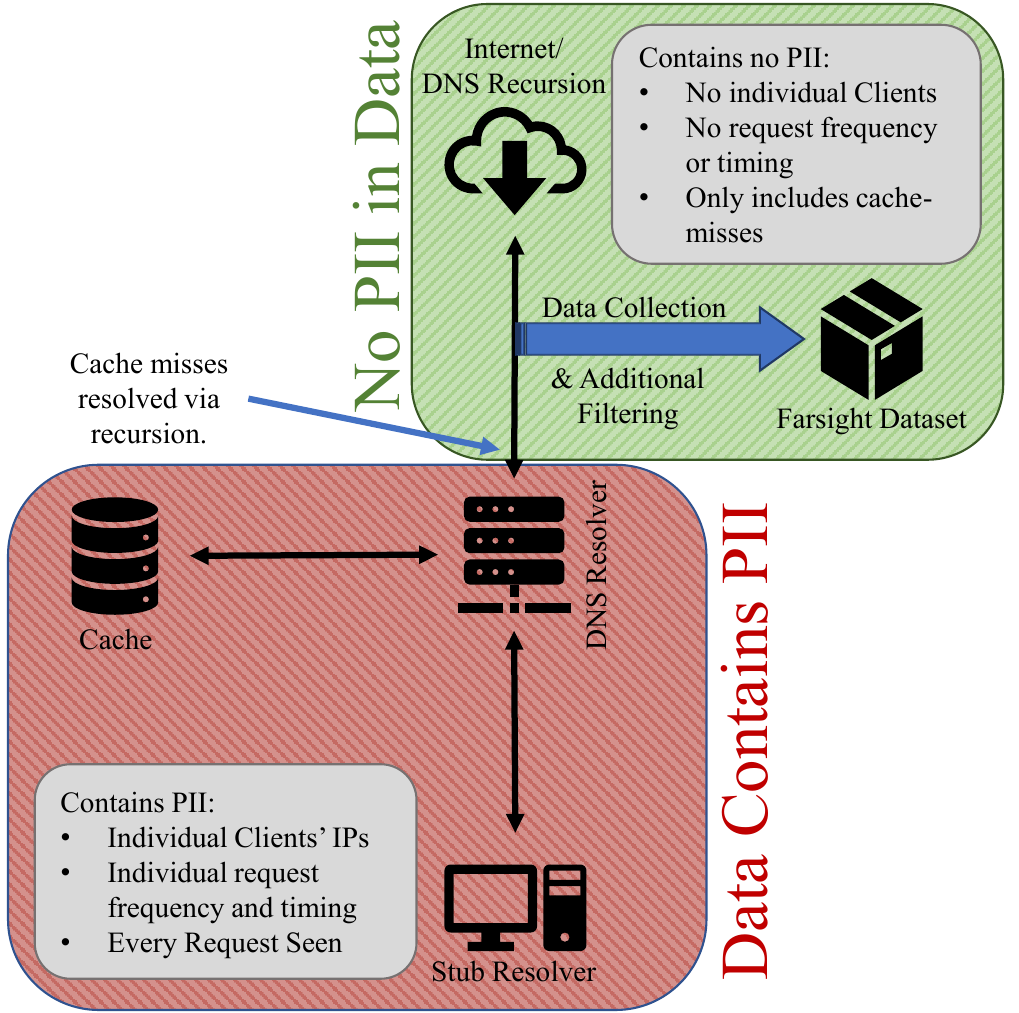}
		\caption{The collection methodology for the Farsight dataset, following \url{https://www.farsightsecurity.com/technical/passive-dns/passive-dns-faq/}. The Farsight dataset collects cache-misses encountered by recursive resolvers participating as sensors. Critical PII (DNS request times and timings, frequency of lookups and individual clients' IP addresses) are not included in the dataset, as they are not part of what is collected from cache-misses. Furthermore, additional filtering takes places for well-known privacy risks and patterns. All data produced by sensors is aggregated to the count of occurrences of specific cache-misses based on the unique touple of \texttt{<rrname, rrtype, bailiwick, rdata>}. Furthermore, additional filtering takes place to remove potential PII from the dataset, as for example, DNS queries and replies for IP-in-DNS tunneling.}
		\label{fig:fs_sensors}
	\end{center}
\end{figure}

\vspace{0.5em}
\noindent\textbf{Collection.}
The dataset is being collected on collaborating recursive resolvers (called
`sensors') around the world.  Each sensor reports all cache-misses along with
the retrieved data to Farsight.  Farsight then further aggregates this data, so
that individual sensors can not be inferred.  Please see \Fref{fig:fs_sensors}
for an overview.  By ensuring that neither individual clients nor specific
sensors can be inferred from the aggregate data view, Farsight prevents the
exposure of personally identifiable information.  More broadly speaking, from
the dataset collected by Farsight it is possible to infer that a specific name
exists, but not \emph{which} user requested it, or even \emph{where} a specific
user requested that name.  Furthermore, due to the use of cache-misses, the
exact popularity of names cannot be inferred.

\vspace{0.5em}
\noindent\textbf{Dataset Structure.}
We use a historic dataset of all cache misses observed by participating DNS
resolvers spanning from January~1, 2015 to \lastdatefull in per-month slices.
A unique cache miss is defined by the tuple of \texttt{<rrname, rrtype,
bailiwick, rdata>}.  See \Fref{tab:data} for an overview of the dataset, and
the description below for a detailed explanation.

\begin{itemize}
	\item \texttt{count:} The aggregate number of times cache misses for this unique tuple of rrname, rrtype, bailiwick, and rdata (sorted) have been observed by Farsight sensors. There is no distinction between one sensor having seen a tuple 10 times or 10 sensors having each seen it once. Furthermore, the count depends on the TTL of the RRset, as a higher TTL leads to less cache misses. Hence, the count only provides an indication of request frequency, which is why we do not rely on it in our study.
Instead, we focus our analysis on establishing a lower bound on the use of cloud resources, or, in more practical terms, we determine \emph{if} an organization uses specific cloud resources, but not \emph{how much} they use it.
	\item \texttt{time\_first:} The first time in a month the unique tuple of rrname, rrtype, bailiwick, and rdata (sorted) was seen.
	\item \texttt{time\_last:} The last time in a month the unique tuple of rrname, rrtype, bailiwick, and rdata (sorted) was seen.
	\item \texttt{rrname:} The name for which the rrset has been requested.
	\item \texttt{rrtype:} The rrtype of the requested rrset.
	\item \texttt{bailiwick:} The zone from which a reply to a query was received, e.g., considering the example from \Fref{fig:dns_res}, the bailiwick would be \texttt{example.com.} for \texttt{ex\-am\-ple.\-com. IN A 198.\-51.\-100.\-142} received from \texttt{a.\-iana-\-ser\-vers.\-net.}, and \texttt{com.} for \texttt{ex\-am\-ple.\-com. IN NS a.\-iana-\-ser\-vers.\-net.} received from \texttt{a.\-gtld-\-ser\-vers.\-net.}.
	\item \texttt{rdata:} A list of the answers received. If only one resource record is part of the RRset, this is a list with one element. If there are more resource records for the requested rrtype, the list contains multiple sorted elements.
\end{itemize}

\begin{table}[t!]
\centering
\caption{List of data fields in the Farsight SIE dataset. In our study, we work with monthly slices of the dataset. For an overview of common \texttt{rrtype} and \texttt{rdata} values, please see \Fref{tab:rrtypes}, and for an overview of DNS terminology, see \Fref{tab:dns_terms}.}
\label{tab:data}
\footnotesize
\setlength{\tabcolsep}{2pt}
\begin{tabularx}{\columnwidth}{lXr}
\toprule
\textbf{Field} & \textbf{Description}                                                               & \textbf{Example}      \\ \midrule
count               & \# of times the unique tuple \verb+<rrname, rrtype, bailiwick, rdata>+ has been seen.  & \texttt{12}                    \\
\rowcolor[HTML]{C0C0C0} 
time\_first         & Unix timestamp of the first occurrence of the unique tuple during the data slice. & \texttt{1422251650}            \\
time\_last          & Unix timestamp of the last occurrence of the unique tuple during the data slice.   & \texttt{1422251650}            \\
\rowcolor[HTML]{C0C0C0} 
rrname              & Requested name in the DNS.                                                         & \texttt{www.\-ex\-am\-ple.\-com}       \\
rrtype              & Requested RRtype of the query.                                                     & \texttt{A}                    \\
\rowcolor[HTML]{C0C0C0} 
bailiwick           & Zone authoritative for the reply.                                              & \texttt{example.com}           \\
rdata               & List of all responses received in a single query.                                  & \texttt{{[}"93.184.216.34"{]}} \\ \bottomrule
\end{tabularx}
\vspace{-0.75em}
\end{table}

\noindent\textbf{Visibility of the Dataset.}
The data collection approach of the Farsight dataset also explains why it is
better suited for our research question than actively collected DNS datasets.
Compared to actively collected DNS datasets, for example
OpenINTEL~\cite{hohlfeld2018operating,van2016high}, the Farsight dataset
enables us to look \emph{deeper} into the DNS tree of individual organizations,
i.e., we will observe a specific name, e.g.,
\texttt{random.subdomain.service.example.com} as soon as that record is
requested at least once by a system using a recursive resolver that acts as a
sensor for Farsight.  In turn, active measurement platforms use a known list of
domains retrieved from the zone files of top-level domains and will regularly
request \emph{specific} records below said domain.  Using \texttt{example.com.}
here, this might be the \texttt{NS}, \texttt{MX}, \texttt{A}, and \texttt{AAAA}
record for \texttt{example.com}.  Furthermore, they may request the \texttt{A}
and \texttt{AAAA} records for \texttt{www.\-ex\-am\-ple.\-com} and a restricted
set of well-known names.  Hence, datasets collected by these platforms will not
contain data on names like \texttt{lms.students.example.com}, because the
subdomain \texttt{students.example.com} is not listed in the authoritative zone
file of the top-level-domain.

\vspace{-0.25em}
Contrary to that, the Farsight SIE dataset will contain data on
\texttt{lms.students.example.com}, if at least one client behind a sensor did
request that name during the measurement period, and data was successfully
returned.  At the same time, this also means that we miss specific names or
institutions if the corresponding DNS resources have not been requested by a
client behind a sensor contributing to the dataset.  However, this does not
pose a problem in the context of our objective to identify a \emph{lower bound}
of cloud usage in universities, as those records we \emph{do} observe are
certainly there. Also, as we discuss in \S\ref{sec:visibility}, our
measurements are not polluted by private cloud usage (e.g., Gmail) on
university campuses.
\vspace{-0.5em}

\section{Online Domain-List \& Artifact}


\noindent\textbf{Hosted on GitHub (Microsoft).}\\
\noindent\emph{Machine readable domain-list:}\\
\url{https://github.com/headsinthecloud/universities} 
\\
\noindent\emph{Artifact to measure cloud-usage for individual universities:}\\
\url{https://github.com/headsinthecloud/cloudheadschecker}

\noindent\textbf{Self-Hosted.}\\
\noindent\emph{Machine readable domain-list:}\\
\url{https://git.aperture-labs.org/Cloudheads/universities}\\
\noindent\emph{Artifact to measure cloud-usage for individual universities:}\\
\url{https://git.aperture-labs.org/Cloudheads/cloudheadschecker}

\section{Investigated Institutions}
\label{app:domains}
\subsection{Austria}
\footnotesize
\begin{description}
\vspace{-0.00em}
\item[1) Akademie der Bildenden Künste Wien:]  \hfill \\\texttt{akbild.ac.at}
\vspace{-0.00em}
\item[2) Anton Bruckner Privatuniversistät:]  \hfill \\\texttt{bruckneruni.at}
\vspace{-0.00em}
\item[3) Bertha von Suttner Privatuniversität:]  \hfill \\\texttt{suttneruni.at}
\vspace{-0.00em}
\item[4) Central European University:]  \hfill \\\texttt{ceu.edu}
\vspace{-0.00em}
\item[5) Danube Private University:]  \hfill \\\texttt{dp-uni.ac.at}
\vspace{-0.00em}
\item[6) Graz University of Technology:]  \hfill \\\texttt{tugraz.at}
\vspace{-0.00em}
\item[7) Jam Music Lab - Privatuniversität für Jazz und Popularmusik Wien:]  \hfill \\\texttt{jammusiclab.com}
\vspace{-0.00em}
\item[8) Johannes Kepler Universität Linz:]  \hfill \\\texttt{jku.at}
\vspace{-0.00em}
\item[9) Karl Landsteiner Privatuniversität für Gedundheitswissenschaften:]  \hfill \\\texttt{kl.ac.at}
\vspace{-0.00em}
\item[10) Katholische Privatuniversistät Linz:]  \hfill \\\texttt{ku-linz.at}
\vspace{-0.00em}
\item[11) Kunst Universität Graz:]  \hfill \\\texttt{kug.ac.at}
\vspace{-0.00em}
\item[12) Kunst Universität Linz:]  \hfill \\\texttt{ufg.ac.at}
\vspace{-0.00em}
\item[13) Medizinische Universität Graz:]  \hfill \\\texttt{medunigraz.at}
\vspace{-0.00em}
\item[14) Medizinische Universität Innsbruck:]  \hfill \\\texttt{i-med.ac.at}
\vspace{-0.00em}
\item[15) Medizinische Universität Wien:]  \hfill \\\texttt{meduniwien.ac.at}
\vspace{-0.00em}
\item[16) Modul University Vienna:]  \hfill \\\texttt{modul.ac.at}
\vspace{-0.00em}
\item[17) Montanuniversität Leoben:]  \hfill \\\texttt{unileoben.ac.at}
\vspace{-0.00em}
\item[18) Paracelsus Medizinische Privatuniversität:]  \hfill \\\texttt{pmu.ac.at}
\vspace{-0.00em}
\item[19) Privatuniversität Schloss Seeburg:]  \hfill \\\texttt{uni-seeburg.at}
\vspace{-0.00em}
\item[20) The Tyrolean Private University:]  \hfill \\\texttt{umit.at}
\vspace{-0.00em}
\item[21) University of Graz:]  \hfill \\\texttt{uni-graz.at}
\vspace{-0.00em}
\item[22) University of Innsbruck:]  \hfill \\\texttt{uibk.ac.at}
\vspace{-0.00em}
\item[23) University of Klagenfurt:]  \hfill \\\texttt{uni-klu.ac.at}
\vspace{-0.00em}
\item[24) University of Vienna:]  \hfill \\\texttt{univie.ac.at}
\vspace{-0.00em}
\item[25) Universität Mozarteum Salzburg:]  \hfill \\\texttt{moz.ac.at}
\vspace{-0.00em}
\item[26) Universität Salzburg:]  \hfill \\\texttt{uni-salzburg.at}
\vspace{-0.00em}
\item[27) Universität für Bodenkultur Wien:]  \hfill \\\texttt{boku.ac.at}
\vspace{-0.00em}
\item[28) Universität für Musik und Darstellende Kunst Wien:]  \hfill \\\texttt{mdw.ac.at}
\vspace{-0.00em}
\item[29) Universität für Weiterbildung Krems:]  \hfill \\\texttt{donau-uni.ac.at}
\vspace{-0.00em}
\item[30) Universität für angewandte Kunst Wien:]  \hfill \\\texttt{dieangewandte.at}
\vspace{-0.00em}
\item[31) Veterinärmedizinische Universität Wien:]  \hfill \\\texttt{vetmeduni.ac.at}
\vspace{-0.00em}
\item[32) Vienna University of Economics and Business:]  \hfill \\\texttt{wu.ac.at}
\vspace{-0.00em}
\item[33) Vienna University of Technology:]  \hfill \\\texttt{tuwien.ac.at}
\vspace{-0.00em}
\item[34) Webster Vienna Private University:]  \hfill \\\texttt{webster.ac.at}
\end{description}

\subsection{France}
\footnotesize
\begin{description}
\vspace{0.05em}
\item[1) Institut Catholique de Lyon:]  \hfill \\\texttt{ucly.fr} \\\texttt{univ-catholyon.fr}
\vspace{0.05em}
\item[2) Institut Catholique de Paris:]  \hfill \\\texttt{icp.fr}
\vspace{0.05em}
\item[3) Institut Catholique de Toulouse:]  \hfill \\\texttt{ict-toulouse.fr}
\vspace{0.05em}
\item[4) Institut National Universitaire Champollion:]  \hfill \\\texttt{univ-jfc.fr}
\vspace{0.05em}
\item[5) La Rochelle Université:]  \hfill \\\texttt{univ-larochelle.fr}
\vspace{0.05em}
\item[6) Sorbonne Université:]  \hfill \\\texttt{paris-sorbonne.fr} \\\texttt{sorbonne-universite.fr} \\\texttt{univ-paris4.fr}
\vspace{0.05em}
\item[7) Universität Paris 8 Vincennes-Saint-Denis:]  \hfill \\\texttt{univ-paris8.fr}
\vspace{0.05em}
\item[8) Universität Paris III  Sorbonne Nouvelle:]  \hfill \\\texttt{univ-paris3.fr}
\vspace{0.05em}
\item[9) Universität des Oberelsass:]  \hfill \\\texttt{uha.fr}
\vspace{0.05em}
\item[10) Université Blaise Pascal Clermont-Ferrand II:]  \hfill \\\texttt{univ-bpclermont.fr}
\vspace{0.05em}
\item[11) Université Bordeaux Montaigne:]  \hfill \\\texttt{u-bordeaux-montaigne.fr}
\vspace{0.05em}
\item[12) Université Bretagne Sud:]  \hfill \\\texttt{univ-ubs.fr}
\vspace{0.05em}
\item[13) Université Catholique de Lille:]  \hfill \\\texttt{univ-catholille.fr}
\vspace{0.05em}
\item[14) Université Catholique de l’Ouest:]  \hfill \\\texttt{uco.fr}
\vspace{0.05em}
\item[15) Université Claude-Bernard-Lyon-I:]  \hfill \\\texttt{univ-lyon1.fr}
\vspace{0.05em}
\item[16) Université François Rabelais de Tours:]  \hfill \\\texttt{univ-tours.fr}
\vspace{0.05em}
\item[17) Université Grenoble Alpes:]  \hfill \\\texttt{u-grenoble3.fr} \\\texttt{ujf-grenoble.fr} \\\texttt{univ-grenoble-alpes.fr} \\\texttt{upmf-grenoble.fr}
\vspace{0.05em}
\item[18) Université Jean-Monnet-Saint-Étienne:]  \hfill \\\texttt{univ-st-etienne.fr}
\vspace{0.05em}
\item[19) Université Jean-Moulin-Lyon-III:]  \hfill \\\texttt{univ-lyon3.fr}
\vspace{0.05em}
\item[20) Université Lille Nord de France:]  \hfill \\\texttt{cue-lillenorddefrance.fr} \\\texttt{univ-lille.fr} \\\texttt{univ-lille1.fr} \\\texttt{univ-lille2.fr} \\\texttt{univ-lille3.fr} \\\texttt{univ-littoral.fr}
\vspace{0.05em}
\item[21) Université Lumière Lyon 2:]  \hfill \\\texttt{univ-lyon2.fr}
\vspace{0.05em}
\item[22) Université Paris 1 Panthéon-Sorbonne:]  \hfill \\\texttt{pantheonsorbonne.fr} \\\texttt{univ-paris1.fr}
\vspace{0.05em}
\item[23) Université Paris 2 Panthéon-Assas:]  \hfill \\\texttt{u-paris2.fr}
\vspace{0.05em}
\item[24) Université Paris-Est Marne-la-Vallée:]  \hfill \\\texttt{univ-mlv.fr}
\vspace{0.05em}
\item[25) Université Paris-Nanterre:]  \hfill \\\texttt{parisnanterre.fr} \\\texttt{u-paris10.fr} \\\texttt{univ-paris10.fr}
\vspace{0.05em}
\item[26) Université Paris-Sud:]  \hfill \\\texttt{u-psud.fr}
\vspace{0.05em}
\item[27) Université Paul Cézanne Aix-Marseille III:]  \hfill \\\texttt{univ-cezanne.fr}
\vspace{0.05em}
\item[28) Université Paul Sabatier Toulouse III:]  \hfill \\\texttt{univ-tlse3.fr} \\\texttt{ups-tlse.fr}
\vspace{0.05em}
\item[29) Université Savoie Mont Blanc:]  \hfill \\\texttt{univ-savoie.fr}
\vspace{0.05em}
\item[30) Université Savoie-Mont-Blanc:]  \hfill \\\texttt{univ-smb.fr}
\vspace{0.05em}
\item[31) Université Sorbonne Paris Nord:]  \hfill \\\texttt{univ-paris12.fr} \\\texttt{u-pec.fr}
\vspace{0.05em}
\item[32) Université Toulouse 1 Sciences Sociales:]  \hfill \\\texttt{univ-tlse1.fr} \\\texttt{ut-capitole.fr}
\vspace{0.05em}
\item[33) Université ToulouseJean Jaurès:]  \hfill \\\texttt{univ-tlse2.fr}
\vspace{0.05em}
\item[34) Université d'Angers:]  \hfill \\\texttt{univ-angers.fr}
\vspace{0.05em}
\item[35) Université d'Artois:]  \hfill \\\texttt{univ-artois.fr}
\vspace{0.05em}
\item[36) Université d'Avignon et des Pays de Vaucluse:]  \hfill \\\texttt{univ-avignon.fr}
\vspace{0.05em}
\item[37) Université d'Orléans:]  \hfill \\\texttt{univ-orleans.fr}
\vspace{0.05em}
\item[38) Université d'Évry:]  \hfill \\\texttt{univ-evry.fr}
\vspace{0.05em}
\item[39) Université de Bordeaux:]  \hfill \\\texttt{u-bordeaux.fr} \\\texttt{u-bordeaux1.fr} \\\texttt{u-bordeaux4.fr}
\vspace{0.05em}
\item[40) Université de Bourgogne:]  \hfill \\\texttt{u-bourgogne.fr}
\vspace{0.05em}
\item[41) Université de Caen Basse-Normandie:]  \hfill \\\texttt{unicaen.fr}
\vspace{0.05em}
\item[42) Université de Cergy-Pontoise:]  \hfill \\\texttt{cyu.fr} \\\texttt{u-cergy.fr}
\vspace{0.05em}
\item[43) Université de Franche-Comté:]  \hfill \\\texttt{univ-fcomte.fr}
\vspace{0.05em}
\item[44) Université de Haute Bretagne Rennes 2:]  \hfill \\\texttt{univ-rennes2.fr}
\vspace{0.05em}
\item[45) Université de La Réunion:]  \hfill \\\texttt{univ-reunion.fr}
\vspace{0.05em}
\item[46) Université de Limoges:]  \hfill \\\texttt{unilim.fr}
\vspace{0.05em}
\item[47) Université de Lorraine:]  \hfill \\\texttt{univ-lorraine.fr}
\vspace{0.05em}
\item[48) Université de Lyon:]  \hfill \\\texttt{universite-lyon.fr}
\vspace{0.05em}
\item[49) Université de Montpellier:]  \hfill \\\texttt{umontpellier.fr} \\\texttt{univ-montp1.fr} \\\texttt{univ-montp2.fr} \\\texttt{univ-montp3.fr}
\vspace{0.05em}
\item[50) Université de Nantes:]  \hfill \\\texttt{univ-nantes.fr}
\vspace{0.05em}
\item[51) Université de Nice Sophia-Antipolis:]  \hfill \\\texttt{unice.fr}
\vspace{0.05em}
\item[52) Université de Nîmes:]  \hfill \\\texttt{unimes.fr}
\vspace{0.05em}
\item[53) Université de Paris:]  \hfill \\\texttt{univ-paris5.fr} \\\texttt{univ-paris7.fr} \\\texttt{univ-paris-diderot.fr} \\\texttt{u-paris.fr}
\vspace{0.05em}
\item[54) Université de Pau et des Pays de l'Adour:]  \hfill \\\texttt{univ-pau.fr}
\vspace{0.05em}
\item[55) Université de Perpignan Via Domitia:]  \hfill \\\texttt{univ-perp.fr}
\vspace{0.05em}
\item[56) Université de Picardie Jules Verne:]  \hfill \\\texttt{u-picardie.fr}
\vspace{0.05em}
\item[57) Université de Poitiers:]  \hfill \\\texttt{univ-poitiers.fr}
\vspace{0.05em}
\item[58) Université de Provence Aix-Marseille I:]  \hfill \\\texttt{univ-provence.fr}
\vspace{0.05em}
\item[59) Université de Reims Champagne-Ardenne:]  \hfill \\\texttt{univ-reims.fr}
\vspace{0.05em}
\item[60) Université de Rennes 1:]  \hfill \\\texttt{univ-rennes1.fr}
\vspace{0.05em}
\item[61) Université de Rouen:]  \hfill \\\texttt{univ-rouen.fr}
\vspace{0.05em}
\item[62) Université de Strasbourg:]  \hfill \\\texttt{u-strasbg.fr} \\\texttt{unistra.fr}
\vspace{0.05em}
\item[63) Université de Toulon:]  \hfill \\\texttt{univ-tln.fr}
\vspace{0.05em}
\item[64) Université de Valenciennes:]  \hfill \\\texttt{univ-valenciennes.fr}
\vspace{0.05em}
\item[65) Université de Versailles-Saint-Quentin-en-Yvelines:]  \hfill \\\texttt{uvsq.fr}
\vspace{0.05em}
\item[66) Université de la Mediterranée Aix-Marseille II:]  \hfill \\\texttt{univmed.fr}
\vspace{0.05em}
\item[67) Université de la Nouvelle-Calédonie:]  \hfill \\\texttt{unc.nc}
\vspace{0.05em}
\item[68) Université de la Polynésie française:]  \hfill \\\texttt{upf.pf}
\vspace{0.05em}
\item[69) Université de technologie de Compiègne:]  \hfill \\\texttt{utc.fr}
\vspace{0.05em}
\item[70) Université des Antilles et de la Guyane:]  \hfill \\\texttt{univ-ag.fr}
\vspace{0.05em}
\item[71) Université du Havre:]  \hfill \\\texttt{univ-lehavre.fr}
\vspace{0.05em}
\item[72) Université du Maine:]  \hfill \\\texttt{univ-lemans.fr}
\vspace{0.05em}
\item[73) Université d’Aix-Marseille:]  \hfill \\\texttt{univ-amu.fr}
\vspace{0.05em}
\item[74) Université d’Auvergne Clermont-Ferrand:]  \hfill \\\texttt{uca.fr} \\\texttt{u-clermont1.fr}
\end{description}
\subsection{Germany}
\footnotesize
\begin{description}
\vspace{-0.06em}
\vspace{-0.065em}
\item[1) Brandenburgische Technische Universität:]  \hfill \vspace{-0.065em} \\ \texttt{b-tu.de}
\vspace{-0.06em}
\vspace{-0.065em}
\item[2) Europa-Universität Viadrina Frankfurt (Oder):]  \hfill \vspace{-0.065em} \\ \texttt{euv-ffo.de}
\vspace{-0.06em}
\vspace{-0.065em}
\item[3) FU Berlin:]  \hfill \vspace{-0.065em} \\ \texttt{fu-berlin.de}
\vspace{-0.06em}
\vspace{-0.065em}
\item[4) FernUniversität Hagen:]  \hfill \vspace{-0.065em} \\ \texttt{fernuni-hagen.de}
\vspace{-0.06em}
\vspace{-0.065em}
\item[5) Friederich-Alexander Universität Erlangen:]  \hfill \vspace{-0.065em} \\ \texttt{fau.de}
\vspace{-0.06em}
\vspace{-0.065em}
\item[6) HU Berlin:]  \hfill \vspace{-0.065em} \\ \texttt{hu-berlin.de}
\vspace{-0.06em}
\vspace{-0.065em}
\item[7) Heinrich-Heine-Universität Düsseldorf:]  \hfill \vspace{-0.065em} \\ \texttt{hhu.de}
\vspace{-0.06em}
\vspace{-0.065em}
\item[8) Jacobs University Bremen:]  \hfill \vspace{-0.065em} \\ \texttt{jacobs-university.de}
\vspace{-0.06em}
\vspace{-0.065em}
\item[9) Karlsruhe Institute of Technology:]  \hfill \vspace{-0.065em} \\ \texttt{kit.edu}
\vspace{-0.06em}
\vspace{-0.065em}
\item[10) Katholische Universität Eichstätt-Ingolstadt:]  \hfill \vspace{-0.065em} \\ \texttt{ku.de}
\vspace{-0.06em}
\vspace{-0.065em}
\item[11) RWTH Aachen:]  \hfill \vspace{-0.065em} \\ \texttt{rwth-aachen.de}
\vspace{-0.06em}
\vspace{-0.065em}
\item[12) Ruhr Universität Bochum:]  \hfill \vspace{-0.065em} \\ \texttt{ruhr-uni-bochum.de}
\vspace{-0.06em}
\vspace{-0.065em}
\item[13) TU Berlin:]  \hfill \vspace{-0.065em} \\ \texttt{tu-berlin.de}
\vspace{-0.06em}
\vspace{-0.065em}
\item[14) TU Braunschweig:]  \hfill \vspace{-0.065em} \\ \texttt{tu-braunschweig.de}
\vspace{-0.06em}
\vspace{-0.065em}
\item[15) TU Chemnitz:]  \hfill \vspace{-0.065em} \\ \texttt{tu-chemnitz.de}
\vspace{-0.06em}
\vspace{-0.065em}
\item[16) TU Clausthal:]  \hfill \vspace{-0.065em} \\ \texttt{tu-clausthal.de}
\vspace{-0.06em}
\vspace{-0.065em}
\item[17) TU Darmstadt:]  \hfill \vspace{-0.065em} \\ \texttt{tu-darmstadt.de}
\vspace{-0.06em}
\vspace{-0.065em}
\item[18) TU Dortmund:]  \hfill \vspace{-0.065em} \\ \texttt{tu-dortmund.de}
\vspace{-0.06em}
\vspace{-0.065em}
\item[19) TU Dresden:]  \hfill \vspace{-0.065em} \\ \texttt{tu-dresden.de}
\vspace{-0.06em}
\item[20) TU Freiberg:]  \hfill \vspace{-0.065em} \\ \texttt{tu-freiberg.de}
\vspace{-0.06em}
\item[21) TU Hamburg:]  \hfill \vspace{-0.065em} \\ \texttt{tuhh.de}
\vspace{-0.06em}
\item[22) TU Ilmenau:]  \hfill \vspace{-0.065em} \\ \texttt{tu-ilmenau.de}
\vspace{-0.06em}
\item[23) TU München:]  \hfill \vspace{-0.065em} \\ \texttt{tum.de}
\vspace{-0.06em}
\item[24) Universität Augsburg:]  \hfill \vspace{-0.065em} \\ \texttt{uni-augsburg.de}
\vspace{-0.06em}
\item[25) Universität Bamberg:]  \hfill \vspace{-0.065em} \\ \texttt{uni-bamberg.de}
\vspace{-0.06em}
\item[26) Universität Bayreuth:]  \hfill \vspace{-0.065em} \\ \texttt{uni-bayreuth.de}
\vspace{-0.06em}
\item[27) Universität Bielefeld:]  \hfill \vspace{-0.065em} \\ \texttt{uni-bielefeld.de}
\vspace{-0.06em}
\item[28) Universität Bonn:]  \hfill \vspace{-0.065em} \\ \texttt{uni-bonn.de}
\vspace{-0.06em}
\item[29) Universität Bremen:]  \hfill \vspace{-0.065em} \\ \texttt{uni-bremen.de}
\vspace{-0.06em}
\item[30) Universität Duisburg/Essen:]  \hfill \vspace{-0.065em} \\ \texttt{uni-due.de}
\vspace{-0.06em}
\item[31) Universität Erfurt:]  \hfill \vspace{-0.065em} \\ \texttt{uni-erfurt.de}
\vspace{-0.06em}
\item[32) Universität Flensburg:]  \hfill \vspace{-0.065em} \\ \texttt{uni-flensburg.de}
\vspace{-0.06em}
\item[33) Universität Frankfurt:]  \hfill \vspace{-0.065em} \\ \texttt{uni-frankfurt.de}
\vspace{-0.06em}
\item[34) Universität Freiburg:]  \hfill \vspace{-0.065em} \\ \texttt{uni-freiburg.de}
\vspace{-0.06em}
\item[35) Universität Giessen:]  \hfill \vspace{-0.065em} \\ \texttt{uni-giessen.de}
\vspace{-0.06em}
\item[36) Universität Greifswald:]  \hfill \vspace{-0.065em} \\ \texttt{uni-greifswald.de}
\vspace{-0.06em}
\item[37) Universität Göttingen:]  \hfill \vspace{-0.065em} \\ \texttt{uni-goettingen.de}
\vspace{-0.06em}
\item[38) Universität Halle (Saale):]  \hfill \vspace{-0.065em} \\ \texttt{uni-halle.de}
\vspace{-0.06em}
\item[39) Universität Hamburg:]  \hfill \vspace{-0.065em} \\ \texttt{uni-hamburg.de}
\vspace{-0.06em}
\item[40) Universität Hannover:]  \hfill \vspace{-0.065em} \\ \texttt{uni-hannover.de}
\vspace{-0.06em}
\item[41) Universität Heidelberg:]  \hfill \vspace{-0.065em} \\ \texttt{uni-heidelberg.de}
\vspace{-0.06em}
\item[42) Universität Hohenheim:]  \hfill \vspace{-0.065em} \\ \texttt{uni-hohenheim.de}
\vspace{-0.06em}
\vspace{-0.010em}
\item[43) Universität Jena:]  \hfill \vspace{-0.065em} \\ \texttt{uni-jena.de}
\vspace{-0.06em}
\vspace{-0.010em}
\item[44) Universität Kaiserslautern:]  \hfill \vspace{-0.065em} \\ \texttt{uni-kl.de}
\vspace{-0.06em}
\vspace{-0.010em}
\item[45) Universität Kassel:]  \hfill \vspace{-0.065em} \\ \texttt{uni-kassel.de}
\vspace{-0.06em}
\vspace{-0.010em}
\item[46) Universität Kiel:]  \hfill \vspace{-0.065em} \\ \texttt{uni-kiel.de}
\vspace{-0.06em}
\vspace{-0.010em}
\item[47) Universität Koblenz:]  \hfill \vspace{-0.065em} \\ \texttt{uni-koblenz-landau.de}
\vspace{-0.06em}
\vspace{-0.010em}
\item[48) Universität Konstanz:]  \hfill \vspace{-0.065em} \\ \texttt{uni-konstanz.de}
\vspace{-0.06em}
\vspace{-0.010em}
\item[49) Universität Köln:]  \hfill \vspace{-0.065em} \\ \texttt{uni-koeln.de}
\vspace{-0.06em}
\vspace{-0.010em}
\item[50) Universität Leipzig:]  \hfill \vspace{-0.065em} \\ \texttt{uni-leipzig.de}
\vspace{-0.06em}
\vspace{-0.010em}
\item[51) Universität Lübeck:]  \hfill \vspace{-0.065em} \\ \texttt{uni-luebeck.de}
\vspace{-0.06em}
\vspace{-0.010em}
\item[52) Universität Lüneburg:]  \hfill \vspace{-0.065em} \\ \texttt{leuphana.de}
\vspace{-0.06em}
\vspace{-0.010em}
\item[53) Universität Magdeburg:]  \hfill \vspace{-0.065em} \\ \texttt{ovgu.de}
\vspace{-0.06em}
\vspace{-0.010em}
\item[54) Universität Mainz:]  \hfill \vspace{-0.065em} \\ \texttt{uni-mainz.de}
\vspace{-0.06em}
\vspace{-0.010em}
\item[55) Universität Mannheim:]  \hfill \vspace{-0.065em} \\ \texttt{uni-mannheim.de}
\vspace{-0.06em}
\vspace{-0.010em}
\item[56) Universität Marburg:]  \hfill \vspace{-0.065em} \\ \texttt{uni-marburg.de}
\vspace{-0.06em}
\vspace{-0.010em}
\item[57) Universität München:]  \hfill \vspace{-0.065em} \\ \texttt{uni-muenchen.de}
\vspace{-0.06em}
\vspace{-0.010em}
\item[58) Universität Münster:]  \hfill \vspace{-0.065em} \\ \texttt{uni-muenster.de}
\vspace{-0.06em}
\vspace{-0.010em}
\item[59) Universität Oldenburg:]  \hfill \vspace{-0.065em} \\ \texttt{uni-oldenburg.de} \\\texttt{uol.de}
\vspace{-0.06em}
\vspace{-0.010em}
\item[60) Universität Osnabrück:]  \hfill \vspace{-0.065em} \\ \texttt{uni-osnabrueck.de} \\\texttt{uos.de}
\vspace{-0.06em}
\vspace{-0.010em}
\item[61) Universität Paderborn:]  \hfill \vspace{-0.065em} \\ \texttt{uni-paderborn.de}
\vspace{-0.06em}
\vspace{-0.010em}
\item[62) Universität Passau:]  \hfill \vspace{-0.065em} \\ \texttt{uni-passau.de}
\vspace{-0.06em}
\vspace{-0.010em}
\item[63) Universität Regensburg:]  \hfill \vspace{-0.065em} \\ \texttt{uni-regensburg.de}
\vspace{-0.06em}
\vspace{-0.010em}
\item[64) Universität Reutlingen:]  \hfill \vspace{-0.065em} \\ \texttt{reutlingen-university.de}
\vspace{-0.06em}
\vspace{-0.010em}
\item[65) Universität Rostock:]  \hfill \vspace{-0.065em} \\ \texttt{uni-rostock.de}
\vspace{-0.06em}
\vspace{-0.010em}
\item[66) Universität Siegen:]  \hfill \vspace{-0.065em} \\ \texttt{uni-siegen.de}
\vspace{-0.06em}
\vspace{-0.010em}
\item[67) Universität Speyer:]  \hfill \vspace{-0.065em} \\ \texttt{uni-speyer.de}
\vspace{-0.06em}
\vspace{-0.010em}
\item[68) Universität Stuttgart:]  \hfill \vspace{-0.065em} \\ \texttt{uni-stuttgart.de}
\vspace{-0.06em}
\vspace{-0.010em}
\item[69) Universität Trier:]  \hfill \vspace{-0.065em} \\ \texttt{uni-trier.de}
\vspace{-0.06em}
\vspace{-0.010em}
\item[70) Universität Tübingen:]  \hfill \vspace{-0.065em} \\ \texttt{uni-tuebingen.de}
\vspace{-0.06em}
\vspace{-0.010em}
\item[71) Universität Ulm:]  \hfill \vspace{-0.065em} \\ \texttt{uni-ulm.de}
\vspace{-0.06em}
\vspace{-0.010em}
\item[72) Universität Vechta:]  \hfill \vspace{-0.065em} \\ \texttt{uni-vechta.de}
\vspace{-0.06em}
\vspace{-0.010em}
\item[73) Universität Weimar:]  \hfill \vspace{-0.065em} \\ \texttt{uni-weimar.de}
\vspace{-0.06em}
\vspace{-0.010em}
\item[74) Universität Witten/Herdecke:]  \hfill \vspace{-0.065em} \\ \texttt{uni-wh.de}
\vspace{-0.06em}
\vspace{-0.010em}
\item[75) Universität Wuppertal:]  \hfill \vspace{-0.065em} \\ \texttt{uni-wuppertal.de}
\vspace{-0.06em}
\vspace{-0.010em}
\item[76) Universität Würzburg:]  \hfill \vspace{-0.065em} \\ \texttt{uni-wuerzburg.de}
\vspace{-0.06em}
\vspace{-0.010em}
\item[77) Universität der Bundeswehr Hamburg:]  \hfill \vspace{-0.065em} \\ \texttt{hsu-hh.de}
\vspace{-0.06em}
\vspace{-0.010em}
\item[78) Universität der Bundeswehr München:]  \hfill \vspace{-0.065em} \\ \texttt{unibw.de}
\vspace{-0.06em}
\vspace{-0.010em}
\item[79) Universität der Künste Berlin:]  \hfill \vspace{-0.065em} \\ \texttt{udk-berlin.de}
\vspace{-0.06em}
\vspace{-0.010em}
\item[80) Universität des Saarlandes:]  \hfill \vspace{-0.065em} \\ \texttt{uni-saarland.de}
\vspace{-0.06em}
\vspace{-0.010em}
\item[81) Zeppelin University:]  \hfill \vspace{-0.065em} \\ \texttt{zu.de}
\end{description}
\subsection{Switzerland}
\footnotesize
\begin{description}
\vspace{0.05em}
\item[1) EPFL:]  \hfill \\\texttt{epfl.ch}
\vspace{0.05em}
\item[2) ETH Zürich:]  \hfill \\\texttt{ethz.ch}
\vspace{0.05em}
\item[3) FernUniversität Schweiz:]  \hfill \\\texttt{fernuni.ch}
\vspace{0.05em}
\item[4) Graduate Institute Geneva:]  \hfill \\\texttt{graduateinstitute.ch}
\vspace{0.05em}
\item[5) Universita della Svizzera italiana:]  \hfill \\\texttt{usi.ch} \\\texttt{unisi.ch}
\vspace{0.05em}
\item[6) University Basel:]  \hfill \\\texttt{unibas.ch}
\vspace{0.05em}
\item[7) University Bern:]  \hfill \\\texttt{unibe.ch}
\vspace{0.05em}
\item[8) University Fribourg:]  \hfill \\\texttt{unifr.ch}
\vspace{0.05em}
\item[9) University Geneva:]  \hfill \\\texttt{unige.ch}
\vspace{0.05em}
\item[10) University Lausanne:]  \hfill \\\texttt{unil.ch} \\\texttt{idheap.ch}
\vspace{0.05em}
\item[11) University Luzern:]  \hfill \\\texttt{unilu.ch}
\vspace{0.05em}
\item[12) University Neuchatel:]  \hfill \\\texttt{unine.ch}
\vspace{0.05em}
\item[13) University St. Gallen:]  \hfill \\\texttt{unisg.ch}
\vspace{0.05em}
\item[14) University Zürich:]  \hfill \\\texttt{uzh.ch}
\end{description}
\subsection{The Netherlands}
\footnotesize
\begin{description}
\vspace{0.05em}
\item[1) Erasmus Universiteit Rotterdam:]  \hfill \\\texttt{eur.nl}
\vspace{0.05em}
\item[2) Maastricht School of Management:]  \hfill \\\texttt{msm.nl}
\vspace{0.05em}
\item[3) Maastricht University:]  \hfill \\\texttt{maastrichtuniversity.nl}
\vspace{0.05em}
\item[4) Nyenrode Business Universiteit:]  \hfill \\\texttt{nyenrode.nl}
\vspace{0.05em}
\item[5) Open Universiteit:]  \hfill \\\texttt{ou.nl}
\vspace{0.05em}
\item[6) Protestantse Theologische Universiteit:]  \hfill \\\texttt{pthu.nl}
\vspace{0.05em}
\item[7) Radboud Universiteit:]  \hfill \\\texttt{ru.nl}
\vspace{0.05em}
\item[8) Rijksuniversiteit Groningen:]  \hfill \\\texttt{rug.nl}
\vspace{0.05em}
\item[9) TIAS School for Business and Society:]  \hfill \\\texttt{tias.edu}
\vspace{0.05em}
\item[10) Technische Universiteit Delft:]  \hfill \\\texttt{tudelft.nl}
\vspace{0.05em}
\item[11) Technische Universiteit Eindhoven:]  \hfill \\\texttt{tue.nl}
\vspace{0.05em}
\item[12) Tilburg University:]  \hfill \\\texttt{uvt.nl}
\vspace{0.05em}
\item[13) Universiteit Leiden:]  \hfill \\\texttt{leidenuniv.nl} \\\texttt{universiteitleiden.nl}
\vspace{0.05em}
\item[14) Universiteit Twente:]  \hfill \\\texttt{utwente.nl}
\vspace{0.05em}
\item[15) Universiteit Utrecht:]  \hfill \\\texttt{uu.nl}
\vspace{0.05em}
\item[16) Universiteit van Amsterdam:]  \hfill \\\texttt{uva.nl}
\vspace{0.05em}
\item[17) Universiteit voor Humanistiek:]  \hfill \\\texttt{uvh.nl}
\vspace{0.05em}
\item[18) Vrije Universiteit Amsterdam:]  \hfill \\\texttt{vu.nl}
\vspace{0.05em}
\item[19) Wageningen Universiteit \& Research:]  \hfill \\\texttt{wur.nl}
\end{description}
\subsection{THE Top100 (Alphabetical)}
\footnotesize
\begin{description}
\item[1) Australian National University:]  \hfill \\\texttt{anu.edu.au}
\vspace{0.16em}
\item[2) Boston University:]  \hfill \\\texttt{bu.edu}
\vspace{0.16em}
\item[3) Brown University:]  \hfill \\\texttt{brown.edu}
\vspace{0.16em}
\item[4) California Institute of Technology:]  \hfill \\\texttt{caltech.edu}
\vspace{0.16em}
\item[5) Carnegie Mellon University:]  \hfill \\\texttt{cmu.edu}
\vspace{0.16em}
\item[6) Charité - Universitätsmedizin Berlin:]  \hfill \\\texttt{charite.de}
\vspace{0.16em}
\item[7) Chinese University of Hong Kong:]  \hfill \\\texttt{cuhk.edu.hk}
\vspace{0.16em}
\item[8) Columbia University:]  \hfill \\\texttt{columbia.edu}
\vspace{0.16em}
\item[9) Cornell University:]  \hfill \\\texttt{cornell.edu}
\vspace{0.16em}
\item[10) Dartmouth College:]  \hfill \\\texttt{dartmouth.edu}
\vspace{0.16em}
\item[11) Delft University of Technology:]  \hfill \\\texttt{tudelft.nl}
\vspace{0.16em}
\item[12) Duke University:]  \hfill \\\texttt{duke.edu}
\vspace{0.16em}
\item[13) ETH Zurich:]  \hfill \\\texttt{ethz.ch}
\vspace{0.16em}
\item[14) Emory University:]  \hfill \\\texttt{emory.edu}
\vspace{0.16em}
\item[15) Erasmus University Rotterdam:]  \hfill \\\texttt{eur.nl}
\vspace{0.16em}
\item[16) Georgia Institute of Technology:]  \hfill \\\texttt{gatech.edu}
\vspace{0.16em}
\item[17) Harvard University:]  \hfill \\\texttt{harvard.edu}
\vspace{0.16em}
\item[18) Heidelberg University:]  \hfill \\\texttt{heidelberg.edu}
\vspace{0.16em}
\item[19) Humboldt University of Berlin:]  \hfill \\\texttt{hu-berlin.de}
\vspace{0.16em}
\item[20) Imperial College London:]  \hfill \\\texttt{imperial.ac.uk}
\vspace{0.16em}
\item[21) Johns Hopkins University:]  \hfill \\\texttt{jhu.edu}
\vspace{0.16em}
\item[22) KU Leuven:]  \hfill \\\texttt{kuleuven.be}
\vspace{0.16em}
\item[23) Karolinska Institute:]  \hfill \\\texttt{ki.se}
\vspace{0.16em}
\item[24) King’s College London:]  \hfill \\\texttt{kcl.ac.uk}
\vspace{0.16em}
\item[25) Kyoto University:]  \hfill \\\texttt{kyoto-u.ac.jp}
\vspace{0.16em}
\item[26) LMU Munich:]  \hfill \\\texttt{uni-muenchen.de}
\vspace{0.16em}
\item[27) Leiden University:]  \hfill \\\texttt{universiteitleiden.nl} \\\texttt{leidenuniv.nl}
\vspace{0.16em}
\item[28) London School of Economics and Political Science:]  \hfill \\\texttt{lse.ac.uk}
\vspace{0.16em}
\item[29) Lund University:]  \hfill \\\texttt{lu.se}
\vspace{0.16em}
\item[30) Massachusetts Institute of Technology:]  \hfill \\\texttt{mit.edu}
\vspace{0.16em}
\item[31) McGill University:]  \hfill \\\texttt{mcgill.ca}
\vspace{0.16em}
\item[32) McMaster University:]  \hfill \\\texttt{mcmaster.ca}
\vspace{0.16em}
\item[33) Michigan State University:]  \hfill \\\texttt{msu.edu}
\vspace{0.16em}
\item[34) Monash University:]  \hfill \\\texttt{monash.edu} \\\texttt{monash.edu.au}
\vspace{0.16em}
\item[35) Nanyang Technological University, Singapore:]  \hfill \\\texttt{ntu.edu.sg}
\vspace{0.16em}
\item[36) National University of Singapore:]  \hfill \\\texttt{nus.edu.sg}
\vspace{0.16em}
\item[37) New York University:]  \hfill \\\texttt{nyu.edu}
\vspace{0.16em}
\item[38) Northwestern University:]  \hfill \\\texttt{northwestern.edu}
\vspace{0.16em}
\item[39) Ohio State University (Main campus):]  \hfill \\\texttt{osu.edu}
\vspace{0.16em}
\item[40) Paris Sciences et Lettres  PSL Research University Paris:]  \hfill \\\texttt{psl.eu}
\vspace{0.16em}
\item[41) Peking University:]  \hfill \\\texttt{pku.edu.cn}
\vspace{0.16em}
\item[42) Penn State (Main campus):]  \hfill \\\texttt{psu.edu}
\vspace{0.16em}
\item[43) Princeton University:]  \hfill \\\texttt{princeton.edu}
\vspace{0.16em}
\item[44) Purdue University West Lafayette:]  \hfill \\\texttt{purdue.edu}
\vspace{0.16em}
\item[45) RWTH Aachen University:]  \hfill \\\texttt{rwth-aachen.de}
\vspace{0.16em}
\item[46) Seoul National University:]  \hfill \\\texttt{snu.ac.kr}
\vspace{0.16em}
\item[47) Sorbonne Université:]  \hfill \\\texttt{univ-paris4.fr} \\\texttt{sorbonne-universite.fr}
\vspace{0.16em}
\item[48) Stanford University:]  \hfill \\\texttt{stanford.edu}
\vspace{0.16em}
\item[49) Sungkyunkwan University (SKKU):]  \hfill \\\texttt{skku.edu}
\vspace{0.16em}
\item[50) Technical University of Munich:]  \hfill \\\texttt{tum.de}
\vspace{0.16em}
\item[51) The Hong Kong University of Science and Technology:]  \hfill \\\texttt{ust.hk}
\vspace{0.16em}
\item[52) The University of Chicago:]  \hfill \\\texttt{uchicago.edu}
\vspace{0.16em}
\item[53) The University of Queensland:]  \hfill \\\texttt{uq.edu.au}
\vspace{0.16em}
\item[54) The University of Tokyo:]  \hfill \\\texttt{u-tokyo.ac.jp}
\vspace{0.16em}
\item[55) Tsinghua University:]  \hfill \\\texttt{tsinghua.edu.cn}
\vspace{0.16em}
\item[56) UCL:]  \hfill \\\texttt{ucl.ac.uk}
\vspace{0.16em}
\item[57) UNSW Sydney:]  \hfill \\\texttt{unsw.edu.au}
\vspace{0.16em}
\item[58) University of Amsterdam:]  \hfill \\\texttt{uva.nl}
\vspace{0.16em}
\item[59) University of Basel:]  \hfill \\\texttt{unibas.ch}
\vspace{0.16em}
\item[60) University of Bristol:]  \hfill \\\texttt{bris.ac.uk}
\vspace{0.16em}
\item[61) University of British Columbia:]  \hfill \\\texttt{ubc.ca}
\vspace{0.16em}
\item[62) University of California, Berkeley:]  \hfill \\\texttt{berkeley.edu}
\vspace{0.16em}
\item[63) University of California, Davis:]  \hfill \\\texttt{ucdavis.edu}
\vspace{0.16em}
\item[64) University of California, Irvine:]  \hfill \\\texttt{uci.edu}
\vspace{0.16em}
\item[65) University of California, Los Angeles:]  \hfill \\\texttt{ucla.edu}
\vspace{0.16em}
\item[66) University of California, San Diego:]  \hfill \\\texttt{ucsd.edu}
\vspace{0.16em}
\item[67) University of California, Santa Barbara:]  \hfill \\\texttt{ucsb.edu}
\vspace{0.16em}
\item[68) University of Cambridge:]  \hfill \\\texttt{cam.ac.uk}
\vspace{0.16em}
\item[69) University of Edinburgh:]  \hfill \\\texttt{ed.ac.uk}
\vspace{0.16em}
\item[70) University of Freiburg:]  \hfill \\\texttt{uni-freiburg.de}
\vspace{0.16em}
\item[71) University of Glasgow:]  \hfill \\\texttt{gla.ac.uk}
\vspace{0.16em}
\item[72) University of Groningen:]  \hfill \\\texttt{rug.nl}
\vspace{0.16em}
\item[73) University of Helsinki:]  \hfill \\\texttt{helsinki.fi}
\vspace{0.16em}
\item[74) University of Hong Kong:]  \hfill \\\texttt{hku.hk}
\vspace{0.16em}
\item[75) University of Illinois, Urbana-Champaign:]  \hfill \\\texttt{illinois.edu}
\vspace{0.16em}
\item[76) University of Manchester:]  \hfill \\\texttt{manchester.ac.uk}
\vspace{0.16em}
\item[77) University of Maryland, College Park:]  \hfill \\\texttt{umd.edu}
\vspace{0.16em}
\item[78) University of Melbourne:]  \hfill \\\texttt{unimelb.edu.au}
\vspace{0.16em}
\item[79) University of Michigan-Ann Arbor:]  \hfill \\\texttt{umich.edu}
\vspace{0.16em}
\item[80) University of Minnesota:]  \hfill \\\texttt{umn.edu}
\vspace{0.16em}
\item[81) University of Montreal:]  \hfill \\\texttt{umontreal.ca}
\vspace{0.16em}
\item[82) University of North Carolina, Chapel Hill:]  \hfill \\\texttt{unc.edu}
\vspace{0.16em}
\item[83) University of Oxford:]  \hfill \\\texttt{ox.ac.uk}
\vspace{0.16em}
\item[84) University of Pennsylvania:]  \hfill \\\texttt{upenn.edu}
\vspace{0.16em}
\item[85) University of Science and Technology of China:]  \hfill \\\texttt{ustc.edu.cn}
\vspace{0.16em}
\item[86) University of Southern California:]  \hfill \\\texttt{usc.edu}
\vspace{0.16em}
\item[87) University of Sydney:]  \hfill \\\texttt{sydney.edu.au}
\vspace{0.16em}
\item[88) University of Texas, Austin:]  \hfill \\\texttt{utexas.edu}
\vspace{0.16em}
\item[89) University of Toronto:]  \hfill \\\texttt{utoronto.ca}
\vspace{0.16em}
\item[90) University of Tübingen:]  \hfill \\\texttt{uni-tuebingen.de}
\vspace{0.16em}
\item[91) University of Warwick:]  \hfill \\\texttt{warwick.ac.uk}
\vspace{0.16em}
\item[92) University of Washington:]  \hfill \\\texttt{wustl.edu}
\vspace{0.16em}
\item[93) University of Wisconsin-Madison:]  \hfill \\\texttt{wisc.edu}
\vspace{0.16em}
\item[94) University of Zurich:]  \hfill \\\texttt{uzh.ch}
\vspace{0.16em}
\item[95) Utrecht University:]  \hfill \\\texttt{uu.nl}
\vspace{0.16em}
\item[96) Wageningen University \& Research:]  \hfill \\\texttt{wur.nl}
\vspace{0.16em}
\item[97) Washington University in St Louis:]  \hfill \\\texttt{uw.edu}
\vspace{0.16em}
\item[98) Yale University:]  \hfill \\\texttt{yale.edu}
\vspace{0.16em}
\item[99) École Polytechnique:]  \hfill \\\texttt{polytechnique.edu}
\vspace{0.16em}
\item[100) École Polytechnique Fédérale de Lausanne:]  \hfill \\\texttt{epfl.ch}
\end{description}
\vspace{2em}
\subsection{The United Kingdom}
\footnotesize
\begin{description}
\vspace{0.16em}
\item[1) Abertay University:]  \hfill \\\texttt{abertay.ac.uk} \\\texttt{tay.ac.uk}
\vspace{0.16em}
\item[2) Aberystwyth University:]  \hfill \\\texttt{aber.ac.uk}
\vspace{0.16em}
\item[3) Anglia Ruskin University:]  \hfill \\\texttt{anglia.ac.uk}
\vspace{0.16em}
\item[4) Aston University:]  \hfill \\\texttt{aston.ac.uk}
\vspace{0.16em}
\item[5) Bangor University:]  \hfill \\\texttt{bangor.ac.uk}
\vspace{0.16em}
\item[6) Bath Spa University:]  \hfill \\\texttt{bathspa.ac.uk}
\vspace{0.16em}
\item[7) Birkbeck University of London:]  \hfill \\\texttt{bbk.ac.uk} \\\texttt{birkbeck.ac.uk}
\vspace{0.16em}
\item[8) Birmingham City University:]  \hfill \\\texttt{bcu.ac.uk} \\\texttt{uce.ac.uk}
\vspace{0.16em}
\item[9) Bournemouth University:]  \hfill \\\texttt{bournemouth.ac.uk}
\vspace{0.16em}
\item[10) Brunel University London:]  \hfill \\\texttt{brunel.ac.uk}
\vspace{0.16em}
\item[11) Canterbury Christ Church University:]  \hfill \\\texttt{cant.ac.uk}
\vspace{0.16em}
\item[12) Cardiff Metropolitan University:]  \hfill \\\texttt{uwic.ac.uk} \\\texttt{cardiffmet.ac.uk}
\vspace{0.16em}
\item[13) Cardiff University:]  \hfill \\\texttt{cardiff.ac.uk} \\\texttt{cf.ac.uk}
\vspace{0.16em}
\item[14) City University of London:]  \hfill \\\texttt{city.ac.uk}
\vspace{0.16em}
\item[15) Coventry University:]  \hfill \\\texttt{coventry.ac.uk}
\vspace{0.16em}
\item[16) Cranfield University:]  \hfill \\\texttt{cranfield.ac.uk}
\vspace{0.16em}
\item[17) De Montfort University:]  \hfill \\\texttt{dmu.ac.uk}
\vspace{0.16em}
\item[18) Durham University:]  \hfill \\\texttt{dur.ac.uk} \\\texttt{durham.ac.uk}
\vspace{0.16em}
\item[19) Edinburgh Napier University:]  \hfill \\\texttt{napier.ac.uk}
\vspace{0.16em}
\item[20) Glasgow Caledonian University:]  \hfill \\\texttt{gcal.ac.uk}
\vspace{0.16em}
\item[21) Goldsmiths University of London:]  \hfill \\\texttt{gold.ac.uk} \\\texttt{goldsmiths.ac.uk}
\vspace{0.16em}
\item[22) Harper Adams University:]  \hfill \\\texttt{harper-adams.ac.uk}
\vspace{0.16em}
\item[23) Heriot-Watt University:]  \hfill \\\texttt{hw.ac.uk}
\vspace{0.16em}
\item[24) Imperial College London:]  \hfill \\\texttt{ic.ac.uk} \\\texttt{imperial.ac.uk}
\vspace{0.16em}
\item[25) Institute of Cancer Research:]  \hfill \\\texttt{icr.ac.uk}
\vspace{0.16em}
\item[26) Keele University:]  \hfill \\\texttt{keele.ac.uk}
\vspace{0.16em}
\item[27) King's College London:]  \hfill \\\texttt{kcl.ac.uk}
\vspace{0.16em}
\item[28) Kingston University:]  \hfill \\\texttt{king.ac.uk} \\\texttt{kingston.ac.uk}
\vspace{0.16em}
\item[29) Lancester University:]  \hfill \\\texttt{lancaster.ac.uk} \\\texttt{lancs.ac.uk}
\vspace{0.16em}
\item[30) Leeds Beckett University:]  \hfill \\\texttt{lmu.ac.uk}
\vspace{0.16em}
\item[31) Liverpool John Moores University:]  \hfill \\\texttt{livjm.ac.uk}
\vspace{0.16em}
\item[32) London Business School:]  \hfill \\\texttt{london.edu}
\vspace{0.16em}
\item[33) London Metropolitan University:]  \hfill \\\texttt{londonmet.ac.uk}
\vspace{0.16em}
\item[34) London School of Economics:]  \hfill \\\texttt{lse.ac.uk}
\vspace{0.16em}
\item[35) London School of Hygiene \& Tropical Medicine:]  \hfill \\\texttt{lshtm.ac.uk}
\vspace{0.16em}
\item[36) London South Bank University:]  \hfill \\\texttt{lsbu.ac.uk}
\vspace{0.16em}
\item[37) Loughborough University:]  \hfill \\\texttt{lboro.ac.uk} \\\texttt{loughborough.ac.uk}
\vspace{0.16em}
\item[38) Manchester Metropolitan University:]  \hfill \\\texttt{mmu.ac.uk}
\vspace{0.16em}
\item[39) Middlesex University:]  \hfill \\\texttt{mdx.ac.uk}
\vspace{0.16em}
\item[40) Newcastle University:]  \hfill \\\texttt{ncl.ac.uk} \\\texttt{newcastle.ac.uk}
\vspace{0.16em}
\item[41) Northumbria University:]  \hfill \\\texttt{northumbria.ac.uk} \\\texttt{unn.ac.uk}
\vspace{0.16em}
\item[42) Nottingham Trent University:]  \hfill \\\texttt{ntu.ac.uk}
\vspace{0.16em}
\item[43) Open University:]  \hfill \\\texttt{open.ac.uk}
\vspace{0.16em}
\item[44) Oxford Brookes University:]  \hfill \\\texttt{brookes.ac.uk}
\vspace{0.16em}
\item[45) Queen Margaret University:]  \hfill \\\texttt{qmuc.ac.uk}
\vspace{0.16em}
\item[46) Queen Mary University:]  \hfill \\\texttt{qmw.ac.uk} \\\texttt{qmul.ac.uk}
\vspace{0.16em}
\item[47) Queen's University Belfast:]  \hfill \\\texttt{qub.ac.uk}
\vspace{0.16em}
\item[48) Robert Gordon University:]  \hfill \\\texttt{rgu.ac.uk}
\vspace{0.16em}
\item[49) Royal Academy of Music:]  \hfill \\\texttt{ram.ac.uk}
\vspace{0.16em}
\item[50) Royal Central School of Speech and Drama:]  \hfill \\\texttt{cssd.ac.uk}
\vspace{0.16em}
\item[51) Royal College of Art:]  \hfill \\\texttt{rca.ac.uk}
\vspace{0.16em}
\item[52) Royal Holloway:]  \hfill \\\texttt{rhbnc.ac.uk} \\\texttt{rhul.ac.uk} \\\texttt{royalholloway.ac.uk}
\vspace{0.16em}
\item[53) School of Oriental and African Studies:]  \hfill \\\texttt{soas.ac.uk}
\vspace{0.16em}
\item[54) Sheffield Hallam University:]  \hfill \\\texttt{shu.ac.uk}
\vspace{0.16em}
\item[55) Staffordshire University:]  \hfill \\\texttt{staffs.ac.uk}
\vspace{0.16em}
\item[56) Swansea Uniiversity:]  \hfill \\\texttt{swan.ac.uk} \\\texttt{swansea.ac.uk}
\vspace{0.16em}
\item[57) Teesside University:]  \hfill \\\texttt{tees.ac.uk}
\vspace{0.16em}
\item[58) Ulster University:]  \hfill \\\texttt{ulst.ac.uk} \\\texttt{ulster.ac.uk}
\vspace{0.16em}
\item[59) University College London:]  \hfill \\\texttt{ioe.ac.uk} \\\texttt{ucl.ac.uk} \\\texttt{ulsop.ac.uk}
\vspace{0.16em}
\item[60) University of Aberdeen:]  \hfill \\\texttt{abdn.ac.uk}
\vspace{0.16em}
\item[61) University of Bath:]  \hfill \\\texttt{bath.ac.uk}
\vspace{0.16em}
\item[62) University of Bedfordshire:]  \hfill \\\texttt{beds.ac.uk}
\vspace{0.16em}
\item[63) University of Birmingham:]  \hfill \\\texttt{bham.ac.uk} \\\texttt{birmingham.ac.uk}
\vspace{0.16em}
\item[64) University of Bradford:]  \hfill \\\texttt{brad.ac.uk} \\\texttt{bradford.ac.uk}
\vspace{0.16em}
\item[65) University of Brighton:]  \hfill \\\texttt{brighton.ac.uk} \\\texttt{bton.ac.uk}
\vspace{0.16em}
\item[66) University of Bristol:]  \hfill \\\texttt{bris.ac.uk} \\\texttt{bristol.ac.uk}
\vspace{0.16em}
\item[67) University of Buckingham:]  \hfill \\\texttt{buckingham.ac.uk}
\vspace{0.16em}
\item[68) University of Cambridge:]  \hfill \\\texttt{cam.ac.uk}
\vspace{0.16em}
\item[69) University of Central Lancashire:]  \hfill \\\texttt{uclan.ac.uk}
\vspace{0.16em}
\item[70) University of Derby:]  \hfill \\\texttt{derby.ac.uk}
\vspace{0.16em}
\item[71) University of Dundee:]  \hfill \\\texttt{dundee.ac.uk}
\vspace{0.16em}
\item[72) University of East Anglia:]  \hfill \\\texttt{uea.ac.uk}
\vspace{0.16em}
\item[73) University of East London:]  \hfill \\\texttt{uel.ac.uk}
\vspace{0.16em}
\item[74) University of Edinburgh:]  \hfill \\\texttt{eca.ac.uk} \\\texttt{ed.ac.uk} \\\texttt{edinburgh.ac.uk}
\vspace{0.16em}
\item[75) University of Essex:]  \hfill \\\texttt{essex.ac.uk} \\\texttt{sx.ac.uk}
\vspace{0.16em}
\item[76) University of Exeter:]  \hfill \\\texttt{ex.ac.uk} \\\texttt{exeter.ac.uk}
\vspace{0.16em}
\item[77) University of Glasgow:]  \hfill \\\texttt{gla.ac.uk} \\\texttt{glasgow.ac.uk}
\vspace{0.16em}
\item[78) University of Gloucestershire:]  \hfill \\\texttt{glos.ac.uk}
\vspace{0.16em}
\item[79) University of Greenwich:]  \hfill \\\texttt{gre.ac.uk} \\\texttt{greenwich.ac.uk}
\vspace{0.16em}
\item[80) University of Hertfordshire:]  \hfill \\\texttt{herts.ac.uk}
\vspace{0.16em}
\item[81) University of Huddersfield:]  \hfill \\\texttt{hud.ac.uk}
\vspace{0.16em}
\item[82) University of Hull:]  \hfill \\\texttt{hull.ac.uk}
\vspace{0.16em}
\item[83) University of Kent:]  \hfill \\\texttt{kent.ac.uk} \\\texttt{ukc.ac.uk}
\vspace{0.16em}
\item[84) University of Leeds:]  \hfill \\\texttt{leeds.ac.uk}
\vspace{0.16em}
\item[85) University of Leicester:]  \hfill \\\texttt{le.ac.uk} \\\texttt{leicester.ac.uk}
\vspace{0.16em}
\item[86) University of Lincoln:]  \hfill \\\texttt{lincoln.ac.uk}
\vspace{0.16em}
\item[87) University of Liverpool:]  \hfill \\\texttt{liv.ac.uk} \\\texttt{liverpool.ac.uk}
\vspace{0.16em}
\item[88) University of Manchester:]  \hfill \\\texttt{man.ac.uk} \\\texttt{manchester.ac.uk} \\\texttt{mcc.ac.uk} \\\texttt{umist.ac.uk}
\vspace{0.16em}
\item[89) University of Northampton:]  \hfill \\\texttt{northampton.ac.uk}
\vspace{0.16em}
\item[90) University of Nottingham:]  \hfill \\\texttt{nott.ac.uk} \\\texttt{nottingham.ac.uk}
\vspace{0.16em}
\item[91) University of Oxford:]  \hfill \\\texttt{ox.ac.uk}
\vspace{0.16em}
\item[92) University of Plymouth:]  \hfill \\\texttt{plym.ac.uk} \\\texttt{plymouth.ac.uk}
\vspace{0.16em}
\item[93) University of Portsmouth:]  \hfill \\\texttt{port.ac.uk}
\vspace{0.16em}
\item[94) University of Reading:]  \hfill \\\texttt{rdg.ac.uk} \\\texttt{reading.ac.uk}
\vspace{0.16em}
\item[95) University of Roehampton:]  \hfill \\\texttt{roehampton.ac.uk}
\vspace{0.16em}
\item[96) University of Salford:]  \hfill \\\texttt{salford.ac.uk}
\vspace{0.16em}
\item[97) University of Sheffield:]  \hfill \\\texttt{shef.ac.uk} \\\texttt{sheffield.ac.uk}
\vspace{0.16em}
\item[98) University of South Wales:]  \hfill \\\texttt{newport.ac.uk}
\vspace{0.16em}
\item[99) University of Southampton:]  \hfill \\\texttt{soton.ac.uk} \\\texttt{southampton.ac.uk}
\vspace{0.16em}
\item[100) University of St Andrews:]  \hfill \\\texttt{st-and.ac.uk} \\\texttt{st-andrews.ac.uk}
\vspace{0.16em}
\item[101) University of Stirling:]  \hfill \\\texttt{stir.ac.uk}
\vspace{0.16em}
\item[102) University of Strathclyde:]  \hfill \\\texttt{strath.ac.uk}
\vspace{0.16em}
\item[103) University of Sunderland:]  \hfill \\\texttt{sund.ac.uk} \\\texttt{sunderland.ac.uk}
\vspace{0.16em}
\item[104) University of Surrey:]  \hfill \\\texttt{surrey.ac.uk}
\vspace{0.16em}
\item[105) University of Sussex:]  \hfill \\\texttt{sussex.ac.uk} \\\texttt{susx.ac.uk}
\vspace{0.16em}
\item[106) University of Wales Trinity Saint David:]  \hfill \\\texttt{lamp.ac.uk}
\vspace{0.16em}
\item[107) University of Warwick:]  \hfill \\\texttt{warwick.ac.uk}
\vspace{0.16em}
\item[108) University of West London:]  \hfill \\\texttt{tvu.ac.uk}
\vspace{0.16em}
\item[109) University of Westminster:]  \hfill \\\texttt{westminster.ac.uk} \\\texttt{wmin.ac.uk}
\vspace{0.16em}
\item[110) University of Wolverhampton:]  \hfill \\\texttt{wlv.ac.uk}
\vspace{0.16em}
\item[111) University of Worcester:]  \hfill \\\texttt{worc.ac.uk}
\vspace{0.16em}
\item[112) University of the Arts London:]  \hfill \\\texttt{arts.ac.uk} \\\texttt{linst.ac.uk}
\vspace{0.16em}
\item[113) University of the West of England Bristol:]  \hfill \\\texttt{uwe.ac.uk}
\vspace{0.16em}
\item[114) University of the West of Scotland:]  \hfill \\\texttt{paisley.ac.uk}
\vspace{0.16em}
\item[115) Universiy of York:]  \hfill \\\texttt{york.ac.uk}
\end{description}
\vspace{2em}
\subsection{The United States}
\footnotesize
\begin{description}
\vspace{0.16em}
\item[1) Air Force Institute of Tech. Graduate School of Eng, \& Mgmt.:]  \hfill \\\texttt{afit.edu}
\vspace{0.16em}
\item[2) American University:]  \hfill \\\texttt{american.edu}
\vspace{0.16em}
\item[3) Arizona State University:]  \hfill \\\texttt{asu.edu}
\vspace{0.16em}
\item[4) Arkansas State University:]  \hfill \\\texttt{astate.edu}
\vspace{0.16em}
\item[5) Auburn University:]  \hfill \\\texttt{auburn.edu}
\vspace{0.16em}
\item[6) Azusa Pacific University:]  \hfill \\\texttt{apu.edu}
\vspace{0.16em}
\item[7) Ball State University:]  \hfill \\\texttt{bsu.edu}
\vspace{0.16em}
\item[8) Baylor University:]  \hfill \\\texttt{baylor.edu}
\vspace{0.16em}
\item[9) Binghamton University:]  \hfill \\\texttt{binghamton.edu}
\vspace{0.16em}
\item[10) Boise State University:]  \hfill \\\texttt{boisestate.edu}
\vspace{0.16em}
\item[11) Boston College:]  \hfill \\\texttt{bc.edu}
\vspace{0.16em}
\item[12) Boston University:]  \hfill \\\texttt{bu.edu}
\vspace{0.16em}
\item[13) Bowling Green State University:]  \hfill \\\texttt{bgsu.edu}
\vspace{0.16em}
\item[14) Brandeis University:]  \hfill \\\texttt{brandeis.edu}
\vspace{0.16em}
\item[15) Brigham Young University:]  \hfill \\\texttt{byu.edu}
\vspace{0.16em}
\item[16) Brown University:]  \hfill \\\texttt{brown.edu}
\vspace{0.16em}
\item[17) CUNY City College:]  \hfill \\\texttt{cuny.edu}
\vspace{0.16em}
\item[18) California Institute of Technology:]  \hfill \\\texttt{caltech.edu}
\vspace{0.16em}
\item[19) Carnegie Mellon University:]  \hfill \\\texttt{cmu.edu}
\vspace{0.16em}
\item[20) Case Western Reserve University:]  \hfill \\\texttt{case.edu}
\vspace{0.16em}
\item[21) Catholic University of America:]  \hfill \\\texttt{catholic.edu}
\vspace{0.16em}
\item[22) Central Michigan University:]  \hfill \\\texttt{cmich.edu}
\vspace{0.16em}
\item[23) Chapman University:]  \hfill \\\texttt{chapman.edu}
\vspace{0.16em}
\item[24) Claremont Graduate University:]  \hfill \\\texttt{cgu.edu}
\vspace{0.16em}
\item[25) Clark Atlanta University:]  \hfill \\\texttt{cau.edu}
\vspace{0.16em}
\item[26) Clark University:]  \hfill \\\texttt{clarku.edu}
\vspace{0.16em}
\item[27) Clarkson University:]  \hfill \\\texttt{clarkson.edu}
\vspace{0.16em}
\item[28) Clemson University:]  \hfill \\\texttt{clemson.edu}
\vspace{0.16em}
\item[29) Cleveland State University:]  \hfill \\\texttt{csuohio.edu}
\vspace{0.16em}
\item[30) College of William and Mary:]  \hfill \\\texttt{wm.edu}
\vspace{0.16em}
\item[31) Colorado School of Mines:]  \hfill \\\texttt{mines.edu}
\vspace{0.16em}
\item[32) Colorado State University:]  \hfill \\\texttt{colostate.edu}
\vspace{0.16em}
\item[33) Columbia University:]  \hfill \\\texttt{columbia.edu}
\vspace{0.16em}
\item[34) Cornell University:]  \hfill \\\texttt{cornell.edu}
\vspace{0.16em}
\item[35) Dartmouth College:]  \hfill \\\texttt{dartmouth.edu}
\vspace{0.16em}
\item[36) DePaul University:]  \hfill \\\texttt{depaul.edu}
\vspace{0.16em}
\item[37) Delaware State University:]  \hfill \\\texttt{desu.edu}
\vspace{0.16em}
\item[38) Drexel University:]  \hfill \\\texttt{drexel.edu}
\vspace{0.16em}
\item[39) Duke University:]  \hfill \\\texttt{duke.edu}
\vspace{0.16em}
\item[40) Duquesne University:]  \hfill \\\texttt{duq.edu}
\vspace{0.16em}
\item[41) East Carolina University:]  \hfill \\\texttt{ecu.edu}
\vspace{0.16em}
\item[42) East Tennessee State University:]  \hfill \\\texttt{etsu.edu}
\vspace{0.16em}
\item[43) Eastern Michigan University:]  \hfill \\\texttt{emich.edu}
\vspace{0.16em}
\item[44) Emory University:]  \hfill \\\texttt{emory.edu}
\vspace{0.16em}
\item[45) Florida Agricultural and Mechanical University:]  \hfill \\\texttt{famu.edu}
\vspace{0.16em}
\item[46) Florida Atlantic University:]  \hfill \\\texttt{fau.edu}
\vspace{0.16em}
\item[47) Florida Institute of Technology:]  \hfill \\\texttt{fit.edu}
\vspace{0.16em}
\item[48) Florida International University:]  \hfill \\\texttt{fiu.edu}
\vspace{0.16em}
\item[49) Florida State University:]  \hfill \\\texttt{fsu.edu}
\vspace{0.16em}
\item[50) Fordham University:]  \hfill \\\texttt{fordham.edu}
\vspace{0.16em}
\item[51) Gallaudet University:]  \hfill \\\texttt{gallaudet.edu}
\vspace{0.16em}
\item[52) George Mason University:]  \hfill \\\texttt{gmu.edu}
\vspace{0.16em}
\item[53) George Washington University:]  \hfill \\\texttt{gwu.edu}
\vspace{0.16em}
\item[54) Georgetown University:]  \hfill \\\texttt{georgetown.edu}
\vspace{0.16em}
\item[55) Georgia Institute of Technology:]  \hfill \\\texttt{gatech.edu}
\vspace{0.16em}
\item[56) Georgia Southern University:]  \hfill \\\texttt{georgiasouthern.edu}
\vspace{0.16em}
\item[57) Georgia State University:]  \hfill \\\texttt{gsu.edu}
\vspace{0.16em}
\item[58) Hampton University:]  \hfill \\\texttt{hamptonu.edu}
\vspace{0.16em}
\item[59) Harvard University:]  \hfill \\\texttt{harvard.edu}
\vspace{0.16em}
\item[60) Howard University:]  \hfill \\\texttt{howard.edu}
\vspace{0.16em}
\item[61) Idaho State University:]  \hfill \\\texttt{isu.edu}
\vspace{0.16em}
\item[62) Illinois Institute of Technology:]  \hfill \\\texttt{iit.edu}
\vspace{0.16em}
\item[63) Illinois State University:]  \hfill \\\texttt{illinoisstate.edu}
\vspace{0.16em}
\item[64) Indiana University  Purdue University Indianapolis:]  \hfill \\\texttt{iupui.edu}
\vspace{0.16em}
\item[65) Indiana University Bloomington:]  \hfill \\\texttt{indiana.edu}
\vspace{0.16em}
\item[66) Iowa State University:]  \hfill \\\texttt{iastate.edu}
\vspace{0.16em}
\item[67) Jackson State University:]  \hfill \\\texttt{jsums.edu}
\vspace{0.16em}
\item[68) Johns Hopkins University:]  \hfill \\\texttt{jhu.edu}
\vspace{0.16em}
\item[69) Kansas State University:]  \hfill \\\texttt{k-state.edu}
\vspace{0.16em}
\item[70) Kennesaw State University:]  \hfill \\\texttt{kennesaw.edu}
\vspace{0.16em}
\item[71) Kent State University at Kent:]  \hfill \\\texttt{kent.edu}
\vspace{0.16em}
\item[72) Lehigh University:]  \hfill \\\texttt{lehigh.edu}
\vspace{0.16em}
\item[73) Louisiana State University and Agricultural \& Mechanical College:]  \hfill \\\texttt{lsu.edu}
\vspace{0.16em}
\item[74) Louisiana Tech University:]  \hfill \\\texttt{latech.edu}
\vspace{0.16em}
\item[75) Loyola Marymount University:]  \hfill \\\texttt{lmu.edu}
\vspace{0.16em}
\item[76) Loyola University Chicago:]  \hfill \\\texttt{luc.edu}
\vspace{0.16em}
\item[77) Marquette University:]  \hfill \\\texttt{marquette.edu}
\vspace{0.16em}
\item[78) Marshall University:]  \hfill \\\texttt{marshall.edu}
\vspace{0.16em}
\item[79) Massachusetts Institute of Technology:]  \hfill \\\texttt{mit.edu}
\vspace{0.16em}
\item[80) Mercer University:]  \hfill \\\texttt{mercer.edu}
\vspace{0.16em}
\item[81) Miami University:]  \hfill \\\texttt{miamioh.edu}
\vspace{0.16em}
\item[82) Michigan State University:]  \hfill \\\texttt{msu.edu}
\vspace{0.16em}
\item[83) Michigan Technological University:]  \hfill \\\texttt{mtu.edu}
\vspace{0.16em}
\item[84) Mississippi State University:]  \hfill \\\texttt{msstate.edu}
\vspace{0.16em}
\item[85) Missouri University of Science and Technology:]  \hfill \\\texttt{mst.edu}
\vspace{0.16em}
\item[86) Montana State University:]  \hfill \\\texttt{montana.edu}
\vspace{0.16em}
\item[87) Montclair State University:]  \hfill \\\texttt{montclair.edu}
\vspace{0.16em}
\item[88) Morgan State University:]  \hfill \\\texttt{morgan.edu}
\vspace{0.16em}
\item[89) New Jersey Institute of Technology:]  \hfill \\\texttt{njit.edu}
\vspace{0.16em}
\item[90) New Mexico State University:]  \hfill \\\texttt{nmsu.edu}
\vspace{0.16em}
\item[91) New York University:]  \hfill \\\texttt{nyu.edu}
\vspace{0.16em}
\item[92) North Carolina A \& T State University:]  \hfill \\\texttt{ncat.edu}
\vspace{0.16em}
\item[93) North Carolina State University:]  \hfill \\\texttt{ncsu.edu}
\vspace{0.16em}
\item[94) North Dakota State University:]  \hfill \\\texttt{ndsu.edu}
\vspace{0.16em}
\item[95) Northeastern University:]  \hfill \\\texttt{northeastern.edu}
\vspace{0.16em}
\item[96) Northern Arizona University:]  \hfill \\\texttt{nau.edu}
\vspace{0.16em}
\item[97) Northern Illinois University:]  \hfill \\\texttt{niu.edu}
\vspace{0.16em}
\item[98) Northwestern University:]  \hfill \\\texttt{northwestern.edu}
\vspace{0.16em}
\item[99) Nova Southeastern University:]  \hfill \\\texttt{nova.edu}
\vspace{0.16em}
\item[100) Oakland University:]  \hfill \\\texttt{oakland.edu}
\vspace{0.16em}
\item[101) Ohio State University:]  \hfill \\\texttt{osu.edu}
\vspace{0.16em}
\item[102) Ohio University-Main Campus:]  \hfill \\\texttt{ohio.edu}
\vspace{0.16em}
\item[103) Oklahoma State University:]  \hfill \\\texttt{okstate.edu}
\vspace{0.16em}
\item[104) Old Dominion University:]  \hfill \\\texttt{odu.edu}
\vspace{0.16em}
\item[105) Oregon State University:]  \hfill \\\texttt{oregonstate.edu}
\vspace{0.16em}
\item[106) Pennsylvania State University:]  \hfill \\\texttt{psu.edu}
\vspace{0.16em}
\item[107) Ponce Health Sciences University:]  \hfill \\\texttt{psm.edu}
\vspace{0.16em}
\item[108) Portland State University:]  \hfill \\\texttt{pdx.edu}
\vspace{0.16em}
\item[109) Princeton University:]  \hfill \\\texttt{princeton.edu}
\vspace{0.16em}
\item[110) Purdue University:]  \hfill \\\texttt{purdue.edu}
\vspace{0.16em}
\item[111) Rensselaer Polytechnic Institute:]  \hfill \\\texttt{rpi.edu}
\vspace{0.16em}
\item[112) Rice University:]  \hfill \\\texttt{rice.edu}
\vspace{0.16em}
\item[113) Rochester Institute of Technology:]  \hfill \\\texttt{rit.edu}
\vspace{0.16em}
\item[114) Rockefeller University:]  \hfill \\\texttt{rockefeller.edu}
\vspace{0.16em}
\item[115) Rowan University:]  \hfill \\\texttt{rowan.edu}
\vspace{0.16em}
\item[116) Rutgers University:]  \hfill \\\texttt{rutgers.edu}
\vspace{0.16em}
\item[117) SUNY College of Environmental Science and Forestry:]  \hfill \\\texttt{esf.edu}
\vspace{0.16em}
\item[118) Saint Louis University:]  \hfill \\\texttt{slu.edu}
\vspace{0.16em}
\item[119) San Diego State University:]  \hfill \\\texttt{sdsu.edu}
\vspace{0.16em}
\item[120) Seton Hall University:]  \hfill \\\texttt{shu.edu}
\vspace{0.16em}
\item[121) South Dakota State University:]  \hfill \\\texttt{sdstate.edu}
\vspace{0.16em}
\item[122) Southern Illinois University:]  \hfill \\\texttt{siue.edu} \\\texttt{siu.edu} \\\texttt{siumed.edu}
\vspace{0.16em}
\item[123) Southern Methodist University:]  \hfill \\\texttt{smu.edu}
\vspace{0.16em}
\item[124) Stanford University:]  \hfill \\\texttt{stanford.edu}
\vspace{0.16em}
\item[125) Stevens Institute of Technology:]  \hfill \\\texttt{stevens.edu}
\vspace{0.16em}
\item[126) Stony Brook University:]  \hfill \\\texttt{stonybrook.edu}
\vspace{0.16em}
\item[127) Syracuse University:]  \hfill \\\texttt{syracuse.edu}
\vspace{0.16em}
\item[128) Temple University:]  \hfill \\\texttt{temple.edu}
\vspace{0.16em}
\item[129) Tennessee State University:]  \hfill \\\texttt{tnstate.edu}
\vspace{0.16em}
\item[130) Tennessee Technological University:]  \hfill \\\texttt{tntech.edu}
\vspace{0.16em}
\item[131) Texas A\&M University:]  \hfill \\\texttt{tamu.edu}
\vspace{0.16em}
\item[132) Texas A\&M UniversityCorpus Christi:]  \hfill \\\texttt{tamucc.edu}
\vspace{0.16em}
\item[133) Texas A\&M UniversityKingsville:]  \hfill \\\texttt{tamuk.edu}
\vspace{0.16em}
\item[134) Texas Christian University:]  \hfill \\\texttt{tcu.edu}
\vspace{0.16em}
\item[135) Texas Southern University:]  \hfill \\\texttt{tsu.edu}
\vspace{0.16em}
\item[136) Texas State University:]  \hfill \\\texttt{txstate.edu}
\vspace{0.16em}
\item[137) Texas Tech University:]  \hfill \\\texttt{ttu.edu}
\vspace{0.16em}
\item[138) The New School:]  \hfill \\\texttt{newschool.edu}
\vspace{0.16em}
\item[139) Thomas Jefferson University:]  \hfill \\\texttt{jefferson.edu}
\vspace{0.16em}
\item[140) Tufts University:]  \hfill \\\texttt{tufts.edu}
\vspace{0.16em}
\item[141) Tulane University:]  \hfill \\\texttt{tulane.edu}
\vspace{0.16em}
\item[142) University of Akron Main Campus:]  \hfill \\\texttt{uakron.edu}
\vspace{0.16em}
\item[143) University of Alabama:]  \hfill \\\texttt{ua.edu}
\vspace{0.16em}
\item[144) University of Alabama in Huntsville:]  \hfill \\\texttt{uah.edu}
\vspace{0.16em}
\item[145) University of Alabama, Birmingham:]  \hfill \\\texttt{uab.edu}
\vspace{0.16em}
\item[146) University of Alaska Fairbanks:]  \hfill \\\texttt{uaf.edu}
\vspace{0.16em}
\item[147) University of Arizona:]  \hfill \\\texttt{arizona.edu}
\vspace{0.16em}
\item[148) University of Arkansas:]  \hfill \\\texttt{uark.edu}
\vspace{0.16em}
\item[149) University of Arkansas, Little Rock:]  \hfill \\\texttt{ualr.edu}
\vspace{0.16em}
\item[150) University of California, Berkeley:]  \hfill \\\texttt{berkeley.edu}
\vspace{0.16em}
\item[151) University of California, Davis:]  \hfill \\\texttt{ucdavis.edu}
\vspace{0.16em}
\item[152) University of California, Irvine:]  \hfill \\\texttt{uci.edu}
\vspace{0.16em}
\item[153) University of California, Los Angeles:]  \hfill \\\texttt{ucla.edu}
\vspace{0.16em}
\item[154) University of California, Merced:]  \hfill \\\texttt{ucmerced.edu}
\vspace{0.16em}
\item[155) University of California, Riverside:]  \hfill \\\texttt{ucr.edu}
\vspace{0.16em}
\item[156) University of California, San Diego:]  \hfill \\\texttt{ucsd.edu}
\vspace{0.16em}
\item[157) University of California, Santa Barbara:]  \hfill \\\texttt{ucsb.edu}
\vspace{0.16em}
\item[158) University of California, Santa Cruz:]  \hfill \\\texttt{ucsc.edu}
\vspace{0.16em}
\item[159) University of Central Florida:]  \hfill \\\texttt{ucf.edu}
\vspace{0.16em}
\item[160) University of Chicago:]  \hfill \\\texttt{uchicago.edu}
\vspace{0.16em}
\item[161) University of Cincinnati:]  \hfill \\\texttt{uc.edu}
\vspace{0.16em}
\item[162) University of Colorado, Boulder:]  \hfill \\\texttt{colorado.edu}
\vspace{0.16em}
\item[163) University of Colorado, Colorado Springs:]  \hfill \\\texttt{uccs.edu}
\vspace{0.16em}
\item[164) University of Colorado, Denver:]  \hfill \\\texttt{ucdenver.edu}
\vspace{0.16em}
\item[165) University of Connecticut:]  \hfill \\\texttt{uconn.edu}
\vspace{0.16em}
\item[166) University of Dayton:]  \hfill \\\texttt{udayton.edu}
\vspace{0.16em}
\item[167) University of Delaware:]  \hfill \\\texttt{udel.edu}
\vspace{0.16em}
\item[168) University of Denver:]  \hfill \\\texttt{du.edu}
\vspace{0.16em}
\item[169) University of Florida:]  \hfill \\\texttt{ufl.edu}
\vspace{0.16em}
\item[170) University of Georgia:]  \hfill \\\texttt{uga.edu}
\vspace{0.16em}
\item[171) University of Hawaii:]  \hfill \\\texttt{hawaii.edu}
\vspace{0.16em}
\item[172) University of Houston:]  \hfill \\\texttt{uh.edu}
\vspace{0.16em}
\item[173) University of Idaho:]  \hfill \\\texttt{uidaho.edu}
\vspace{0.16em}
\item[174) University of Illinois, Chicago:]  \hfill \\\texttt{uic.edu}
\vspace{0.16em}
\item[175) University of Illinois, Urbana Champaign:]  \hfill \\\texttt{illinois.edu}
\vspace{0.16em}
\item[176) University of Iowa:]  \hfill \\\texttt{uiowa.edu}
\vspace{0.16em}
\item[177) University of Kansas:]  \hfill \\\texttt{ku.edu}
\vspace{0.16em}
\item[178) University of Kentucky:]  \hfill \\\texttt{uky.edu}
\vspace{0.16em}
\item[179) University of Louisiana, Lafayette:]  \hfill \\\texttt{louisiana.edu}
\vspace{0.16em}
\item[180) University of Louisville:]  \hfill \\\texttt{louisville.edu}
\vspace{0.16em}
\item[181) University of Maine:]  \hfill \\\texttt{umaine.edu}
\vspace{0.16em}
\item[182) University of Maryland, Baltimore County:]  \hfill \\\texttt{umbc.edu}
\vspace{0.16em}
\item[183) University of Maryland, College Park:]  \hfill \\\texttt{umd.edu}
\vspace{0.16em}
\item[184) University of Maryland, Eastern Shore:]  \hfill \\\texttt{umes.edu}
\vspace{0.16em}
\item[185) University of Massachusetts Amherst:]  \hfill \\\texttt{umass.edu}
\vspace{0.16em}
\item[186) University of Massachusetts Boston:]  \hfill \\\texttt{umb.edu}
\vspace{0.16em}
\item[187) University of Massachusetts Dartmouth:]  \hfill \\\texttt{umassd.edu}
\vspace{0.16em}
\item[188) University of Massachusetts Lowell:]  \hfill \\\texttt{uml.edu}
\vspace{0.16em}
\item[189) University of Memphis:]  \hfill \\\texttt{memphis.edu}
\vspace{0.16em}
\item[190) University of Miami:]  \hfill \\\texttt{miami.edu}
\vspace{0.16em}
\item[191) University of Michigan:]  \hfill \\\texttt{umich.edu}
\vspace{0.16em}
\item[192) University of Minnesota:]  \hfill \\\texttt{umn.edu}
\vspace{0.16em}
\item[193) University of Mississippi:]  \hfill \\\texttt{olemiss.edu}
\vspace{0.16em}
\item[194) University of Missouri:]  \hfill \\\texttt{missouri.edu}
\vspace{0.16em}
\item[195) University of Missouri, Kansas City:]  \hfill \\\texttt{umkc.edu}
\vspace{0.16em}
\item[196) University of Missouri, St. Louis:]  \hfill \\\texttt{umsl.edu}
\vspace{0.16em}
\item[197) University of Montana:]  \hfill \\\texttt{umt.edu}
\vspace{0.16em}
\item[198) University of Nebraska, Lincoln:]  \hfill \\\texttt{unl.edu}
\vspace{0.16em}
\item[199) University of Nebraska, Omaha:]  \hfill \\\texttt{unomaha.edu}
\vspace{0.16em}
\item[200) University of Nevada, Las Vegas:]  \hfill \\\texttt{unlv.edu}
\vspace{0.16em}
\item[201) University of Nevada, Reno:]  \hfill \\\texttt{unr.edu}
\vspace{0.16em}
\item[202) University of New England:]  \hfill \\\texttt{une.edu}
\vspace{0.16em}
\item[203) University of New Hampshire:]  \hfill \\\texttt{unh.edu}
\vspace{0.16em}
\item[204) University of New Mexico:]  \hfill \\\texttt{unm.edu}
\vspace{0.16em}
\item[205) University of New Orleans:]  \hfill \\\texttt{uno.edu}
\vspace{0.16em}
\item[206) University of North Carolina, Chapel Hill:]  \hfill \\\texttt{unc.edu}
\vspace{0.16em}
\item[207) University of North Carolina, Charlotte:]  \hfill \\\texttt{uncc.edu}
\vspace{0.16em}
\item[208) University of North Carolina, Greensboro:]  \hfill \\\texttt{uncg.edu}
\vspace{0.16em}
\item[209) University of North Carolina, Wilmington:]  \hfill \\\texttt{uncw.edu}
\vspace{0.16em}
\item[210) University of North Dakota:]  \hfill \\\texttt{und.edu}
\vspace{0.16em}
\item[211) University of North Texas:]  \hfill \\\texttt{unt.edu}
\vspace{0.16em}
\item[212) University of Notre Dame:]  \hfill \\\texttt{nd.edu}
\vspace{0.16em}
\item[213) University of Oklahoma:]  \hfill \\\texttt{ou.edu}
\vspace{0.16em}
\item[214) University of Oregon:]  \hfill \\\texttt{uoregon.edu}
\vspace{0.16em}
\item[215) University of Pennsylvania:]  \hfill \\\texttt{upenn.edu}
\vspace{0.16em}
\item[216) University of Pittsburgh:]  \hfill \\\texttt{pitt.edu}
\vspace{0.16em}
\item[217) University of Puerto Rico:]  \hfill \\\texttt{upr.edu}
\vspace{0.16em}
\item[218) University of Rhode Island:]  \hfill \\\texttt{uri.edu}
\vspace{0.16em}
\item[219) University of Rochester:]  \hfill \\\texttt{rochester.edu}
\vspace{0.16em}
\item[220) University of San Diego:]  \hfill \\\texttt{sandiego.edu}
\vspace{0.16em}
\item[221) University of South Alabama:]  \hfill \\\texttt{southalabama.edu}
\vspace{0.16em}
\item[222) University of South Carolina:]  \hfill \\\texttt{sc.edu}
\vspace{0.16em}
\item[223) University of South Dakota:]  \hfill \\\texttt{usd.edu}
\vspace{0.16em}
\item[224) University of South Florida:]  \hfill \\\texttt{usf.edu}
\vspace{0.16em}
\item[225) University of Southern California:]  \hfill \\\texttt{usc.edu}
\vspace{0.16em}
\item[226) University of Southern Mississippi:]  \hfill \\\texttt{usm.edu}
\vspace{0.16em}
\item[227) University of Tennessee:]  \hfill \\\texttt{utk.edu}
\vspace{0.16em}
\item[228) University of Texas Rio Grande Valley:]  \hfill \\\texttt{utrgv.edu}
\vspace{0.16em}
\item[229) University of Texas, Arlington:]  \hfill \\\texttt{uta.edu}
\vspace{0.16em}
\item[230) University of Texas, Austin:]  \hfill \\\texttt{utexas.edu}
\vspace{0.16em}
\item[231) University of Texas, Dallas:]  \hfill \\\texttt{utdallas.edu}
\vspace{0.16em}
\item[232) University of Texas, El Paso:]  \hfill \\\texttt{utep.edu}
\vspace{0.16em}
\item[233) University of Texas, San Antonio:]  \hfill \\\texttt{utsa.edu}
\vspace{0.16em}
\item[234) University of Toledo:]  \hfill \\\texttt{utoledo.edu}
\vspace{0.16em}
\item[235) University of Tulsa:]  \hfill \\\texttt{utulsa.edu}
\vspace{0.16em}
\item[236) University of Utah:]  \hfill \\\texttt{utah.edu}
\vspace{0.16em}
\item[237) University of Vermont:]  \hfill \\\texttt{uvm.edu}
\vspace{0.16em}
\item[238) University of Virginia:]  \hfill \\\texttt{virginia.edu}
\vspace{0.16em}
\item[239) University of Washington:]  \hfill \\\texttt{washington.edu}
\vspace{0.16em}
\item[240) University of Wisconsin, Madison:]  \hfill \\\texttt{wisc.edu}
\vspace{0.16em}
\item[241) University of Wisconsin, Milwaukee:]  \hfill \\\texttt{uwm.edu}
\vspace{0.16em}
\item[242) University of Wyoming:]  \hfill \\\texttt{uwyo.edu}
\vspace{0.16em}
\item[243) University, Albany:]  \hfill \\\texttt{albany.edu}
\vspace{0.16em}
\item[244) University, Buffalo:]  \hfill \\\texttt{buffalo.edu}
\vspace{0.16em}
\item[245) Utah State University:]  \hfill \\\texttt{usu.edu}
\vspace{0.16em}
\item[246) Vanderbilt University:]  \hfill \\\texttt{vanderbilt.edu}
\vspace{0.16em}
\item[247) Villanova University:]  \hfill \\\texttt{villanova.edu}
\vspace{0.16em}
\item[248) Virginia Commonwealth University:]  \hfill \\\texttt{vcu.edu}
\vspace{0.16em}
\item[249) Virginia Polytechnic Institute and State University:]  \hfill \\\texttt{vt.edu}
\vspace{0.16em}
\item[250) Wake Forest University:]  \hfill \\\texttt{wfu.edu}
\vspace{0.16em}
\item[251) Washington State University:]  \hfill \\\texttt{wsu.edu}
\vspace{0.16em}
\item[252) Washington University in St. Louis:]  \hfill \\\texttt{wustl.edu}
\vspace{0.16em}
\item[253) Wayne State University:]  \hfill \\\texttt{wayne.edu}
\vspace{0.16em}
\item[254) West Virginia University:]  \hfill \\\texttt{wvu.edu}
\vspace{0.16em}
\item[255) Western Michigan University:]  \hfill \\\texttt{wmich.edu}
\vspace{0.16em}
\item[256) Wichita State University:]  \hfill \\\texttt{wichita.edu}
\vspace{0.16em}
\item[257) Worcester Polytechnic Institute:]  \hfill \\\texttt{wpi.edu}
\vspace{0.16em}
\item[258) Wright State University:]  \hfill \\\texttt{wright.edu}
\vspace{0.16em}
\item[259) Yale University:]  \hfill \\\texttt{yale.edu}
\vspace{0.16em}
\item[260) Yeshiva University:]  \hfill \\\texttt{yu.edu}
\end{description}

\begin{figure*}[p!]
	\begin{center}
		\includegraphics[width=\textwidth]{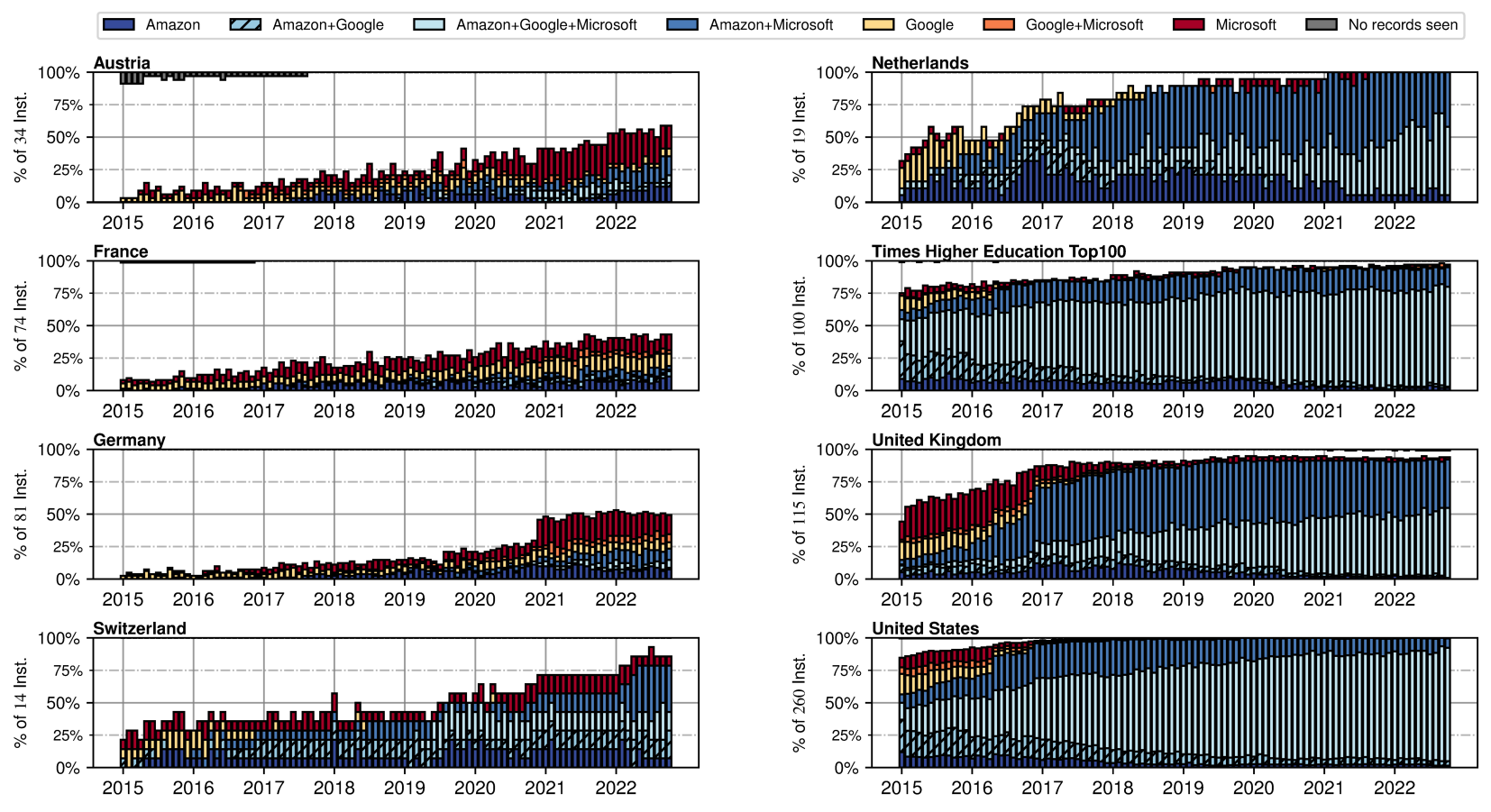}
		\caption{Universities' use of the `Big Three' cloud providers (Amazon, Google, Microsoft) from January 2015 to \lastdate.}
		\label{fig:full_public_clouds}
	\end{center}
\end{figure*}

\begin{figure*}[p!]
	\begin{center}
		\includegraphics[width=\textwidth]{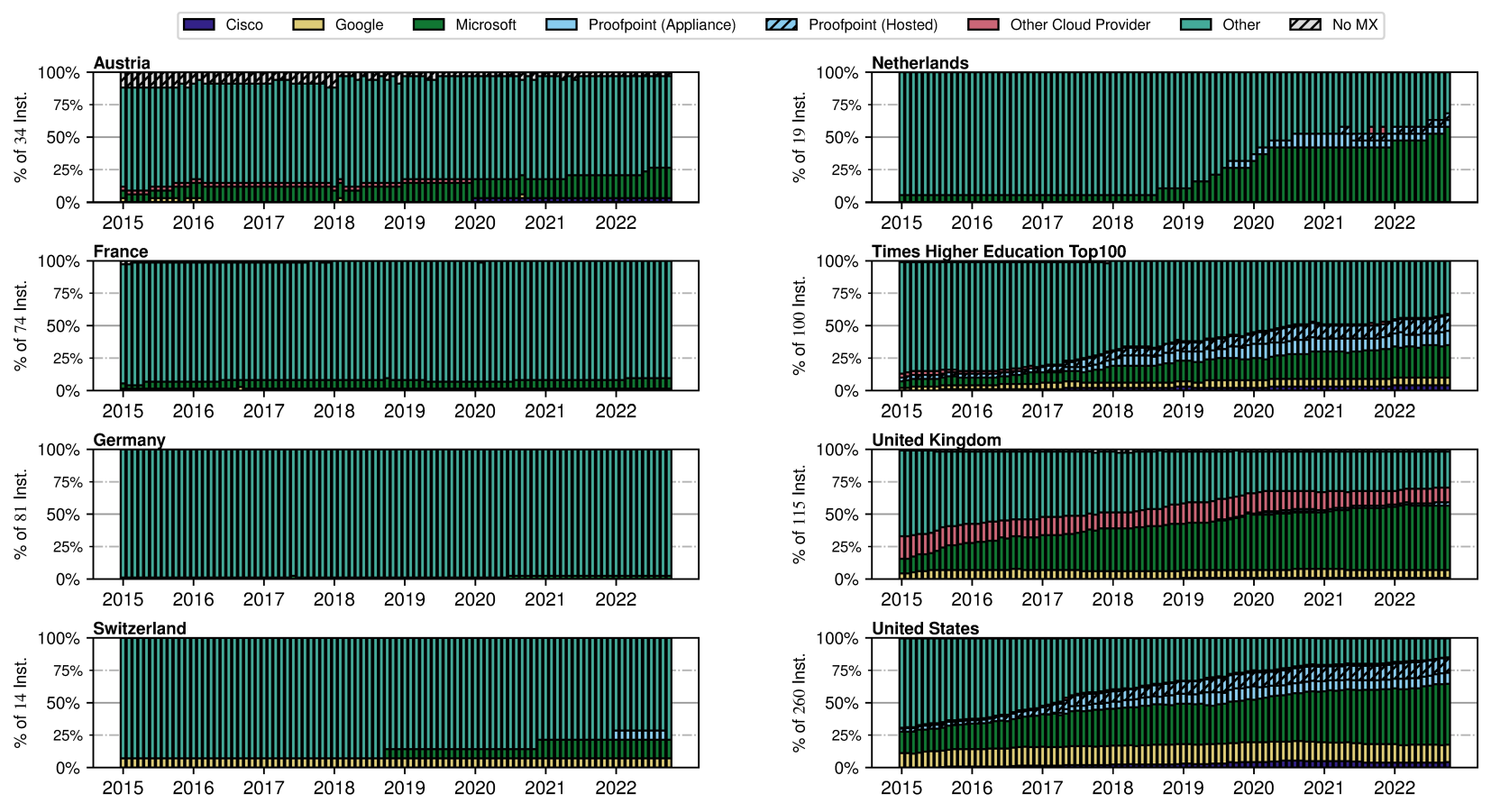}
		\caption{Email providers used by universities from January 2015 to \lastdate.}
		\label{fig:full_mail}
	\end{center}
\end{figure*}

\begin{figure*}[p!]
	\begin{center}
		\includegraphics[width=\textwidth]{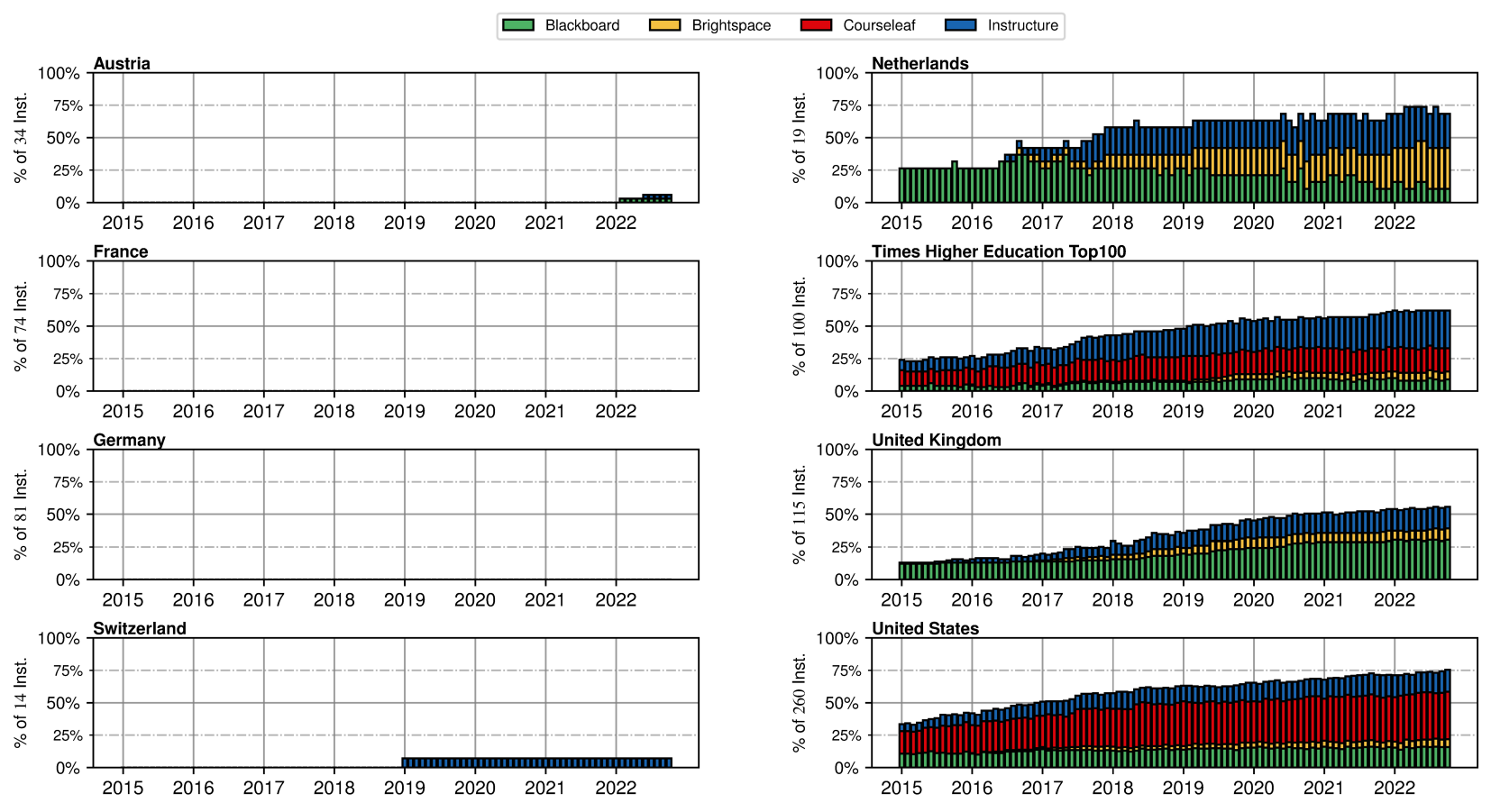}
		\caption{Cloud-hosted Learning Management Systems (LMS) used by universities from January 2015 to \lastdate.}
		\label{fig:full_lms}
	\end{center}
\end{figure*}

\begin{figure*}[p!]
	\begin{center}
		\includegraphics[width=\textwidth]{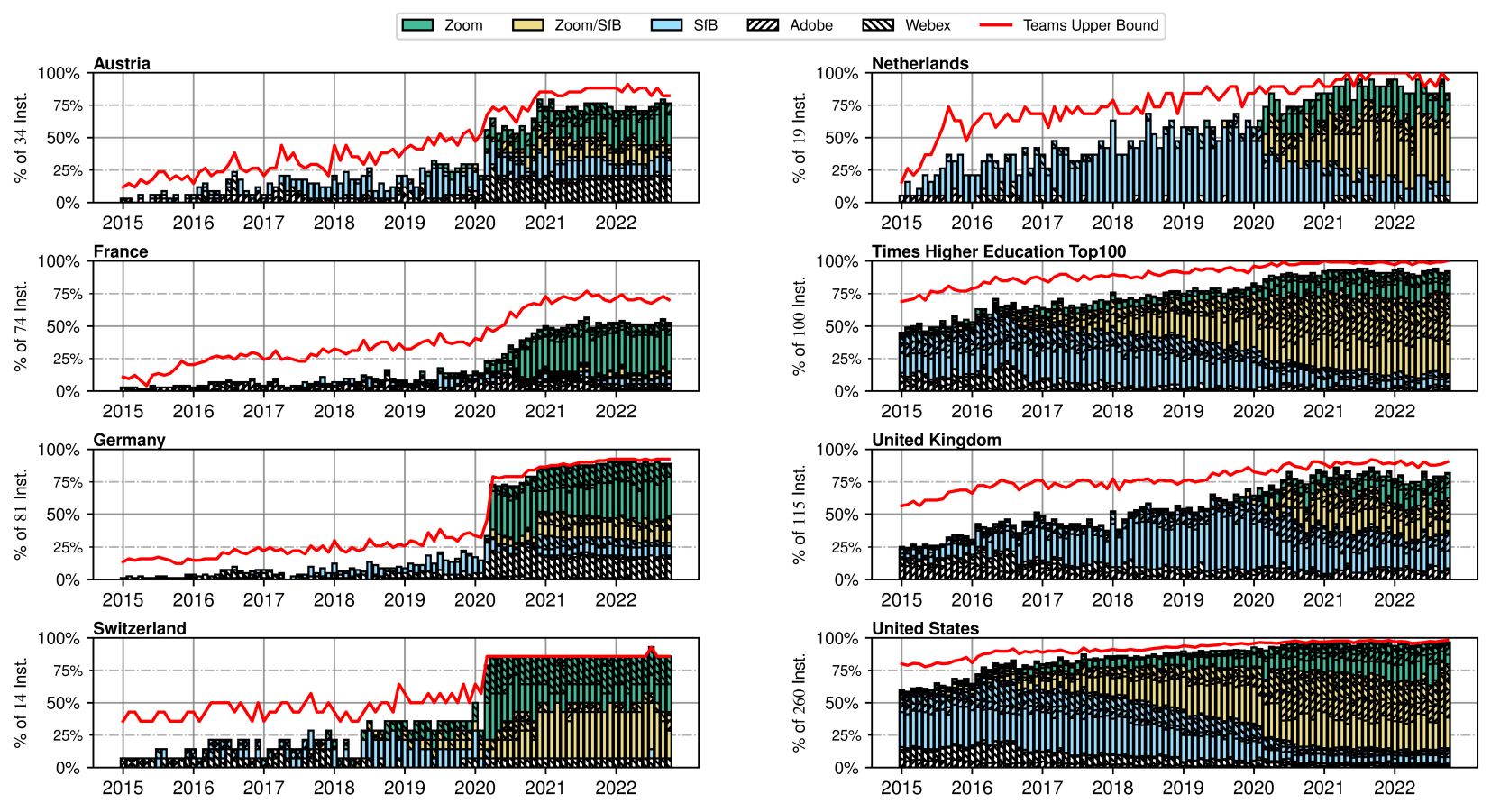}
		\caption{Video chat tools used by universities from January 2015 to \lastdate.}
		\label{fig:full_zoom}
	\end{center}
\end{figure*}

\begin{figure*}[p!]
	\begin{center}
		\includegraphics[width=\textwidth]{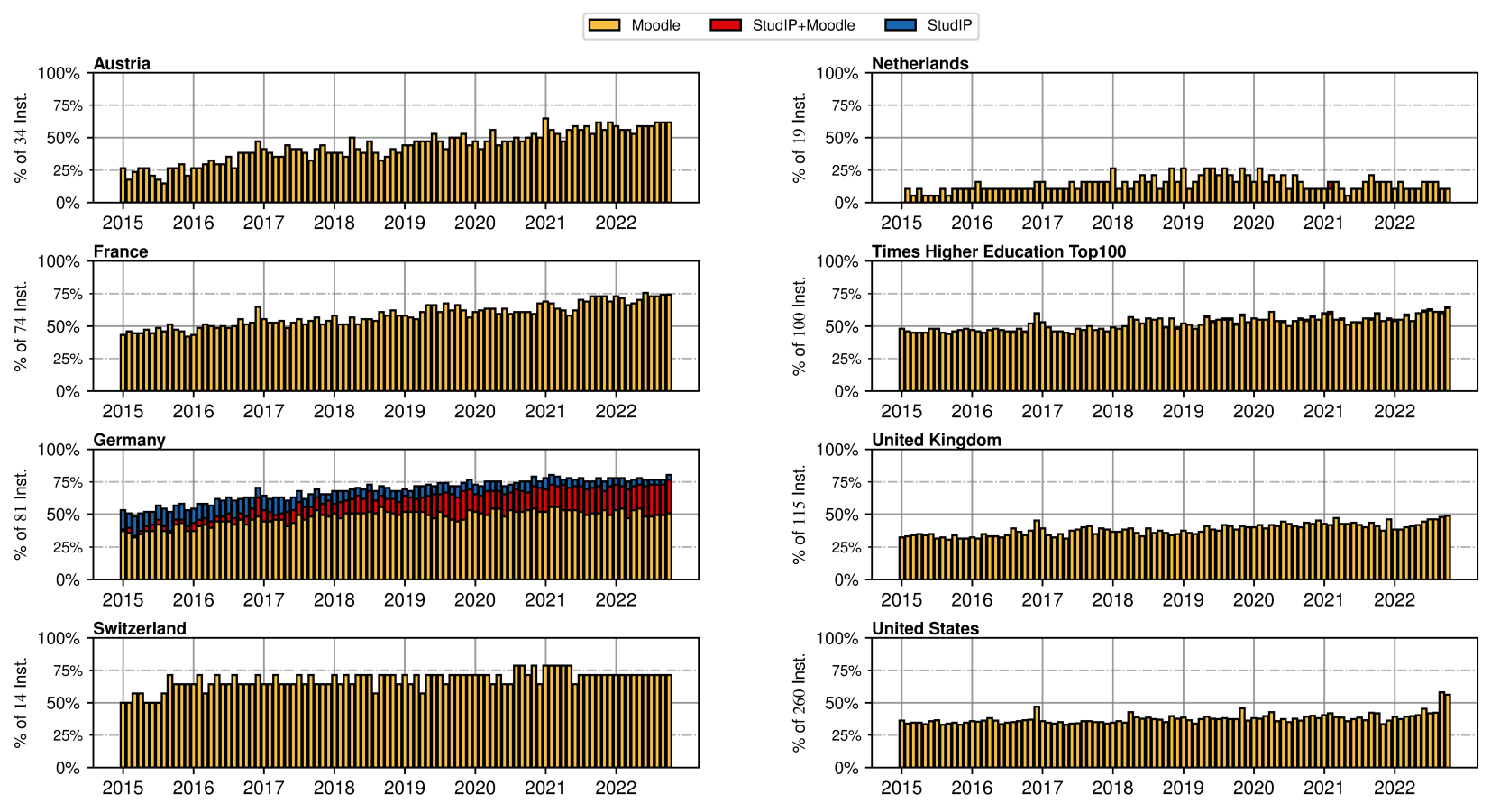}
		\caption{Universities with at least one name containing `moodle' or `studip' from January 2015 to \lastdate.}
		\label{fig:full_studip}
	\end{center}
\end{figure*}

\begin{figure*}[p!]
	\begin{center}
		\includegraphics[width=\textwidth]{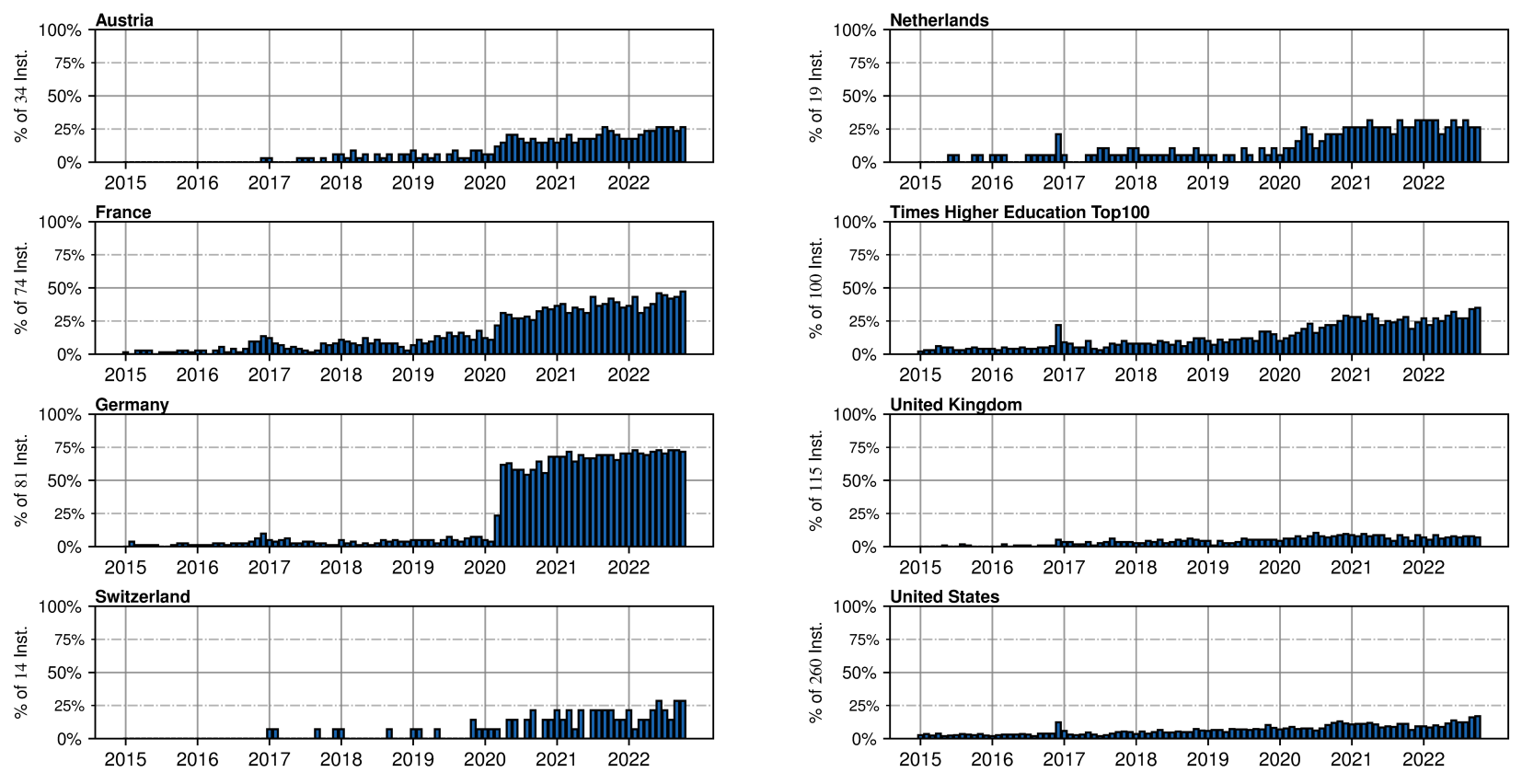}
		\caption{Universities with at least one BigBlueButton-related DNS entry from January 2015 to \lastdate.}
		\label{fig:full_bbb}
	\end{center}
\end{figure*}

\end{document}